\documentclass[preprint,showpacs,showkeys,preprintnumbers,amsmath,amssymb,nofootinbib,aps]{revtex4}

\usepackage{amssymb}
\usepackage{amsmath}
\usepackage{graphicx}
\usepackage{dcolumn}
\usepackage{bm}
\usepackage{epsfig}
\usepackage{hyperref}

\newcommand\beq{\begin{eqnarray}}
\newcommand\eeq{\end{eqnarray}}

\newcommand\figwidth{.48\textwidth}
\newcommand\Eq[1]{Eq.~\ref{eq:#1}}
\newcommand\Fig[1]{Fig.~\ref{fig:#1}}
\newcommand\Sec[1]{Sec.~\ref{sec:#1}}
\newcommand\Appendix[1]{Appendix~\ref{sec:#1}}

\newcommand\Tab[1]{Table~\ref{tab:#1}}
\newcommand\bfx{\mathbf x}
\newcommand\bfy{\mathbf y}
\newcommand\bfz{\mathbf z}
\newcommand\bfw{\mathbf w}

\newcommand\bfj{\mathbf j}

\newcommand\bfp{\mathbf p}

\newcommand\bfq{\mathbf q}

\newcommand\calA{\mathcal A}
\newcommand\calB{\mathcal B}

\newcommand\calL{\mathcal L}
\newcommand\calC{\mathcal C}
\newcommand\calO{\mathcal O}
\newcommand\calT{\mathcal T}
\newcommand\calU{\mathcal U}

\newcommand\calP{\mathcal P}

\begin{document}

\title{Lattice Monte Carlo calculations for unitary fermions in a finite box} 

\begin{abstract}
We perform lattice Monte Carlo simulations for up to $66$ unitary fermions in a finite box using a highly improved lattice action for nonrelativistic spin $1/2$ fermions.
We obtain a value of $0.366^{+0.016}_{-0.011}$ for the Bertsch parameter, defined as the energy of the unitary Fermi gas measured in units of the free gas energy in the thermodynamic limit.
In addition, for up to four unitary fermions, we compute the spectrum of the lattice theory by exact diagonalization of the transfer matrix projected onto irreducible representations of the octahedral group for small to moderate size lattices, providing an independent check of our few-body simulation results.
We compare our exact numerical and simulation results for the spectrum to benchmark studies of other research groups, as well as perform an extended analysis of our lattice action improvement scheme, including an analysis of the errors associated with higher partial waves and finite temporal discretization.
\end{abstract}

\author{Michael G. Endres}
\email{endres@riken.jp}
\affiliation{Theoretical Research Division, RIKEN Nishina Center, Wako, Saitama 351-0198, Japan}
\author{David B. Kaplan}
\email{dbkaplan@uw.edu}
\affiliation{Institute for Nuclear Theory, Box 351550, Seattle, WA 98195-1550, USA}
\author{Jong-Wan Lee}
\email{jongwan@post.kek.jp}
\affiliation{KEK Theory Center, High Energy Accelerator Research Organization (KEK), Tsukuba 305-0801, Japan}
\author{Amy N. Nicholson}
\email{amynn@umd.edu}
\affiliation{Department of Physics, University of Maryland, College Park MD 20742-4111, USA}


\preprint{INT-PUB-12-011}
\preprint{KEK-CP-266}
\preprint{RIKEN-QHP-16}
\preprint{UM-DOE/ER/40762-517}


\date{\today}

\maketitle

\section{Introduction}
\label{sec:introduction}

Unitary fermions have gained widespread attention from theorists, particularly since their successful creation in experiments involving trapped, ultracold atoms.
The universal nature of this system promises applications to many fields; for example, it has been suggested as an expansion point for an effective field theory for nuclear physics.
Numerical studies have been essential to the progress in our knowledge about unitary fermions due to the strongly coupled nature of the interaction, which renders standard perturbative techniques unreliable.

Unitarity corresponds to the idealized limit in which the $s$-wave scattering length becomes infinite and the interaction range vanishes, or equivalently, the two-particle s-wave scattering phase shift $\delta_0=\pi/2$.
For a homogeneous system of two-component fermions at unitarity, the only relevant scale is the density $\rho = N/V$, with $N=N^\downarrow+N^\uparrow$.
Consequently, the ground state energy is related to that of noninteracting fermions by
\begin{eqnarray}
E(\rho)=\xi E_{Free}(\rho),
\label{eq:bertsch}
\end{eqnarray}
where $E_{Free}(\rho) = 3N E_{F}(\rho)/5$, $E_{F}(\rho) = k_F^2/(2M)$ is the Fermi energy, and $k_{F} = (3\pi^2\rho)^{1/3}$ the Fermi momentum.
The dimensionless parameter $\xi$, known as the Bertsch parameter \cite{baker2000mbx}, is of particular interest because it is the unique parameter which relates zero temperature thermodynamic quantities between the unitary and free Fermi gas.
Several experimental groups have measured the Bertsch parameter using a variety techniques involving ultra-cold trapped atoms \cite{%
O'Hara13122002,%
PhysRevA.68.011401,%
PhysRevLett.93.050401,%
PhysRevLett.92.120401,%
Kinast25022005,%
Partridge27012006,%
PhysRevLett.95.250404,%
PhysRevLett.97.220406,%
springerlink:10.1007/s10909-008-9850-2,%
Navon07052010,%
2010Natur.463.1057N,Ku12012012%
}. 
On the theoretical side, in addition to analytical calculations \cite{%
PhysRevB.55.15153,%
PhysRevC.60.054311,%
2000nucl.th..10066S,%
PhysRevA.63.043606,%
PhysRevLett.93.100404,%
2005NuPhA.762...82S,%
PhysRevA.72.041603,%
PhysRevLett.97.050403,%
PhysRevA.75.023610,%
PhysRevA.75.043614,%
PhysRevA.75.043605,%
PhysRevA.75.063617,%
0256-307X-24-7-011,%
PhysRevA.79.013627%
}, there has also been a substantial number of numerical studies of unitary fermions from the microscopic theory using quantum Monte Carlo and other techniques \cite{%
PhysRevLett.91.050401,%
PhysRevA.70.043602,%
PhysRevLett.93.200404,%
PhysRevLett.95.060401,%
PhysRevC.73.015202,%
PhysRevLett.96.090404,%
PhysRevB.73.115112,%
1367-2630-9-6-163,%
springerlink:10.1140/epja/i2008-10537-2,%
PhysRevC.78.024001,%
PhysRevA.78.023625,%
PhysRevC.79.054003,%
PhysRevLett.103.210403,%
PhysRevA.83.041601,%
PhysRevLett.106.235303,%
PhysRevA.84.023615,%
PhysRevA.84.061602%
}.
Many of these studies use the Bertsch parameter as a benchmark calculation.

At very low energies and large scattering length the detailed structure of the inter-particle potential become irrelevant, and the system is well-described by an effective theory for spin $1/2$ fermions with a zero-range contact interaction:
\begin{eqnarray}
\calL = \psi^{\dagger} \left( \partial_\tau  - \frac{\nabla^2}{2M}\right) \psi + C_0 \left( \psi^\dagger \psi \right)^2\ .
\label{eq:eft}
\end{eqnarray}
Here, the field $\psi = (\psi_\uparrow, \psi_\downarrow)$ is a two-component spinor and the coupling $C_0$ is tuned to an $\calO(1)$ critical value, determined nonperturbatively by an exact, analytic evaluation of the two-particle scattering amplitude $\calA^{-1}(p) = p\cot\delta(p) -i p$ at zero external momentum $p = \sqrt{M E}$ \cite{Kaplan:1998tg,Kaplan:1998we}.

Perhaps the simplest lattice construction of \Eq{eft} at finite chemical potential was developed by Ref. \cite{Chen:2003vy}, and employs a nonpropagating real scalar field $\phi$ of mass $m^2 = 1/C_0$  to induce two-body interactions between fermions of opposite spin through the type of interaction $\phi \psi^\dagger \psi$.
\footnote{This technique is often used in lattice simulations involving quartic fermion interactions, and is commonly known as the Hubbard-Stratonovich transformation \cite{1957SPhD....2..416S,PhysRevLett.3.77}, named after its inventors.}
This lattice construction was shown to be free of the fermion ``sign problem'' at finite density, a problem that is notorious for rendering numerical simulations of certain fermionic theories at finite density impractical, due to the presence of a complex effective action for the scalar field $\phi$.

We have since developed a highly improved lattice theory based upon the construction of Ref. \cite{Chen:2003vy}, allowing us to study up to 70 unitary fermions confined to a harmonic trap \cite{Endres:2011er} and up to 66 unitary fermions confined to a finite box.
Results of the latter study are described in detail in the proceeding sections. 
Here, we summarize some of the salient features of our construction:
\begin{enumerate}
\item We employ open boundary conditions in the time direction, preventing fermion propagation from wrapping ``around the world.''
This choice eliminates $\phi$-dependence in the fermion-determinant obtained upon integrating out the fermion degrees of freedom in the path-integral, and therefore yields a trivial effective action for the scalar auxiliary field and eliminates the need for importance sampling in the simulation.
\item Due to our choice of temporal boundary conditions, simulations must be performed at zero temperature and zero chemical potential.
The energy of systems at finite density are obtained by studying the long-time exponential fall-off of multi-fermion correlation functions.
\item We use a continuum single particle dispersion relation for fermions, thus reducing lattice discretization errors.
\item We introduce Galilean invariant derivative interactions which allow us to eliminate higher order terms in the effective range expansion for $p\cot \delta_0(p)$.
We are thus able to simulate fermions close to the unitary limit even at small lattice volumes.
\end{enumerate}

Several recent papers have indicated that lattice Monte Carlo methods can be affected by large systematic errors due to a finite filling factor \cite{2011arXiv1109.5327C, 2011arXiv1111.5079D, 2012PhRvA..85a3640P}.
For a given number of particles, this systematic error corresponds to a dependence of the Bertsch parameter on the number of lattice sites.
The improvements referred to above are crucial in reducing these errors.
In this paper, we provide an extensive discussion of the discretization errors which remain after improvement, based upon an analysis of the Symanzik action \cite{Symanzik1983187,Symanzik1983205}.

The organization of this paper is as follows:
In \Sec{lattice_construction}, we summarize our highly improved lattice construction for numerically simulating untrapped unitary fermions and provide details regarding the construction of multi-fermion correlation functions used to extract the ground state energy of the system.
In addition, the method used to tune two-body couplings to the unitary point is briefly reviewed.
The details of the lattice construction are discussed at greater length in Ref. \cite{Endres:2011er}; here we only provide the main ingredients.
In \Sec{exact_results}, we present exact spectrum results for the two- and three-fermion systems on a lattice at finite volume, and use those results to try to understand the systematic errors associated with our construction due to interactions from higher partial waves and temporal discretization errors; a description of how the multi-fermion transfer matrix is constructed is provided in the appendix, along with the construction of projection operators onto the irreducible representations (irreps) of the octahedral group.
In \Sec{simulation_results}, we summarize the techniques used for extracting the energies of up to 66 unitary fermions in a finite box, and present simulation results for the few- and many-body system, including an estimate for the Bertsch parameter.

\section{Lattice construction}
\label{sec:lattice_construction}

\subsection{Action}
\label{sec:lattice_construction.action}

We consider a highly improved lattice theory for an interacting system of nonrelativistic spin $1/2$ fermions of mass $M$ on a $T\times L^3$ Euclidean space-time lattice with temporal extent $T$ and spatial extent $L$.
The sites of the lattice are labeled by integers $\tau\in[0,T)$ in the time direction and $x_j \in [0,L)$ in the space directions with $j=(1,2,3)$.
Throughout this work we impose open boundary conditions in the time direction and periodic boundary conditions in the space directions.
Unless otherwise noted, we measure all quantities with dimensions of energy in units of the inverse temporal lattice constant $b_\tau^{-1}$ and all quantities with dimensions of length in units of the spatial lattice constant $b_s$.

The lattice action for this theory is given by \cite{Endres:2011er}:
\begin{eqnarray}
S = \sum_\sigma \psi_\sigma^\dagger K \psi_{\sigma}\ ,
\label{eq:lat_action}
\end{eqnarray}
where $\psi_\sigma$ and  $\psi^\dagger_\sigma$ are Grassmann valued $(T\times L^3)$-dimensional vectors associated with each spin component $\sigma = (\uparrow,\downarrow)$, and $K$ is a $(T\times L^3)$-dimensional matrix of commuting numbers.
The matrix elements of the fermion operator $K$ are given by
\begin{eqnarray}
K_{\bfp, \bfp'}(\tau,\tau') = \delta_{\tau,\tau'} D_{\bfp,\bfp'} + \delta_{\tau,\tau'-1} X_{\bfp,\bfp'}(T - \tau')\ ,
\label{eq:Kmat}
\end{eqnarray}
where
\begin{eqnarray}
D_{\bfp,\bfp'} = \delta_{\bfp,\bfp'} \times \left\{ \begin{array}{ll}  e^{\bfp^2/2M}\ , & |\bfp| < \Lambda \\ \infty\ , & |\bfp| \ge \Lambda  \end{array}  \right.
\label{eq:Dmat}
\end{eqnarray}
and
\begin{eqnarray}
X_{\bfp,\bfp'}(\tau) = \delta_{\bfp,\bfp'} + C^{1/2}(\bfp-\bfp') \tilde\phi_{\bfp-\bfp'}(\tau)\ .
\label{eq:Xmat}
\end{eqnarray}
The matrix elements are labeled by a time coordinate $\tau$ and by a three-momentum $p_j = 2\pi n_j /L$, where $n_j \in [-L/2,L/2)$ for a periodic spatial lattice (assuming even $L$).

Two-body interactions are induced by the periodic field $\tilde\phi_\bfp(\tau)$, defined as the spatial Fourier transform of a random auxiliary field $\phi_\bfx(\tau)$ in position space, which satisfies the conditions:
\begin{eqnarray}
\langle \phi_\bfx(\tau) \rangle = 0 \ ,\qquad \langle \phi_\bfx(\tau) \phi_{\bfx'}(\tau') \rangle = \delta_{\bfx,\bfx'} \delta_{\tau,\tau'}\ .
\label{eq:auxvevs}
\end{eqnarray}
Throughout this work, we take $\phi_\bfx(\tau)$ to be a $Z_2$-valued field with probability distribution $\rho(\phi) = (\delta_{1,\phi} + \delta_{-1,\phi})/2$ for all $\bfx$ and $\tau$.
\footnote{Although we use $Z_2$ auxiliary fields in this work, Gaussian distributed $\phi$ fields with probability distribution $\rho(\phi) = e^{-\phi^2/2}$ for every $\bfx$ and $\tau$ would work equally well.}
The two-body coupling $C(\bfp)$ is a periodic function of momenta and is given by the operator expansion:
\begin{eqnarray}
C(\bfp) = \frac{4\pi}{M} \sum_{n=0}^{N_\calO-1} C_{2n} \calO_{2n}(\bfp)\ ,
\label{eq:Cfunc}
\end{eqnarray}
up to some fixed order $N_\calO-1$.
Throughout this work we use the operator basis:
\begin{eqnarray}
\calO_{2n}(\bfp) =  M_0^n \left(1-e^{-\hat\bfp^2 /M_0} \right)^n\ ,
\label{eq:Ofunc}
\end{eqnarray}
where $\hat\bfp$ is taken to be a periodic function of $\bfp$ and satisfies the relation $\hat\bfp^2 = \bfp^2 \theta(\Lambda-|\bfp|) + \Lambda^2  \theta(|\bfp| - \Lambda)$ for $\bfp\in BZ$, with $BZ$ denoting the first Brillouin zone.
Note that at low momenta, the operators satisfy the low energy expansion $\calO_{2n}(\bfp) = \bfp^{2n} \left[ 1 + \calO(\bfp^2) \right]$,  irrespective of the mass parameter $M_0$.
For simplicity, we choose $M=M_0$, although these mass parameters need not be the same.

The partition function for the lattice theory is obtained by integrating out the fermionic degrees of freedom, yielding a path-integral over the auxiliary fields given by $Z = \int [d\phi] \rho(\phi) \det K$.
Due to the upper block tri-diagonal form of the fermion operator, one may show that the fermion determinant is given by $\det K = \det D^T$, and is independent of the auxiliary field.
Importance sampling for the field $\phi$ in a Monte Carlo simulation thus reduces to generating random field configurations distributed according to the trivial distribution $\rho(\phi)$.
Hence, a full simulation of the theory is the same as a quenched simulation.

Upon integrating out the fermion degrees of freedom, expectation values of operators involving the fermion fields $\psi_\sigma$ and $\psi^\dagger_\sigma$, such as multi-fermion correlation functions, reduce to the expectation values of appropriately contracted fermion propagators $K^{-1}$.
A fermion propagator, which evolves a single fermion state from time slice zero to time slice $\tau$ over a given background auxiliary field, may be expressed using the simple recursive formula:
\begin{eqnarray}
K^{-1}(\tau;0) = D^{-1} X(\tau-1) K^{-1}(\tau-1;0)
\label{eq:prop}
\end{eqnarray}
with $K^{-1}(0;0) = D^{-1}$.

\subsection{Observables}
\label{sec:lattice_construction.observables}

Multi-fermion correlation functions are constructed using sources formed from a direct product of single particle states $|\alpha_i, \sigma \rangle$ labeled by spin $\sigma = (\uparrow, \downarrow)$ and wavefunction quantum numbers $\alpha_i$, where $i=1\ldots,N$.
Separable sources of this form are typically favored in Monte Carlo simulations due to the nature of the algorithms used.
In this study, we choose eigenstates of the noninteracting Hamiltonian (i.e., $\alpha = \bfp$) as our single particle states, and use the free $N$-fermion ground state as our source by filling states in momentum space up to the Fermi surface.
Specifically, we consider an initial state of the form $| N/2,\uparrow \rangle \otimes \calP | N/2, \downarrow\rangle$, where
\begin{eqnarray}
| N, \sigma \rangle = \epsilon_{i_1,\ldots,i_N} |\bfp_{i_1} ,\sigma \rangle \otimes \ldots \otimes  |\bfp_{i_N}, \sigma \rangle\ ,
\label{eq:source_state}
\end{eqnarray}
and $\calP$ is a parity flip operator satisfying $\calP|\bfp_i, \sigma\rangle = |-\bfp_i, \sigma \rangle $.
The specific momenta $\bfp_i$ used in this construction are tabulated in \Tab{sources}.
Note that the parity flip operator is necessary to ensure that the total momentum of the initial state vanishes when $N/2$ is not equal to one of the closed shell values (e.g.,  1, 7, 19, 27, 33, \ldots).

\begin{table}
\caption{%
\label{tab:sources}%
Single fermion sources used in untrapped fermion simulations (choosing $\alpha=\bfp$).
}
\begin{ruledtabular}
\begin{tabular}{c|c|| c|c|| c|c|| c|c|| c|c}
$i$ & $\bfp_i$ & $i$ & $\bfp_i$ & $i$ & $\bfp_i$ & $i$ & $\bfp_i$ &  $i$ & $\bfp_i$  \\
\hline
  1 & (  0,  0,  0) & 2 & (  0,  0,  1) &  8 & (  0,  1,  1) & 20 & (  1,  1,  1) & 28   & (  0,  0,  2) \\
    &               & 3 & (  0,  0, -1) &  9 & (  0,  1, -1) & 21 & (  1,  1, -1) & 29   & (  0,  0, -2) \\
    &               & 4 & (  0,  1,  0) & 10 & (  0, -1,  1) & 22 & (  1, -1,  1) & 30   & (  0,  2,  0) \\
    &               & 5 & (  0, -1,  0) & 11 & (  0, -1, -1) & 23 & (  1, -1, -1) & 31   & (  0, -2,  0) \\
    &               & 6 & (  1,  0,  0) & 12 & (  1,  0,  1) & 24 & ( -1,  1,  1) & 32   & (  2,  0,  0) \\
    &               & 7 & ( -1,  0,  0) & 13 & (  1,  0, -1) & 25 & ( -1,  1, -1) & 33   & ( -2,  0,  0) \\
    &               &   &               & 14 & (  1,  1,  0) & 26 & ( -1, -1,  1) &      &               \\
    &               &   &               & 15 & (  1, -1,  0) & 27 & ( -1, -1, -1) &      &               \\
    &               &   &               & 16 & ( -1,  0,  1) &    &               &      &               \\
    &               &   &               & 17 & ( -1,  0, -1) &    &               &      &               \\
    &               &   &               & 18 & ( -1,  1,  0) &    &               &      &               \\
    &               &   &               & 19 & ( -1, -1,  0) &    &               &      &               \\
\end{tabular}
\end{ruledtabular}
\end{table}

Although the form of source wavefunctions is constrained by separability due to the nature of our algorithm, greater freedom is allowed for the construction of sink wavefunctions.
In order to maximize the overlap with the unitary Fermi gas ground state, sink wavefunctions are constructed by considering a direct product of $N/2$ spin-paired two particle states, $|\Psi\rangle\otimes\ldots\otimes|\Psi\rangle$, following the approach of Refs. \cite{PhysRevLett.91.050401,Endres:2011er}.
The two-particle states used in such a construction are given by 
\begin{eqnarray}
|\Psi \rangle = \frac{1}{V} \sum_{\bfp\in BZ} \tilde \Psi(\bfp) \, |\bfp, \uparrow \rangle \otimes |-\bfp, \downarrow\rangle\ ,
\label{eq:sink_state}
\end{eqnarray}
with the two-particle pairing wave-function given by
\begin{eqnarray}
\tilde\Psi(\bfp) = \left\{ \begin{array}{ll}
                   \frac{ e^{-\beta |\bfp|}}{\bfp^2}\ , & \bfp \neq 0 \\
                   \Psi_0\ ,                            & \bfp = 0 \ .
                   \end{array} \right. 
\label{eq:sink_wavefunc}
\end{eqnarray}
The functional form of the pairing wave function $\tilde\Psi(\bfp)$ in momentum space was obtained by considering the Fourier transform of the continuum pairing wavefunction $\Psi(\bfx_{rel}) = e^{-\beta^{-1}|\bfx_{rel}|}/|\bfx_{rel}|$ for two fermions in the unitary regime in position space, where $\bfx_{rel} = \bfx_\uparrow - \bfx_\downarrow$ is the relative coordinate between the pair.
Note that in the continuum and away from unitarity, $\beta$ is identified with the scattering length $a$.
Here, we introduce $\beta$ and $\Psi_0$, the zero momentum component of the wave function, as tunable free parameters which may be varied in order to improve the overlap with the ground-state wave function; the specific values used in our simulations will be discussed in \Sec{simulation_results}.

Given the sources and sinks defined above, the multi-fermion correlation function for  $N^\uparrow = N^\downarrow = N/2$ fermions used in our simulations is finally given by:
\begin{eqnarray}
\calC_{N}(\tau) = \langle \det{ S^{\downarrow\uparrow}(\tau)} \rangle\ ,
\label{eq:slater2}
\end{eqnarray}
where
\begin{eqnarray}
S^{\downarrow\uparrow}_{i,j}(\tau) = \sum_{\bfq\in BZ} \tilde\Psi(\bfq) \langle \bfq | K^{-1}(\tau,0) | \bfp_i, \downarrow \rangle \langle -\bfq | K^{-1}(\tau,0) |\bfp_j, \uparrow \rangle\ ,
\label{eq:slater_mat}
\end{eqnarray}
and $S^{\downarrow\uparrow}$ is an $N/2$-dimensional matrix with indices $i,j = 1,\ldots,N/2$.
Note that the determinant appearing in \Eq{slater2} ensures that fermions of the same species are properly antisymmetrized.

\subsection{Parameter tuning}
\label{sec:lattice_construction.parameter_tuning}

The lattice action defined in \Eq{lat_action} contains $N_\calO$ couplings $C_{2n}$ ($n=0,\ldots,N_\calO-1$) which must be tuned to scattering data.
The details of our tuning procedure are described at length in Ref. \cite{Endres:2011er}, and is similar to the method used in Ref. \cite{2005PhRvC..72b4006L}; here we summarize the main points.
The couplings $C_{2n}$ are tuned by matching the lowest $N_\calO$ $s$-wave eigenvalues $\lambda = e^{-E}$ of the two-body transfer matrix defined on the lattice at finite volume onto the lowest $N_\calO$ solutions to L\"uscher's formula, given by \cite{Luscher:1985dn,Luscher:1986pf,Luscher:1990ux,Beane:2003da}:
\begin{eqnarray}
p\cot \delta_0(p) = \frac{1}{\pi L} S(\eta)\ ,\qquad \eta = \left( \frac{p L}{2\pi} \right)^2\ ,
\label{eq:luschers_formula}
\end{eqnarray}
where $S(\eta)$ is the three-dimensional Zeta function:
\begin{eqnarray}
S(\eta) = \lim_{\Lambda\to\infty} \left[  \sum_{|\bfj|<\Lambda} \frac{1}{|\bfj|^2-\eta} - 4\pi \Lambda \right]\ .
\end{eqnarray}
In the unitary limit, the solutions to L\"uscher's formula are just the roots of the function $S(\eta)$, which may be easily calculated numerically and are related to the energies by $p = \sqrt{ME}$.
It was shown in Ref. \cite{Endres:2011er} that this tuning procedure may be used to systematically eliminate the leading $N_\calO$ terms in the effective range expansion
\beq
p\cot\delta_0(p) = -\frac{1}{a} + \frac{1}{2} \sum_{n=1} r_{n-1} p^{2n}
\label{eq:effective_range_expansion}
\eeq
up to negligible residual contributions to $r_{n-1}$ for $n<N_\calO$.

\section{Exact results for few-body states}
\label{sec:exact_results}

In \Appendix{transfer_matrices} we derive an exact expression for the $N$-particle transfer matrix, as well as projection operators onto the center of mass (c.m.) frame and the irreducible representations (irreps) of the octahedral group $O_h$.
Using these results, we have performed exact diagonalization of the $N=1+1$, $N=2+1$ and $N=2+2$ unitary fermion transfer matrices on small to moderate lattice volumes.
\footnote{%
The dimensionality of the transfer matrix after projecting onto the c.m. frame and irrep $r$ scales roughly like $ \left(\Lambda L/2\pi\right)^{3(N-1)}/48$.
For a fixed amount of computer memory or computing time, the maximum allowable lattice size decreases sharply with increasing $N$.
}
Armed with exact numerical results for the eigenstates and energies, we investigate the systematic errors associated with partial wave scattering from nonzero angular momentum interactions in the two-body sector, as well as finite volume and lattice spacing artifacts in the three-body sector.
In addition to these studies, we perform consistency checks with our numerical simulation in both the three- and four-body sectors.
Those results, however, will be presented in \Sec{simulation_results}.

\subsection{Two unitary fermions}

As discussed in \Sec{lattice_construction.parameter_tuning}, the unitary limit corresponds to the limit that $p \cot\delta_0 = 0$ for all momenta, where $\delta_0$ is the $s$-wave scattering phase shift.
One tacitly assumes that in addition to this limit, the effects of scattering in higher partial waves are negligible compared to those of $s$-wave scattering.
The latter is a condition that becomes arbitrarily valid at low energies (or equivalently, in the dilute limit), when $s$-wave scattering becomes the dominant contribution to the total scattering cross-section.

In its simplest form, our lattice action involves only a single four-fermion contact interaction (i.e., taking $N_\calO=1$ in \Eq{Cfunc}), which allows for scattering in $s$-waves, but not in the higher partial waves.
In this case, there is only a single parameter $C_0$, which may be used to tune the $s$-wave scattering length to infinity.
However, since the effective range and higher order shape parameters appearing in the effective range expansion for $p\cot\delta_0$ are typically $\calO(1)$ in lattice units, extrapolations in the lattice volume are required to eliminate systematic errors associated with those parameters.
One may reduce the systematic errors by introducing higher derivative lattice operators (i.e., taking $N_\calO>1$) as described in \Sec{lattice_construction.parameter_tuning}.
In doing so, however, one in turn introduces systematic errors associated with higher partial-wave interactions which are attributed to the fact that the lattice interactions no longer involve fermions at a single lattice site, but also all the neighboring sites as well.

In light of these considerations, it is important for us to estimate the size of systematic effects attributed to higher partial wave interactions in our lattice theory when $N_\calO>1$.
We proceed by studying the energy eigenvalues associated with higher partial-waves in the two-body sector, and particularly study their deviation from the energies expected for two noninteracting fermions.
It is well known that the eigenstates of the lattice theory transform as irreps of the octahedral group, and that those irreps contain specific angular momentum components in the infinite volume and continuum limits \cite{Mandula:1983ut}.
The decomposition of orbital angular momentum eigenstates into the various irreps of the hypercubic group are provided for reference in \Tab{decomposition}.
Note that even $\ell$ correspond to the positive parity irreps, whereas odd $\ell$ correspond to negative parity irreps.

\begin{table}
\caption{%
\label{tab:decomposition}%
Decomposition of angular momentum eigenstates into irreps of $O_h$.
Also indicated are the $O_h$ irreps containing $\ell$ as their lowest-lying state (LLS).
}
\begin{ruledtabular}
\begin{tabular}{ccc}
$\ell$ & decomposition & irreps containing $l$ as LLS \\ 
\hline
0 & $A_1^+$                                                                                          & $A_1^+$        \\
1 & $T_1^-$                                                                                          & $T_1^-$        \\
2 & $E^+\oplus T_2^+$                                                                                & $E^+, T_2^+$   \\
3 & $A_2^-\oplus T_1^-\oplus T_2^-$                                                                  & $A_2^-, T_2^-$ \\
4 & $A_1^+\oplus E^+\oplus T_1^+\oplus T_2^+$                                                        & $T_1^+$        \\
5 & $E^- \oplus T_1^- \oplus T_1^- \oplus T_2^-$                                                     & $E^-$          \\
6 & $A_1^+\oplus A_2^+ \oplus E^+ \oplus T_1^+ \oplus T_2^+\oplus T_2^+$                             & $A_2^+$        \\
9 & $A_1^- \oplus A_2^- \oplus E^- \oplus T_1^- \oplus T_1^- \oplus T_1^- \oplus T_2^- \oplus T_2^-$ & $A_1^-$        \\
\end{tabular}
\end{ruledtabular}
\end{table}

\begin{figure}
\includegraphics[width=\figwidth]{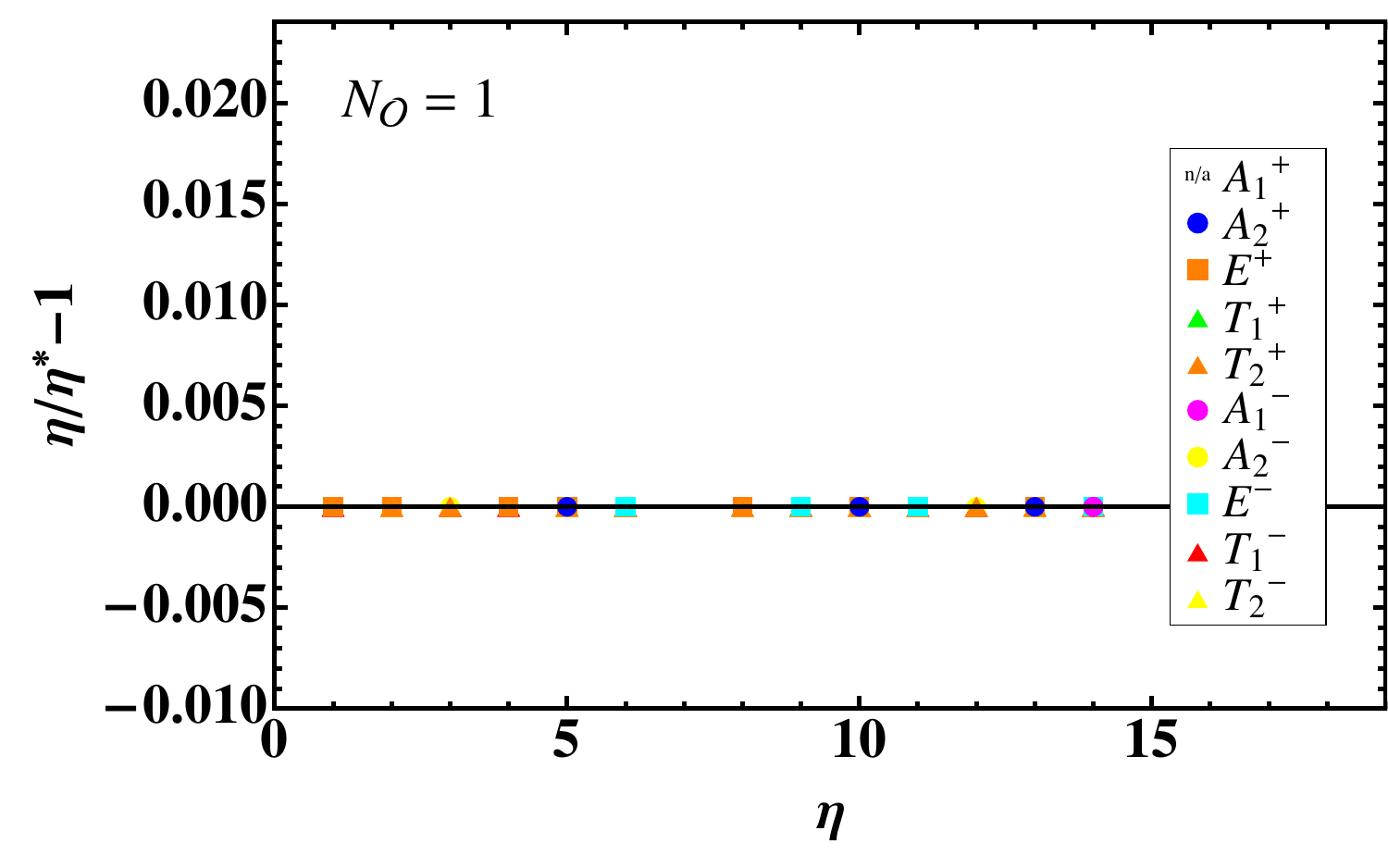}
\includegraphics[width=\figwidth]{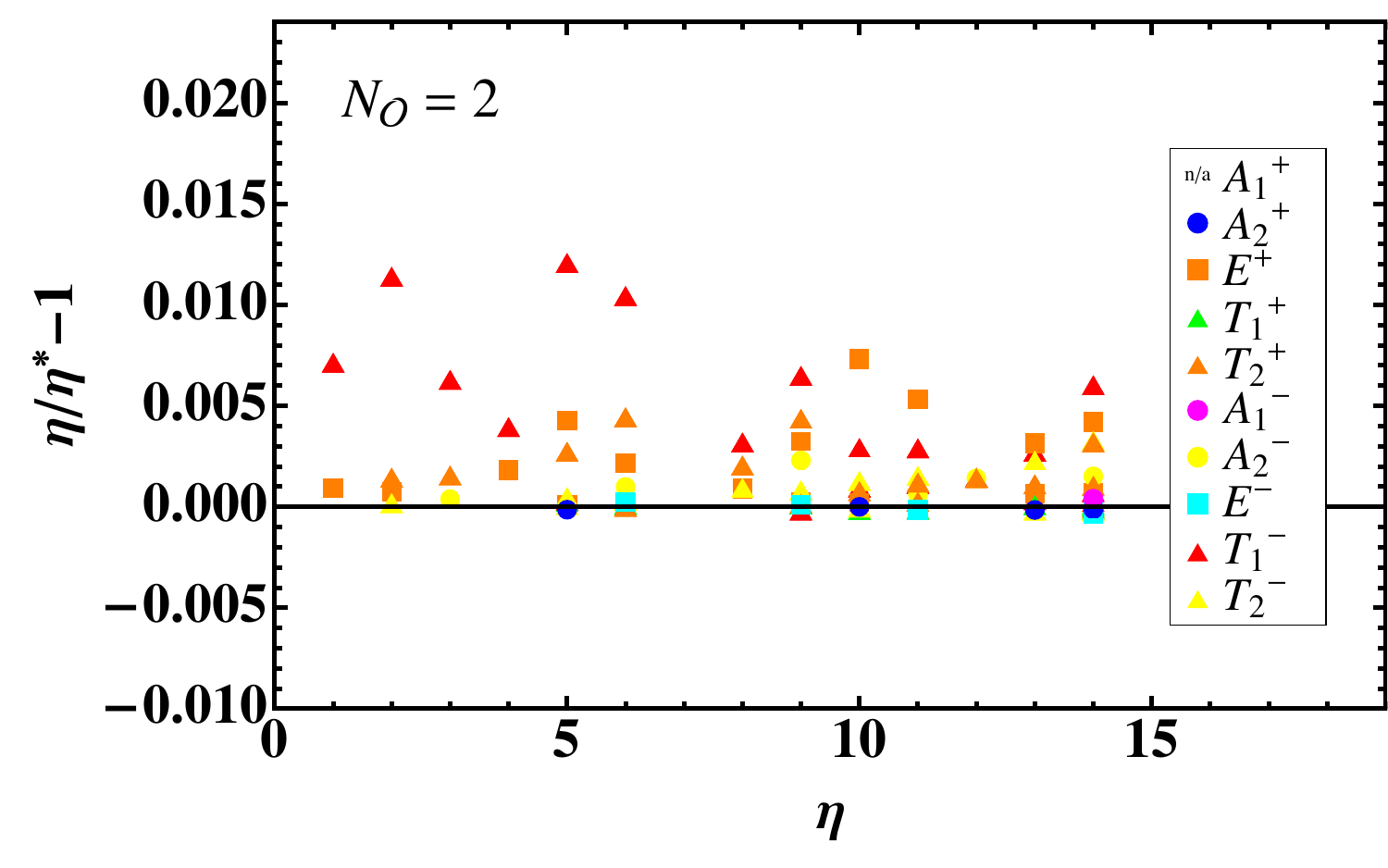}
\includegraphics[width=\figwidth]{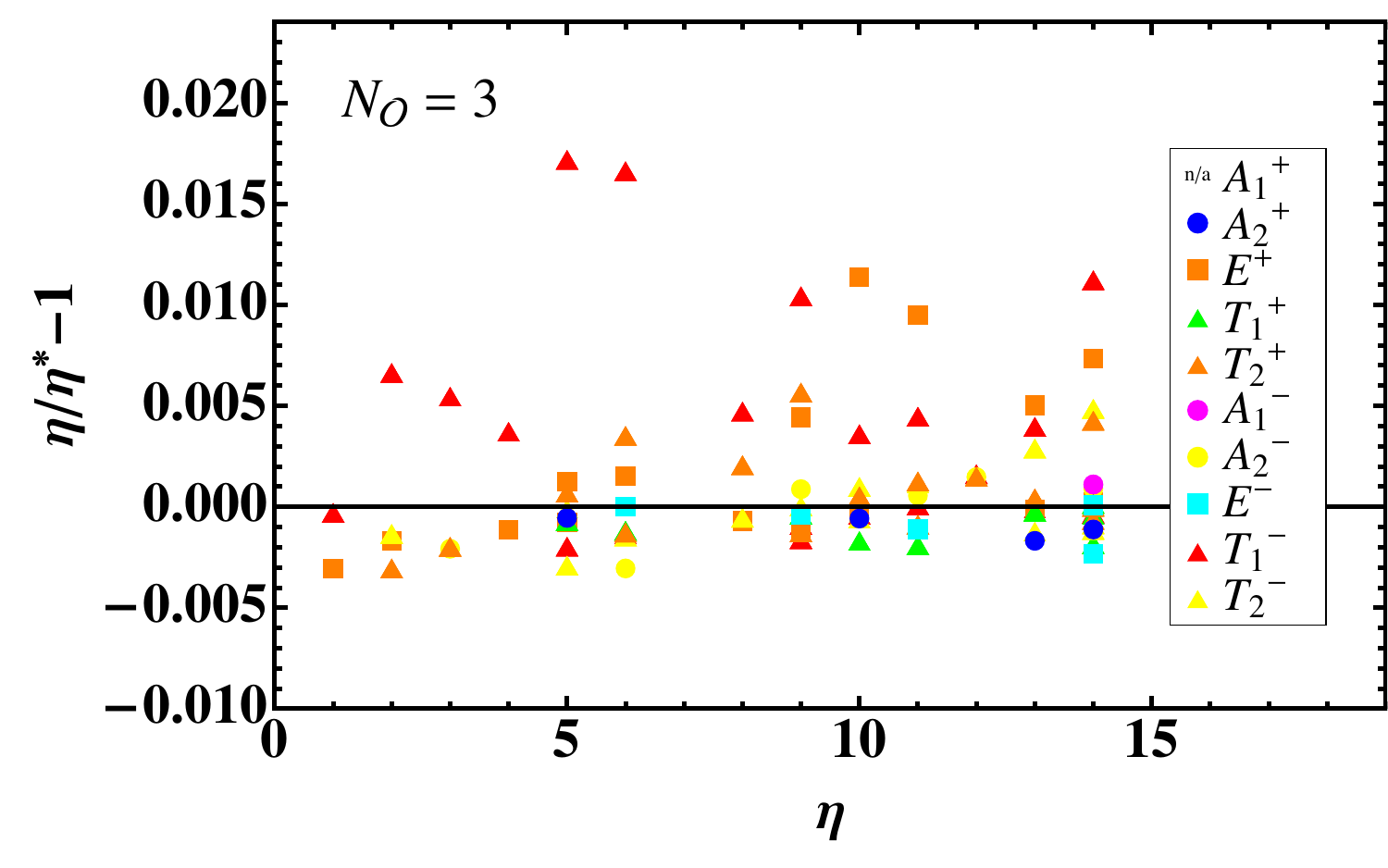}
\includegraphics[width=\figwidth]{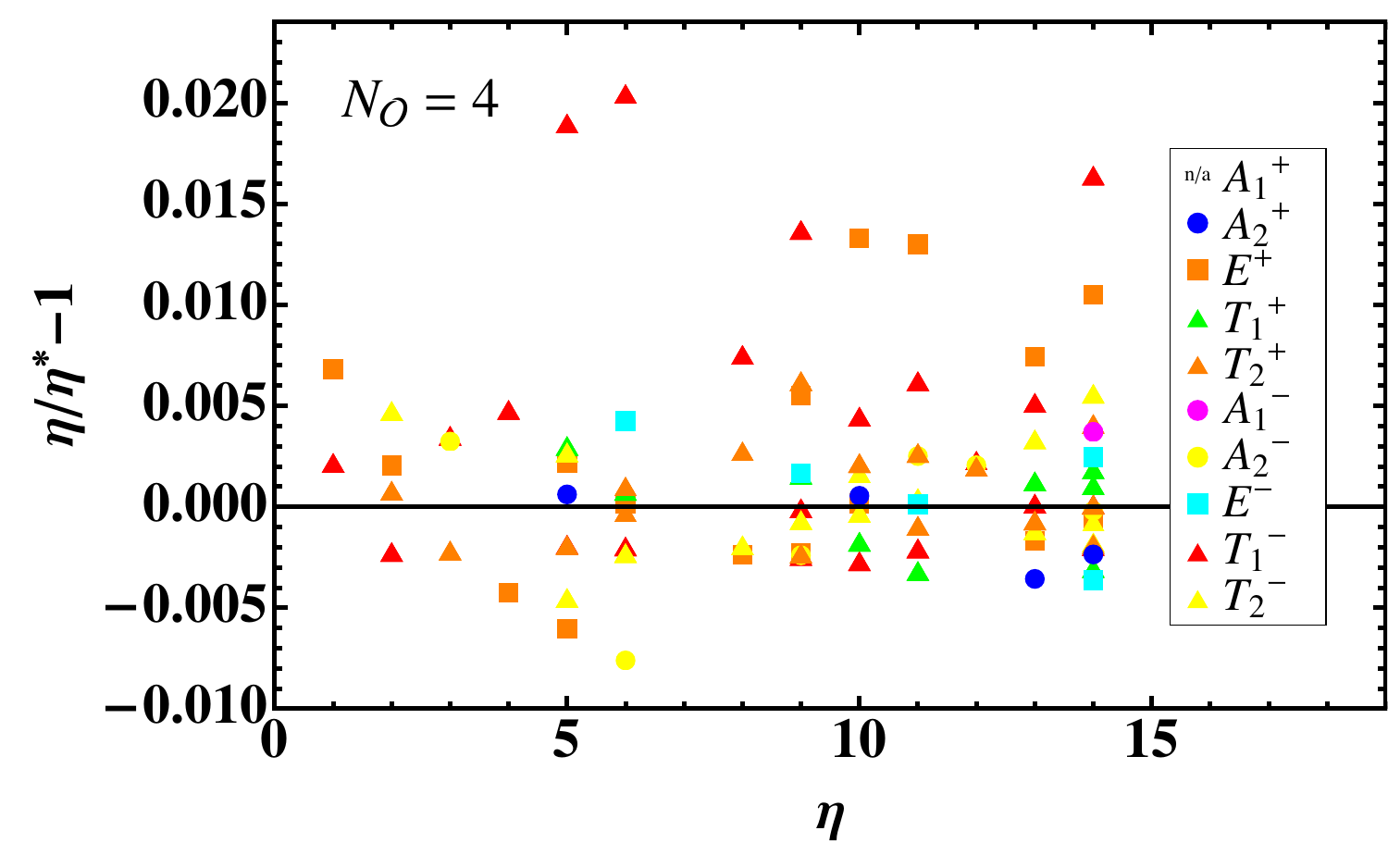}
\caption{%
\label{fig:two_fermions_spectrum_L8}%
Energy spectrum, given by $-\log \lambda = \frac{1}{M} \left( \frac{2\pi}{L}\right)^2 \eta $, of two unitary fermions of mass $M=5$ in a finite box of size $L=8$ for up to four tuned couplings; $\eta^*$ are the corresponding eigenvalues of the noninteracting theory.
Eigenvalues are labeled by dimensionality of the irrep to which they belong: circle (A),  square (E), and triangle (T), as well as color coded according to the lowest orbital angular momentum component contained in each irrep: red ($\ell=1$), orange ($\ell=2$), yellow ($\ell=3$), green ($\ell=4$), cyan ($\ell=5$), blue ($\ell=6$), and violet ($\ell=9$).
}
\end{figure}

\begin{figure}
\includegraphics[width=\figwidth]{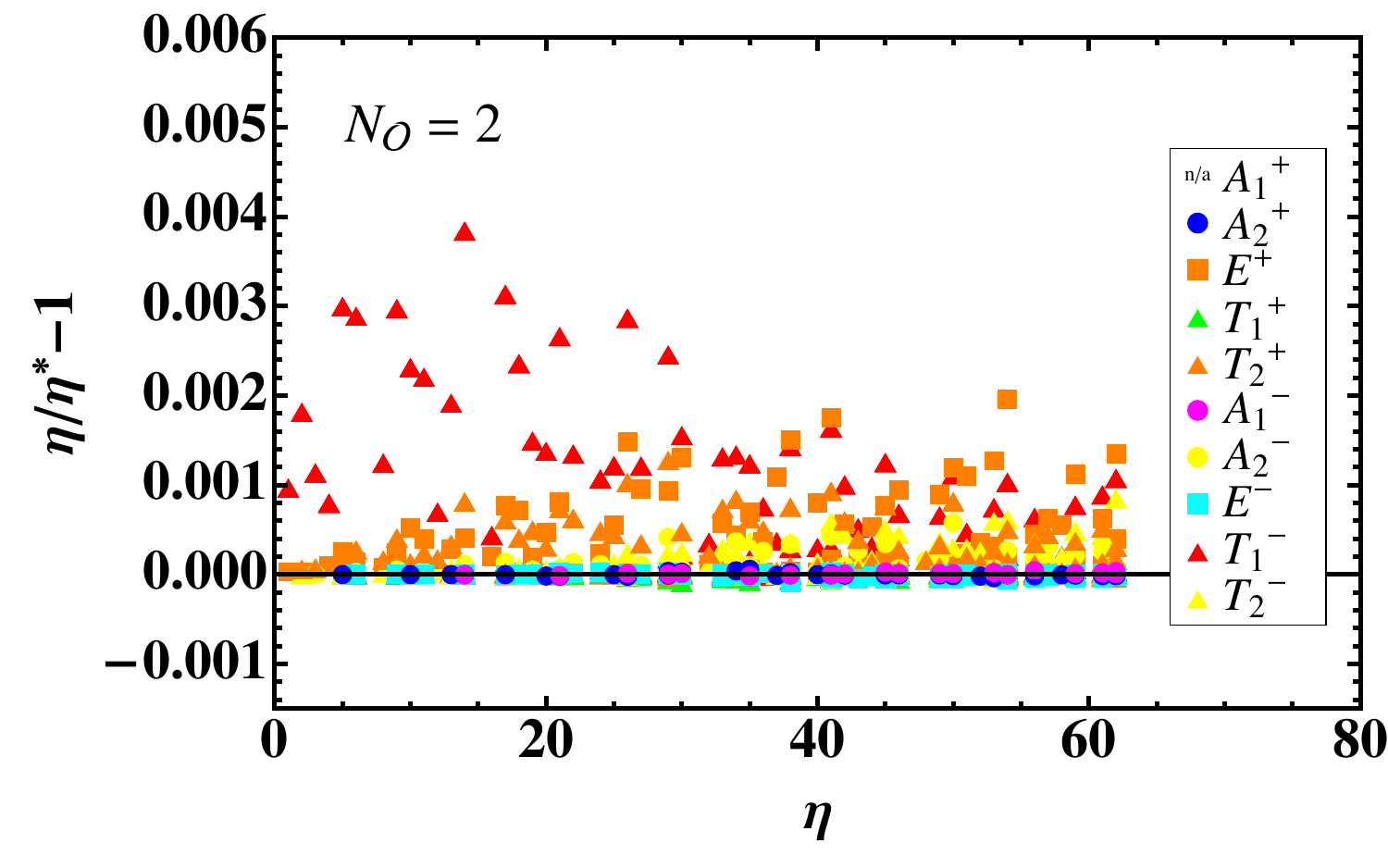}
\includegraphics[width=\figwidth]{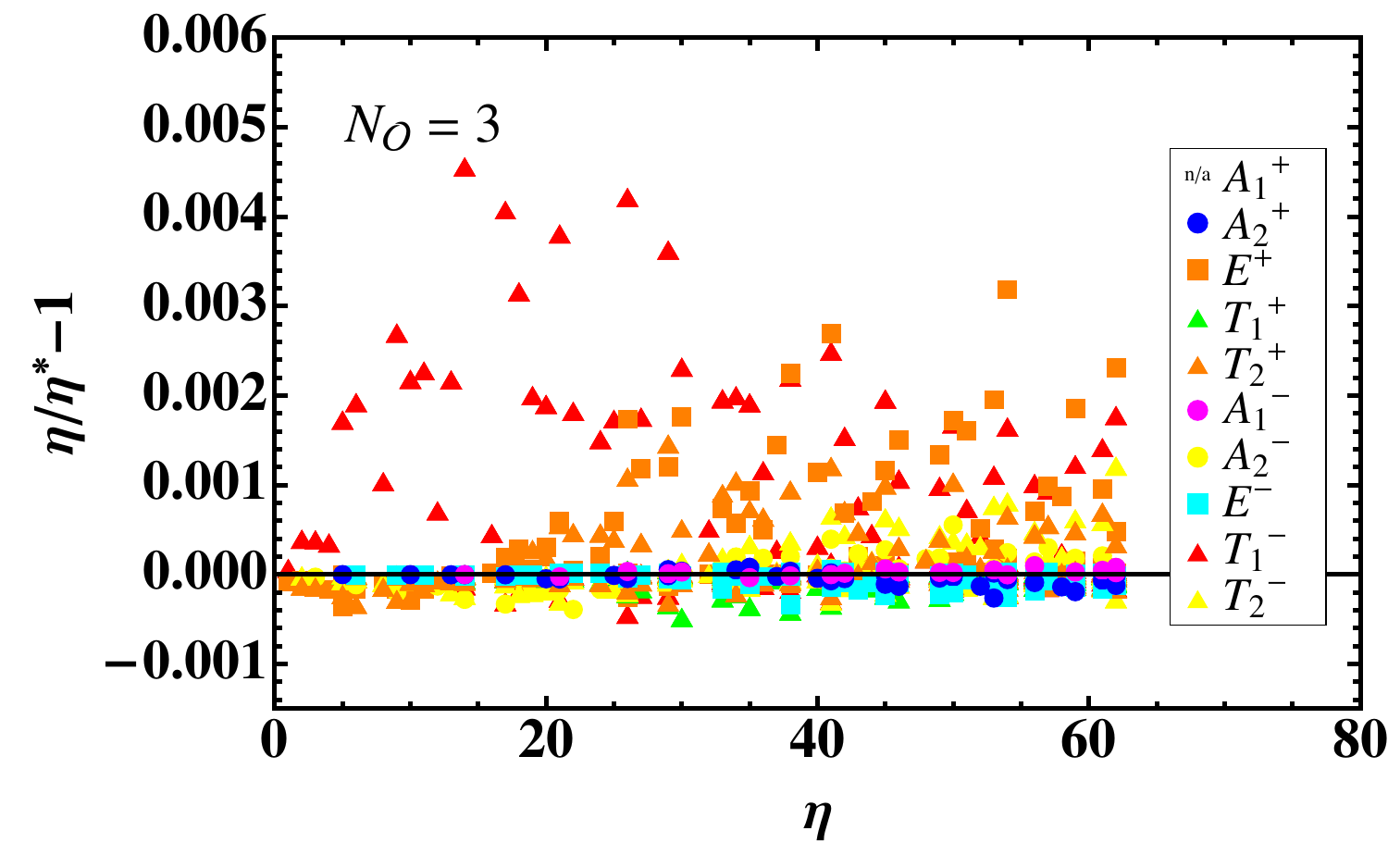}
\includegraphics[width=\figwidth]{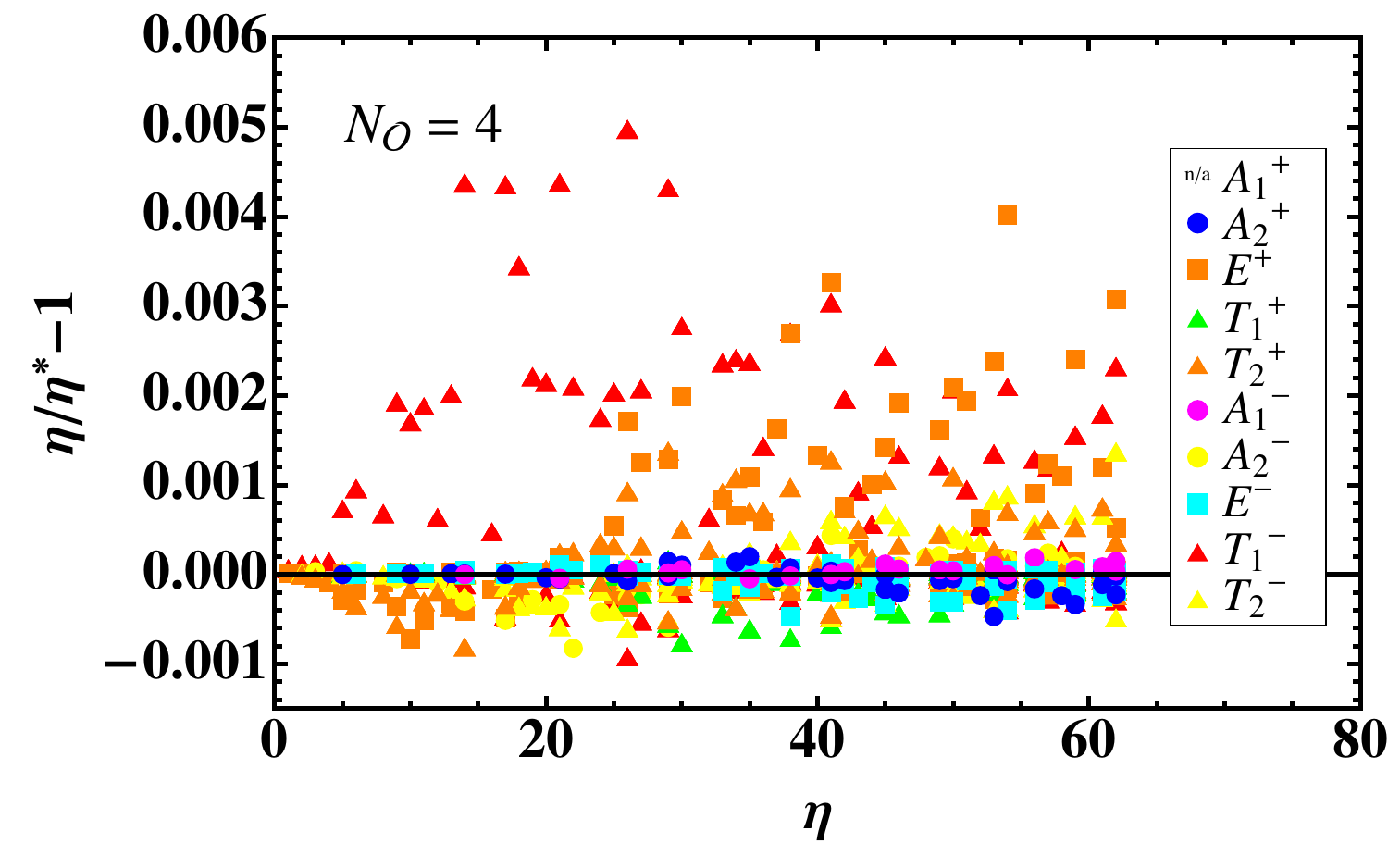}
\includegraphics[width=\figwidth]{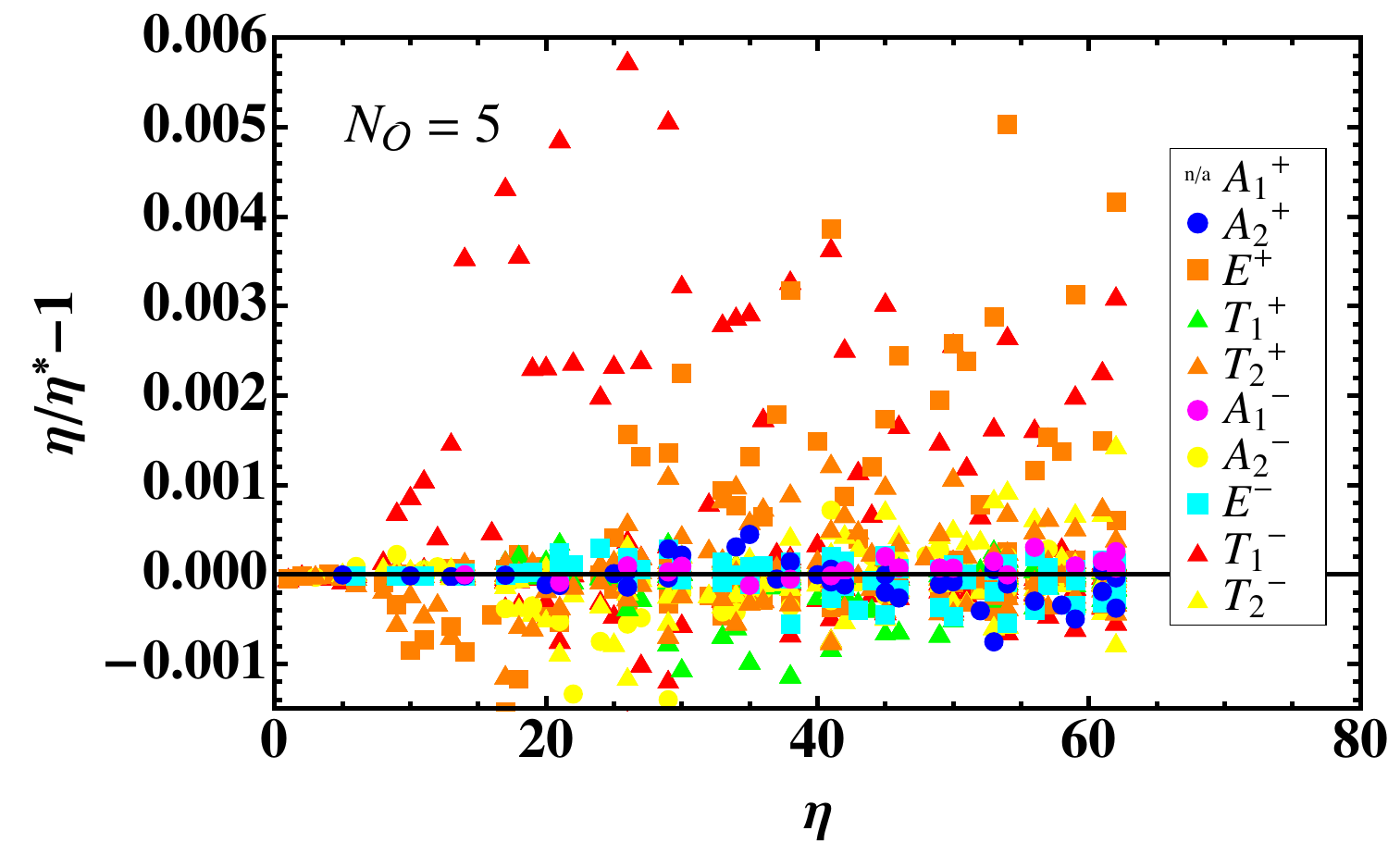}
\caption{%
\label{fig:two_fermions_spectrum_L16}%
Energy spectrum, given by $-\log \lambda = \frac{1}{M} \left( \frac{2\pi}{L}\right)^2 \eta $, of two unitary fermions of mass $M=5$ in a finite box of size $L=16$ for up to five tuned couplings; $\eta^*$ are the corresponding eigenvalues of the noninteracting theory.
Eigenvalues labeled according to irrep, as described in \Fig{two_fermions_spectrum_L8}.
In the case $N_\calO=1$ (shown in \Fig{two_fermions_spectrum_L8} for $L=8$, but omitted here), the deviation $\eta/\eta^*-1$ is exactly zero for all nontrivial irreps.
}
\end{figure}

A cursory examination of \Tab{decomposition} shows that the $A_1^+$ lattice eigenstates contain angular momenta components $\ell=0, 4, 6, 8,\ldots$ in the continuum and infinite volume limits.
Similarly, the $T_1^-$ eigenstates possess the components $\ell=1, 5, 7, 9,\ldots$.
Generally speaking, the size of the effects of interactions with angular momenta $\ell<6$ and $\ell=9$ may be deduced by studying the energies of the two-body lattice eigenstates classified by their irreps under $O_h$.
By studying the deviations in the lattice eigenvalues $\eta$, defined by $-\log \lambda = \frac{1}{M} \left( \frac{2\pi}{L}\right)^2 \eta $, where $\lambda$ is an eigenvalue of the two-body transfer matrix, in a given irrep from those of noninteracting fermions $\eta^*$, which take integer values, we may estimate the size of the effects from scattering in higher partial waves.
Figures~\ref{fig:two_fermions_spectrum_L8} and \ref{fig:two_fermions_spectrum_L16} show the deviation $\eta/\eta^*-1$ as a function of $\eta$ for the entire spectrum for unitary fermions on an $L=8$ lattice with  $1\le N_\calO\le 4$  and an $L=16$ lattice with $2\le N_\calO\le 5$.
We find that the largest deviation is approximately $2\%$ in the smallest box size considered, while the deviations are considerably smaller for the $L=16$ box ($\lesssim 0.6\%$). 

\begin{figure}
\includegraphics[width=\figwidth]{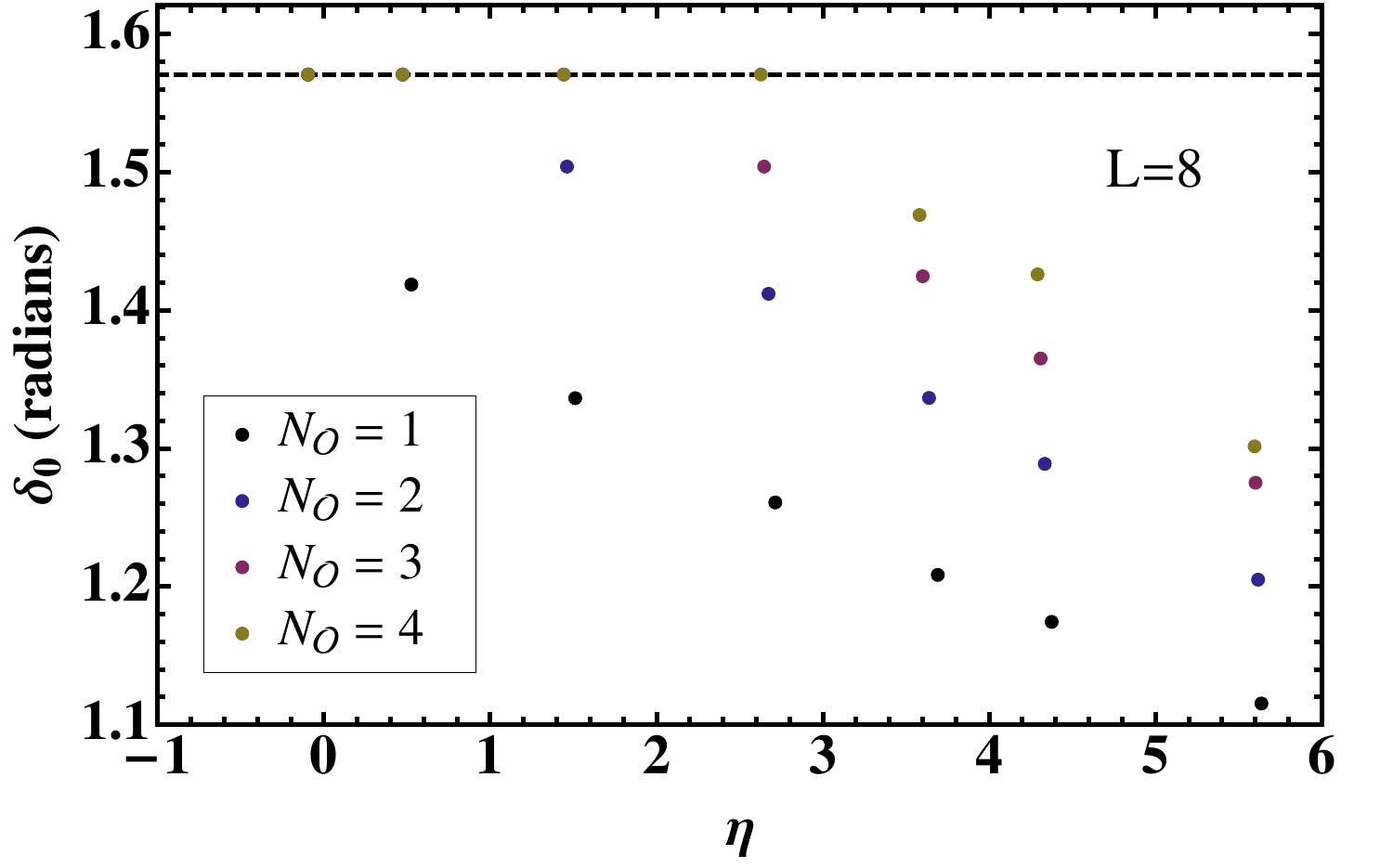}
\includegraphics[width=\figwidth]{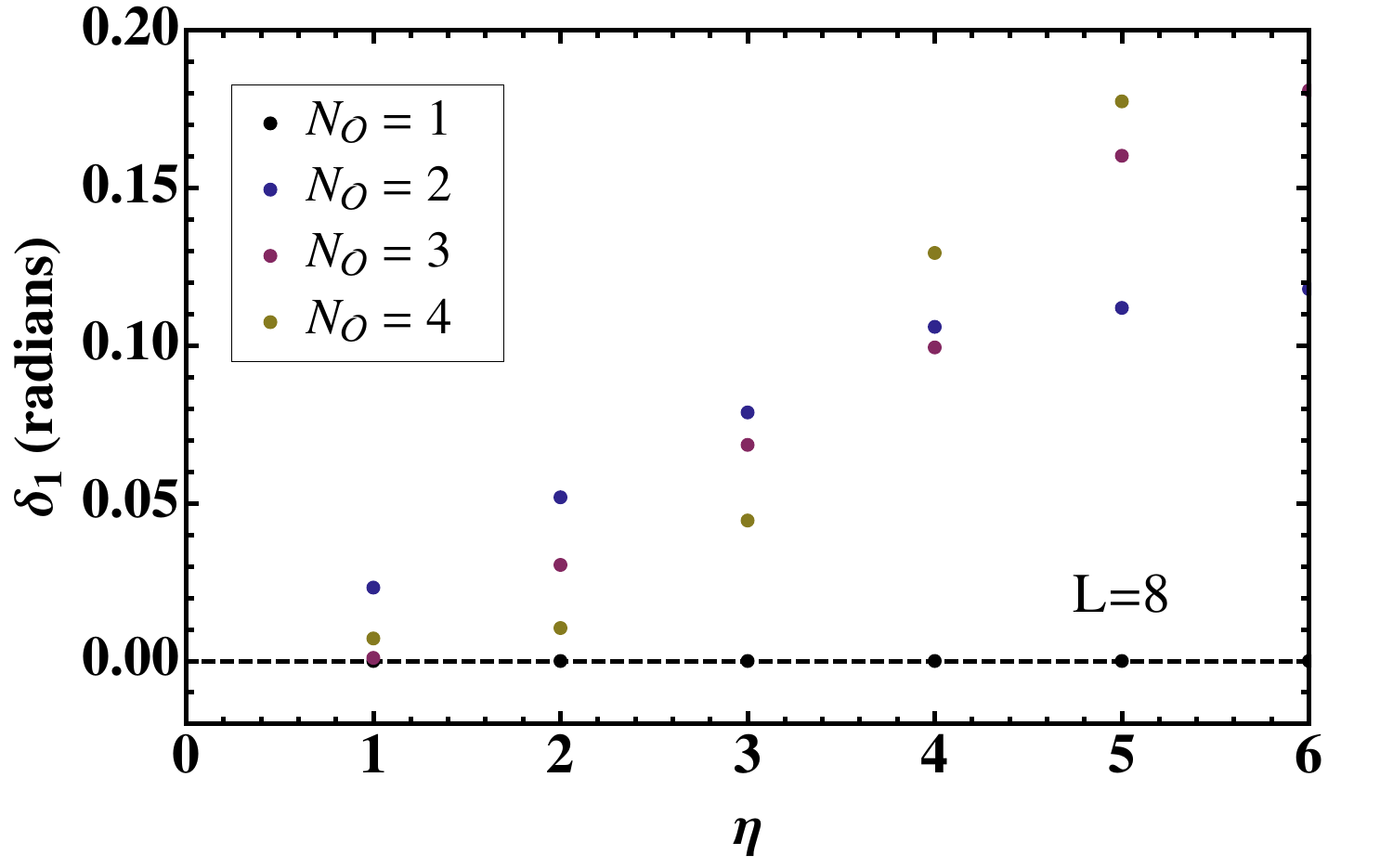}
\includegraphics[width=\figwidth]{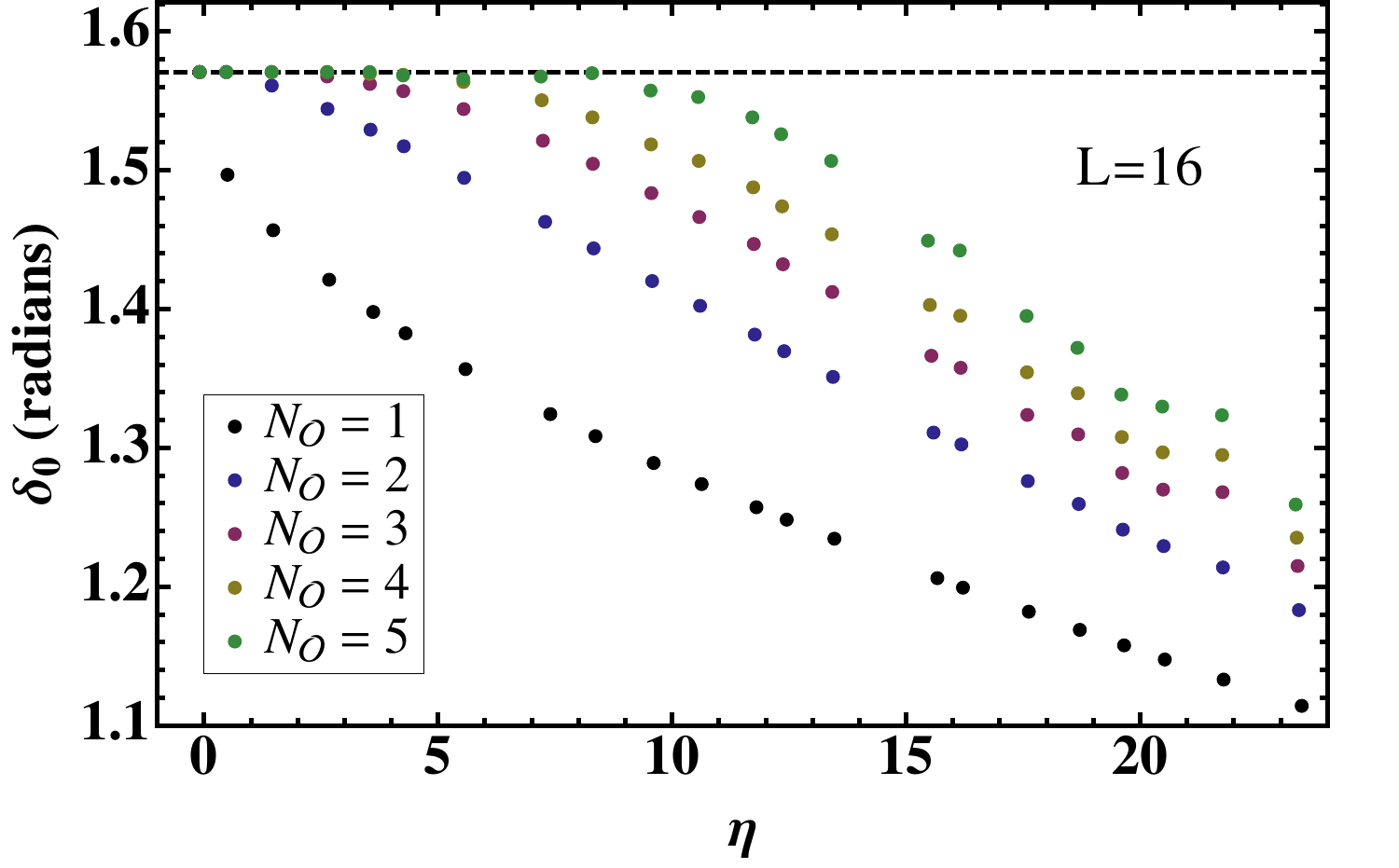}
\includegraphics[width=\figwidth]{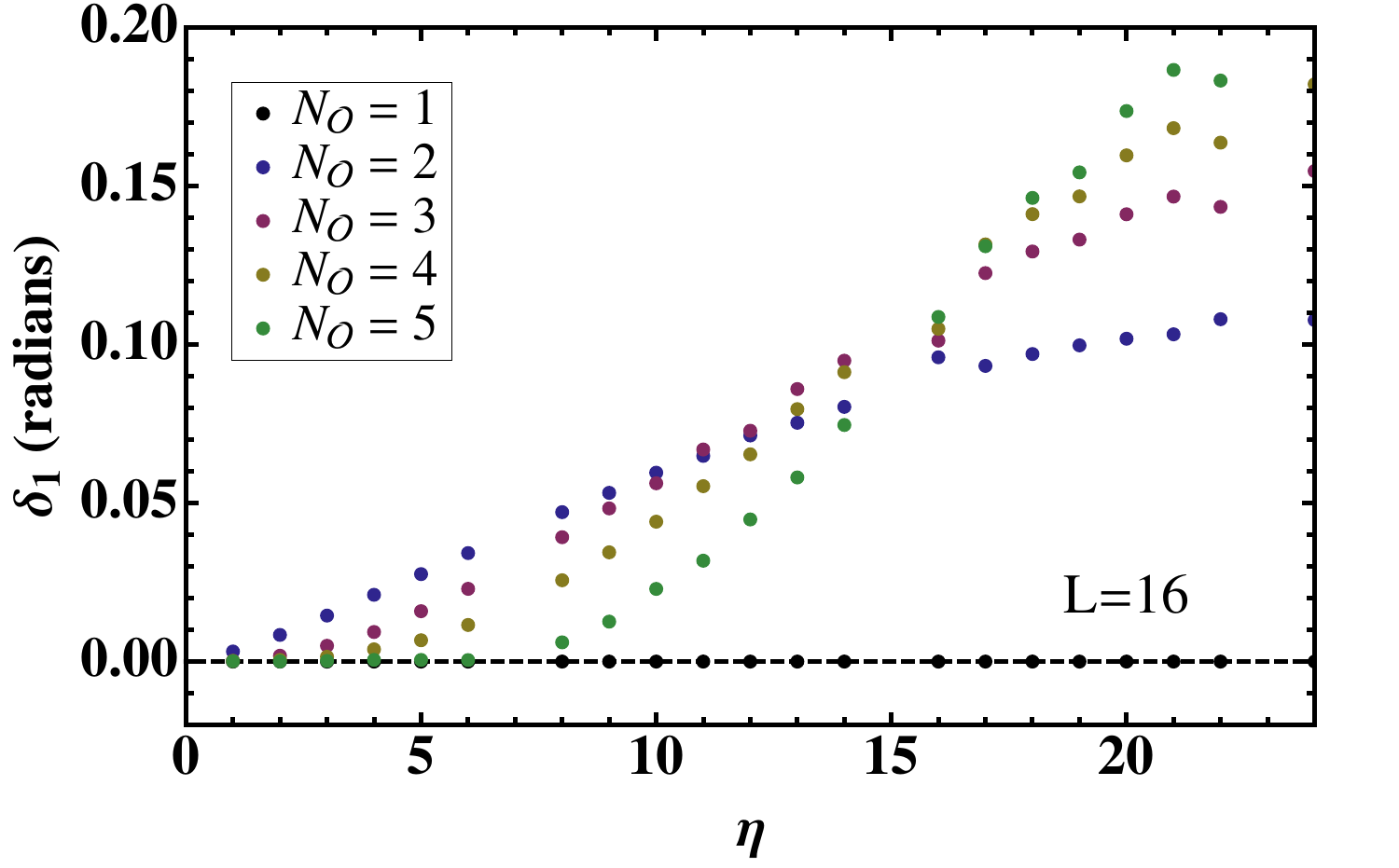}
\caption{%
\label{fig:deltas}%
$\delta_0$ and $\delta_1$ as a function of the dimensionless parameter $\eta$ for unitary fermions of mass $M=5$ on an $L=8$ and  $L=16$ lattice, obtained from \Eq{luschers_formula} and \Eq{luschers_formula_pwave}, respectively.
Dashed lines correspond to the phases $\pi/2$ (left) and zero (right).
}
\end{figure}

Using the generalization of L\"uscher's formula for $s$-wave scattering, we may determine the scattering phase shifts for the higher partial waves.
For $p$-wave scattering, if one assumes $\tan\delta_4 \ll \tan\delta_1$, one finds (see, for example, \cite{Luu:2011ep}):
\begin{eqnarray}
p^3 \cot\delta_1(p) = \left(\frac{2\pi}{L}\right)^3 \frac{1}{2\pi^2} \, \eta\, S(\eta)\ , 
\label{eq:luschers_formula_pwave}
\end{eqnarray}
where $\eta$ and  $S(\eta)$ are defined in \Eq{luschers_formula}.
Plugging the lattice eigenvalues obtained for the $T_1^-$ irrep shown in \Fig{two_fermions_spectrum_L16} into the right-hand side of \Eq{luschers_formula_pwave}, we obtain a lattice prediction for $p^3 \cot\delta_1$.
In \Fig{deltas}, the scattering phase shift $\delta_1$ obtained by this procedure is plotted as a function of $\eta$ for $L=8$ and $L=16$, and for $N_\calO = 1,2,3,4$ and $5$.
For reference, also shown in this figure is the scattering phase shift $\delta_0$ obtained from \Eq{luschers_formula}.

More generally, one can show from the results of Ref. \cite{Luu:2011ep} that if $\eta$ is sufficiently close to $\eta^*$, then for the partial waves $\ell=0,1,2,3,4,5,6$ and $9$ one finds:
\begin{eqnarray}
\delta_\ell(p)  \approx  \frac{ (\eta^*)^{3/2} }{ g_\ell({\eta^*}) } (\eta/\eta^* - 1) + \calO(\eta/\eta^*-1)^3\ ,
\label{eq:luschers_approxformula_pwave}
\end{eqnarray}
where $g_\ell({\eta^*})$ is some non-zero calculable numerical factor.
For $s$- and $p$-waves, $g_0({\eta^*}) = g_1({\eta^*}) = d(\eta^*)/(2\pi^2)$, where $d(\eta^*)$ equals the number of times the integer triplet $\bfj$ satisfies $|\bfj| = \eta^*$ for a given pole $\eta^*$ (i.e., an integer taking the value: 1, 6, 8, 12, 24, or 48 depending on $\eta^*$); for $\ell>1$, the expression for $g_\ell({\eta^*})$ is more complicated (involving spherical harmonics), and therefore is not provided here.
From \Fig{two_fermions_spectrum_L8} and \Fig{two_fermions_spectrum_L16}, it is evident that the deviation of higher angular momentum modes diminish with $\ell$, and likewise based on \Eq{luschers_approxformula_pwave} so must the corresponding phase shifts.

\subsection{Three unitary fermions}
\label{sec:three_unitary_fermions}

Here we present results from exact numerical diagonalization of the $N=2+1$ unitary fermion transfer matrix in an effort to better understand the effectiveness of our parameter tuning method.
In addition, we use the exact numerical results to investigate the  finite spatial and temporal discretization errors of the lattice theory.
Exact numerical results for the $N=2+1$ system were obtained for the $A_1^+$ irrep for $L=4,6,8$ and $10$ and for up to $N_\calO=5$ tuned couplings.
At small lattice volumes, we were limited to fewer than five tuned couplings for reasons discussed in Ref. \cite{Endres:2011er}.

\begin{figure}
\includegraphics[width=\figwidth]{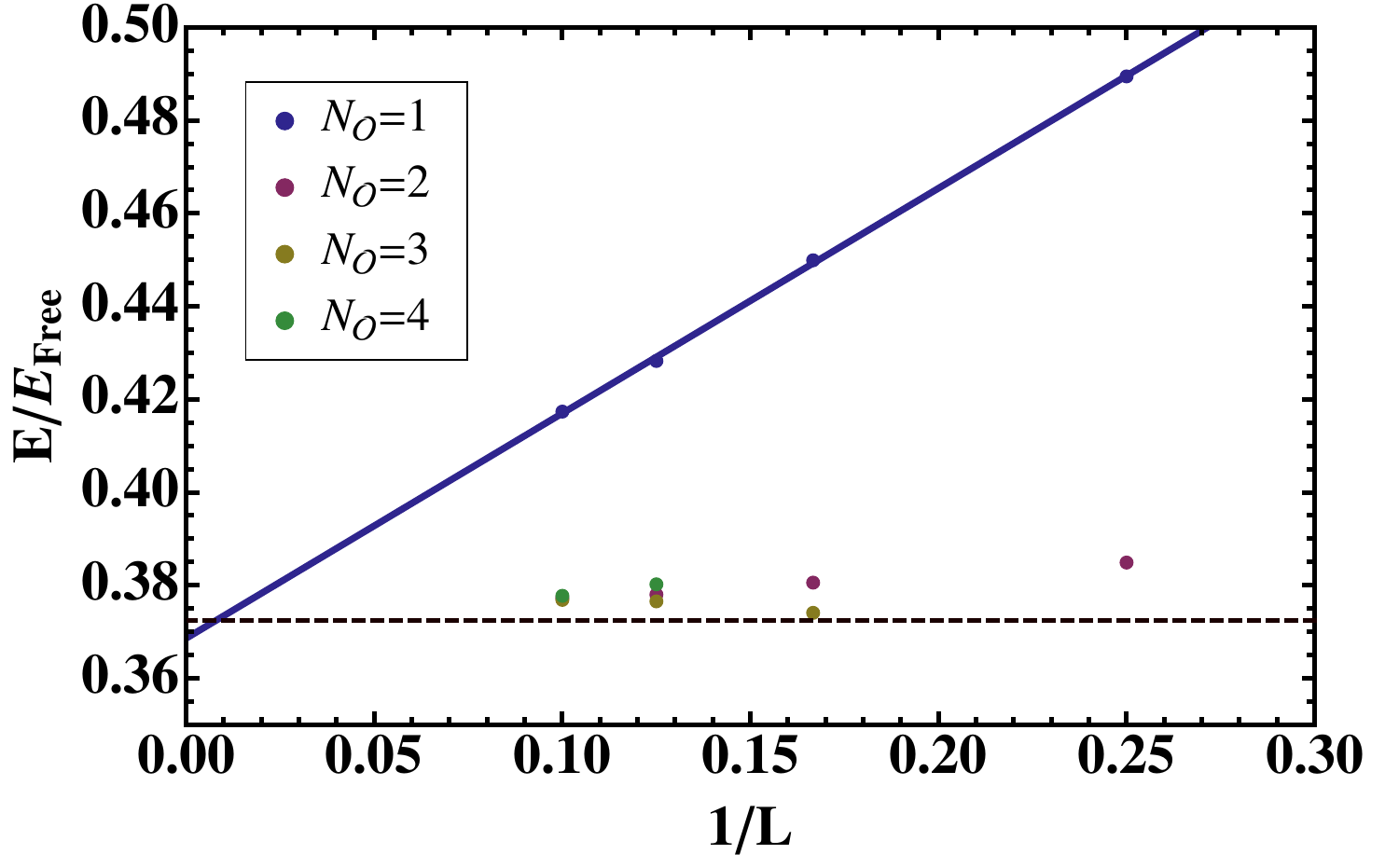} 
\includegraphics[width=\figwidth]{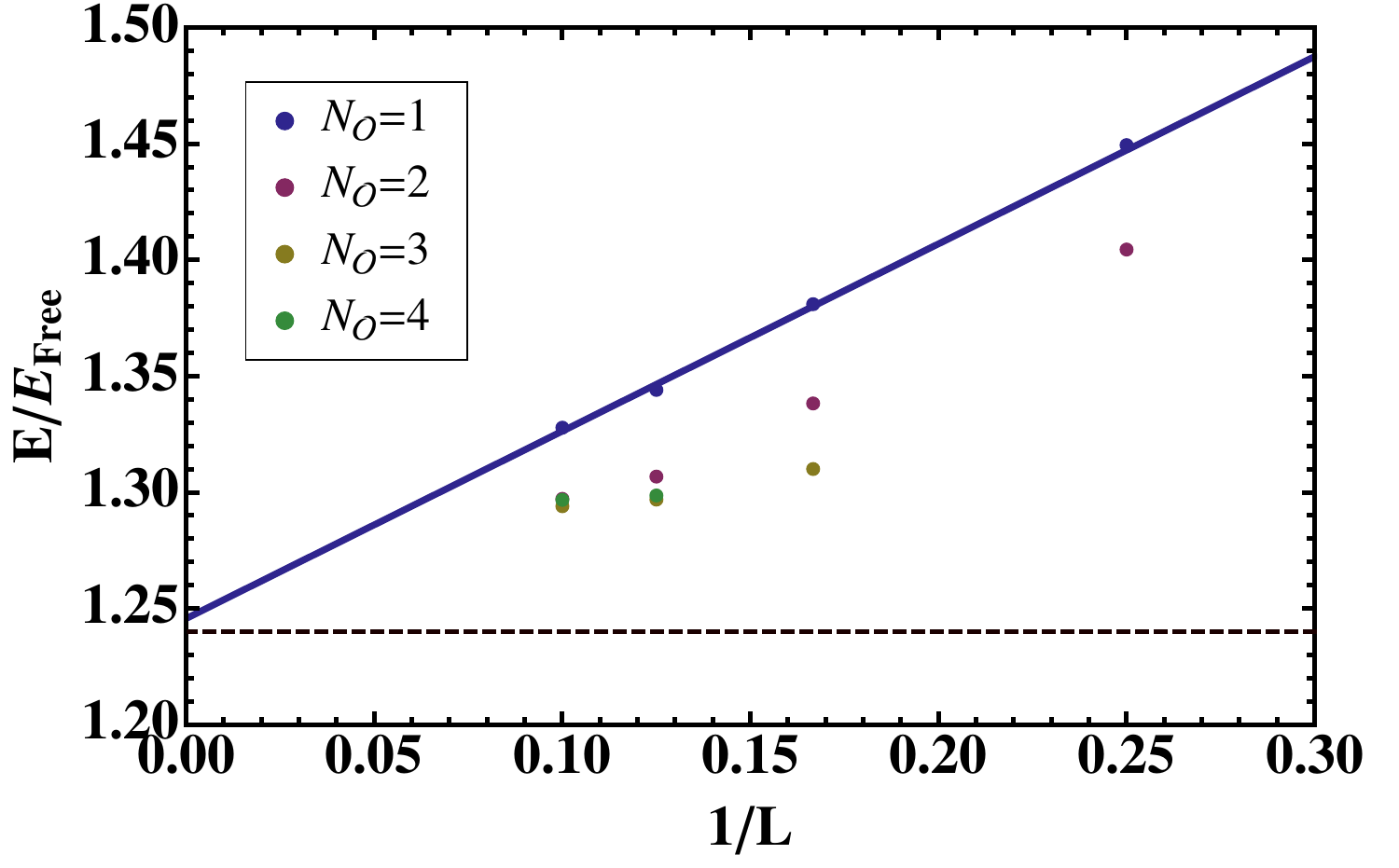} 
\caption{%
\label{fig:three_fermions_lowest}%
Left: ground-state energy of $N=2+1$ unitary fermions of mass $M=5$ in the $A_1^+$ irrep as a function of $1/L$ for up to four tuned couplings.
Right: first excited state energy of $N=2+1$ unitary fermions of mass $M=5$ in the $A_1^+$ irrep as a function of $1/L$ for up to four tuned couplings.
Dashed lines in both plots correspond to the infinite volume exact results of Ref. \cite{2007JPhA...4012863P}; solid lines are infinite volume extrapolations based on a linear fit to all $N_\calO=1$ data (deviations in the extrapolation reflect systematic effects of unaccounted for higher order corrections).
}
\end{figure}

\Fig{three_fermions_lowest} shows the energies of the ground and first excited states in the $A_1^+$ irrep as a function of volume, along with the extrapolated continuum limit, infinite-volume results obtained in Ref. \cite{2007JPhA...4012863P}.
Note that these energies are measured in units of the non-interacting energy for three zero-momentum fermions, $E_{Free} = (2\pi/L)^2/M$.
Extrapolation of the $N_\calO=1$ results using a linear fit in $L^{-1}$ to the data yields better than 1\% agreement with results from Ref. \cite{2007JPhA...4012863P}; deviations may be attributed to the long extrapolation performed on our exact finite volume energies.
Note that for $N_\calO>1$ tuned couplings, the lattice energies show substantially improved agreement with exact results for the ground state energy--even at small to moderate volumes--compared to the $N_\calO=1$ results.
The improvement with $N_\calO$ is less pronounced in the excited $A_1^+$ state, however, for which we do not have a rigorous explanation.

Although tuning more than one $s$-wave two-body operator results in increased improvement in ground state energies for $L=4$ and $6$, little improvement in energies is evident among $N_\calO =2,3,4$ and $5$ results at $L=8$ and  $L=10$.
It is possible that this peculiar behavior may be due to the effects of an untuned $\ell=1$ two-body operator (giving rise to $L^{-3}$ scaling) or three-body operators (which are expected to contribute starting at $L^{-4.33}$).
On the other hand, the leading volume corrections from untuned two-body $\ell=0$ operators scale as $1/L^{2 N_\calO-1}$.
At large $L$, the former volume corrections may dominate irrespective of $N_\calO>1$, whereas for small $L$ the latter corrections becomes non-negligible even for large $N_\calO$, giving rise to stronger $N_\calO$-dependence of the energies at small $L$.

\begin{figure}
\includegraphics[width=\figwidth]{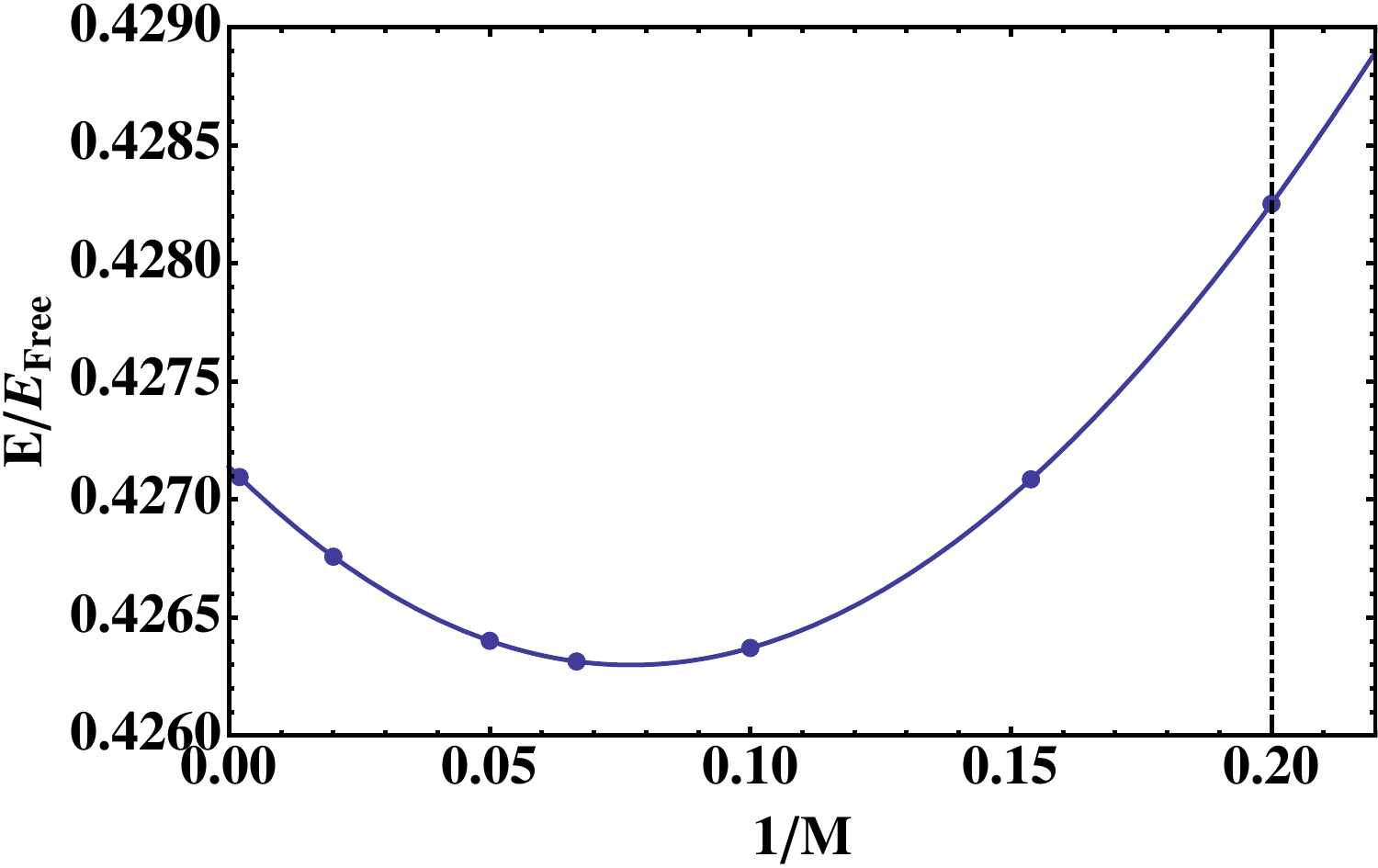} 
\includegraphics[width=\figwidth]{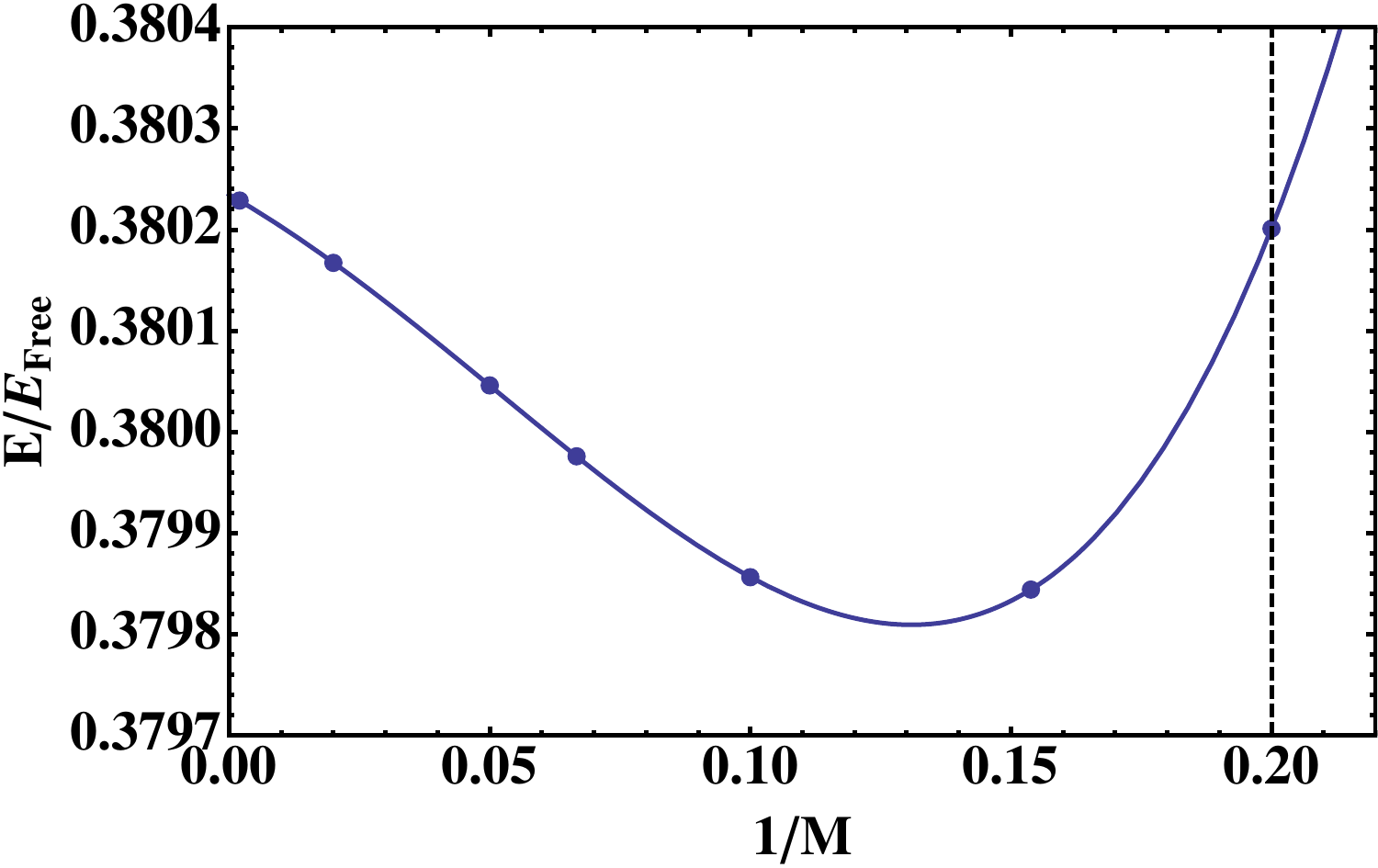} 
\caption{%
\label{fig:three_fermions_time_discretization}%
Ground-state energy of $N=2+1$ unitary fermions in the $A_1^+$ irrep as a function of $1/M$ for one (left) and four (right) tuned couplings with $L=8$.
Blue dashed line corresponds to a fourth order polynomial fit in $1/M$; black dashed line indicates the mass value at which all simulations have been performed.
}
\end{figure}

In \Fig{three_fermions_time_discretization}, we show the fermion mass-dependence of the $N=2+1$ ground state energy for $N_\calO=1$ and $N_\calO=4$ on an $L=8$ lattice.
The data was fit using a fourth-order polynomial in $1/M$ and extrapolated to the $M\to\infty$ limit.
Since the physical (dimensionful) mass is equal to $ M b_\tau/ b_s^2$, taking $M\to\infty$ is equivalent to taking the temporal continuum limit $b_\tau\to 0$ while keeping the spatial lattice spacing $b_s$ and physical mass held constant (in other words, the lattice mass parameter may be viewed as an anisotropy factor).
From \Fig{three_fermions_time_discretization}, we see that the temporal discretization effects on the three fermions system at $M=5$, the mass value used in our few- and many-body simulations, is roughly 0.5\% for $N_\calO=1$ and 0.1\% for $N_\calO=4$ tuned operators.
Time discretization errors are therefore likely negligible compared to other systematic and statistical uncertainties in our few- and  many-body simulations.

\section{Simulation results for few- and many-body states}
\label{sec:simulation_results}

\subsection{Ensembles and parameters}
\label{sec:analysis_and_results.unitary_fermions_in_a_finite_box.ensembles_and_parameters}

\begingroup
\squeezetable
\begin{table}
\caption{%
\label{tab:untrapped_ensembles}%
Simulation parameters for untrapped fermions ($M=5$) using the pairing wave function given by \Eq{sink_state} with $\Psi_0=100$.
For ensembles denoted by an asterisk, we have used $\bar{\Psi} ({\bf p})=1/(\beta-e^{-{\bf p}^2/(2M)})$ for the pairing wave function rather than \Eq{sink_wavefunc}.
$N_{conf}$ is the total size of the ensemble; observable data was block-averaged in blocks of size $N_{conf}/N_{\calB}$ prior to analysis.
}
\begin{ruledtabular}
\begin{tabular}{ccc|ccccc|cc}
$N$ & $L$ & $T$ & $N_\calO$ & $\beta$ & $N_{conf}$ & $N_\calB$ \\
\hline
$4$      & 4 & 24 & 1 & 0.1 / 0.4   & 190M / 330M & 190 / 330 \\
$4$      & 6 & 36 & 1 & 0.2 / 0.35 / 0.5   & 300M / 300M / 300M & 300 / 300 / 300 \\
$4$      & 8 & 48 & 1 & 0.15 / 0.25   & 500M\footnote{For this ensemble, we have used $\Psi_0=50$.} / 500M & 250 / 250 \\
$4$      & 10 & 64 & 1 & 0.25 / 0.3 / 0.32   & 400M / 400M / 400M & 400 / 400 / 400 \\
$4$      & 12 & 54 & 1 & 0.25 / 0.31 / 0.35   & 400M / 400M / 400M & 400 / 400 / 400 \\
$4$      & 14 & 54 & 1 & 0.37 / 0.43   & 680M / 680M & 340 / 340 \\
$4$      & 4 & 24 & 2 & 1.01* / 0.1   & 2.67B / 1.29B & 300 / 300 \\
$4$      & 6 & 36 & 2 & 0.1   & 200M & 200 \\
$4$      & 10 & 42 & 2 & 0.15 / 0.19   & 400M / 300M & 400 / 300 \\
$4$      & 8 & 48 & 5 & 0.05 / 0.15   & 110M / 110M & 110 / 110 \\
$4$      & 10 & 64 & 5 & 1.01* / 0.07 / 0.1   & 150M / 100M / 100M & 300 / 200 /200 \\
$4$      & 12 & 64 & 5 & 0.15 / 0.2 / 0.25   & 150M / 60M / 180M & 300 / 120 / 360 \\
$4$      & 14 & 64 & 5 & 0.2 / 0.25 / 0.3   & 250M / 250M / 130M & 250 / 250 / 130 \\
$4$      & 16 & 64 & 5 & 0.25 / 0.3 / 0.35   & 390M / 390M / 390M & 390 / 390 / 390 \\
$4$      & 18 & 64 & 5 & 0.25 / 0.3   & 350M / 350M & 350 / 350 \\
\hline
$\le66$  & 10 & 54 & 5 & 1.0   & 40M & 400 \\
$\le66$  & 12 & 64 & 5 & 0.5   & 40M & 400 \\
$\le66$  & 12 & 54 & 5 & 0.75  & 40M & 400 \\
$\le66$  & 12 & 64 & 5 & 1.0   & 40M & 400 \\
$\le66$  & 14 & 36 & 5 & 0.5   & 19M & 190 \\
$\le66$  & 14 & 36 & 5 & 0.75  & 35M & 350 \\
$\le66$  & 14 & 72 & 5 & 1.0   & 39M & 390 \\
$\le66$  & 16 & 54 & 5 & 0.9   & 20M & 200 \\
$\le66$  & 16 & 30 & 5 & 0.9   & 20M & 200 \\
$\le66$  & 16 & 30 & 5 & 0.75   & 63M & 315 \\
\end{tabular}
\end{ruledtabular}
\end{table}
\endgroup

\begin{table}
\caption{%
\label{tab:untrapped_couplings}%
Simulation parameters (i.e., tuned couplings) for untrapped unitary fermions ($M=5$) for various lattice volumes and $N_\calO$ values.
}
\begin{ruledtabular}
\begin{tabular}{c|ccccc}
$L$ & $C_0$ & $C_2$ & $C_4$ & $C_6$ & $C_8$ \\
\hline
 4 & 0.673068 &   --       &    --     &   --       &    --      \\
 6 & 0.689184 &   --       &    --     &   --       &    --      \\
 8 & 0.680971 &   --       &    --     &   --       &    --      \\
10 & 0.684858 &   --       &    --     &   --       &    --      \\
12 & 0.679787 &   --       &    --     &   --       &    --      \\
14 & 0.684345 &   --       &    --     &   --       &    --      \\
\hline
 4 & 0.333477 &  0.1552055 & --        & --         & --         \\
 6 & 0.428091 &  0.1128065 & --        & --         & --         \\
 10 & 0.455289 &  0.0939424 & --        & --         & --         \\
\hline
8 & 0.931735 & -2.1243485 & 2.2200002 &  -0.7798253 & 0.08856646 \\
10 & 0.585273 & -0.1507720 & 0.2120923 &  0.0974153 & 0.01455297 \\
12 & 0.544064 & -0.0354881 & 0.0770458 & -0.0433194 & 0.00783886 \\
14 & 0.547526 & -0.0218146 & 0.0489023 & -0.0291435 & 0.00588451 \\
16 & 0.537953 & -0.0042753 & 0.0284083 & -0.0211698 & 0.00492156 \\
18 & 0.547918 & -0.0111534 & 0.0375212 & -0.0279792 & 0.00613497 \\
\end{tabular}
\end{ruledtabular}
\end{table}

Our numerical studies of untrapped unitary fermions consisted of two parts: 1) high precision calculations for $N\le4$ fermions, intended for investigating the systematic errors associated with finite volume artifacts in the few-body system, and 2) simulations for up to $N=66$ fermions in order to extract a thermodynamic limit value for the Bertsch parameter.
Ensemble details for each of these studies are provided in \Tab{untrapped_ensembles}; the $C_{2n}$ values used for a given $L$ and $N_\calO$ at $M=5$ are provided in \Tab{untrapped_couplings}.
Throughout our studies, we have used an ultra-violet cutoff of $\Lambda = 0.99999 \pi$.
For our few-body studies we generated ensembles of size $N_{conf} \approx 100M$-$1B$ on lattices ranging from $L=4-18$ in size and $N_\calO =1, 2,$ and $5$ tuned couplings.
The total CPU time required for all fifteen few-body simulations was approximately 100K CPU-hours.
In our many-body studies of up to 66 unitary fermions, we generated ensembles of size $N_{conf}\approx 20M$-$60M$ on lattices ranging from $L=10-16$ and with $N_\calO=5$ tuned couplings.
A total of approximately 450K CPU-hours was required to generate all ten many-body ensembles.

Observables were measured on each ensemble and the results were averaged into $N_\calB$ blocks of size $N_{conf}/N_\calB$ prior to analysis.
For even $N$, correlation functions were measured using \Eq{slater2}, and appeared to be insensitive to the free parameter $\Psi_0$ in the sink wave-function provided that the parameter is sufficiently large.
Throughout this work we therefore fixed $\Psi_0=100$ in \Eq{sink_wavefunc} but considered multiple values for the free parameter $\beta$ on each ensemble.
Details regarding the ensembles and correlation functions used for the case when $N=3$ are provided in \cite{Endres:2011er}

The lattice action possesses one additional free parameter, the mass term $M$, which controls the anisotropy of the lattice (i.e., a conversion factor between space and time).
The temporal discretization errors in the many-body problem are controlled by the quantity $k_F^2/M \sim  (N^{1/3}/L)^2/M$.
For fixed $b_\tau$, the temporal discretization errors are subleading in the density compared to spatial discretization errors, which are controlled by $k_F \sim N^{1/3} /L$.
The temporal discretizations are therefore expected to be under control provided $M$ is larger than $\calO(1)$ in lattice units.
Since the decay rate of correlation functions are proportional to $1/M$, we may obtain earlier plateaus in $\tau$ by decreasing $M$, however, this comes at the cost of increased temporal discretization errors.
With these considerations in mind, we find that $M=5$ is an ideal compromise and use this value throughout all of our studies.

\subsection{Analysis technique}
\label{sec:analysis_technique}

Throughout this work, we study the behavior of multifermion correlation functions at late time in order to extract information about the low-lying spectrum of the system at unitarity. 
Specifically, correlators have the late-time behavior:
\begin{eqnarray}
\calC(\tau) = Z_0 e^{-E_0 \tau} + \textrm{excited state contributions}\ ,
\end{eqnarray}
where $Z_0$ is a complex number which quantifies the overlap between our source and sink wave functions and the multi-fermion ground state, and $E_0$ is the corresponding ground state energy of the system.
For small numbers of fermions, we may use a conventional approach for extracting $E_0$ by studying the plateau region of the effective mass, defined as:
\begin{eqnarray}
\label{eq:eff_mass}
m_{eff}(\tau) = \frac{1}{\Delta\tau} \log \frac{\calC(\tau)}{\calC(\tau+\Delta\tau)}\ .
\end{eqnarray}
The correlator $\calC(\tau)$ is typically estimated from an ensemble average of correlators measured on random background field configurations and $\Delta\tau$ is an arbitrary positive integer, usually chosen to be unity.
In the late time limit one finds $ m_{eff}(\tau) \approx  E_0 $ up to corrections that are exponentially small in the energy splittings.
Using more sophisticated analysis methods, one may extract excited states from the effective mass as well.

For larger $N$, however, correlators measured using our simulation algorithm generally possess a distribution overlap problem, rendering conventional estimates of the effective mass unreliable.
The problem is particularly severe when the number of configurations is less than on the order of $e^{ 40 E_{Free}(N) \tau /(3\pi)}$, where $E_{Free}(N)$ is the free gas energy of $N$ fermions \cite{Endres:2011er}.
For small numbers of fermions this problem may be overcome with brute force by generating very large ensembles, but for large numbers of fermions, brute force becomes impractical and one must resort to alternative techniques for reliably estimating the effective mass.

The approach we take for estimating effective masses for large numbers of fermions exploits properties of the distribution for multifermion correlators.
Particularly, we have demonstrated through numerical studies as well as a mean-field calculation that the correlator distribution function is log-normal in character, thus motivating a method for extracting effective masses based on the properties of cumulant expansions.
Defined in terms of a cumulant expansion, the effective mass is given by \cite{PhysRevLett.107.201601}:
\begin{eqnarray}
\label{eq:eff_mass_cum}
m^{(N_\kappa)}_{eff}(\tau) = \frac{1}{\Delta\tau} \sum_{n=1}^{N_\kappa} \frac{1}{n!} \left[ \kappa_n(\tau) - \kappa_n(\tau+\Delta\tau) \right]\ ,
\end{eqnarray}
where $\kappa_n(\tau)$ in the $n$th cumulant of the distribution of the logarithm of the correlator.
For perfectly log-normal distributed correlation functions, the above expansion truncates exactly at second order, whereas for distributions that deviate from log-normal, such deviations are encoded in nonzero but presumably small contributions to the sum at order $n>2$.
In practice, the cumulants are estimated from the moments of the logarithm of the correlation function, and one must carefully study the effective masses as a function of the truncation order $N_\kappa$ to determine the ideal value at which statistical errors in the estimate of $\kappa_n(\tau)$ are comparable to the systematic error associated with the truncation of the expansion.

In each of our our studies, the ground state energy of the multifermion system has been estimated by performing correlated $\chi^2$ fits of the effective masses to a constant over the plateau region at late Euclidean time.
In most cases, we considered either two or three values of the sink parameter $\beta$ for each ensemble in order to gauge possible systematic errors associated with excited state contamination, which may arise due to poor overlap with the ground state for a given interpolating operator or due to energy splittings which are smaller than the typical inverse time considered.
Simultaneous correlated $\chi^2$ fits to the correlation functions were performed using all available correlators on a given ensemble.

In cases where a plateau in the effective mass plot failed to appear before the onset of noise, we instead fit the ground plus excited state using a constant plus an exponential fit function.
In all cases, the statistical uncertainties were obtained by resampling data using the bootstrapping method.
In order to take into account systematic effects due to temporal correlations and excited state contamination, we varied the end points of the fitting region by up to $\pm 3$ time steps and regarded the maximum and minimum fit values as our fitting systematic errors.
The total fitting uncertainty is determined by combining both statistical and systematic errors in quadrature.

\subsection{Few-body Results}
\label{sec:analysis_and_results.unitary_fermions_in_a_finite_box.few-body_results}

Numerical simulations of $N=2+1$ and $N=2+2$ untrapped unitary fermions at zero total momentum were performed in order to study finite volume effects as a function of $1/L$, as well as to make direct comparisons with precision benchmark results of previously reported studies.
In addition, a comparison with exact diagonalization results of the $N=2+1$ and $N=2+2$ unitary fermion transfer matrices on small volumes provide a nontrivial check for our lattice simulations.
As was the case in \Sec{three_unitary_fermions}, all few-body energies are measured in units of the non-interacting few-body energies, e.g., $E_{Free} = (2\pi/L)^2/M$ for both $N=2+1$ and $N=2+2$ fermions at zero total momentum.

\begin{figure}
\begin{center}
\includegraphics[width=\figwidth]{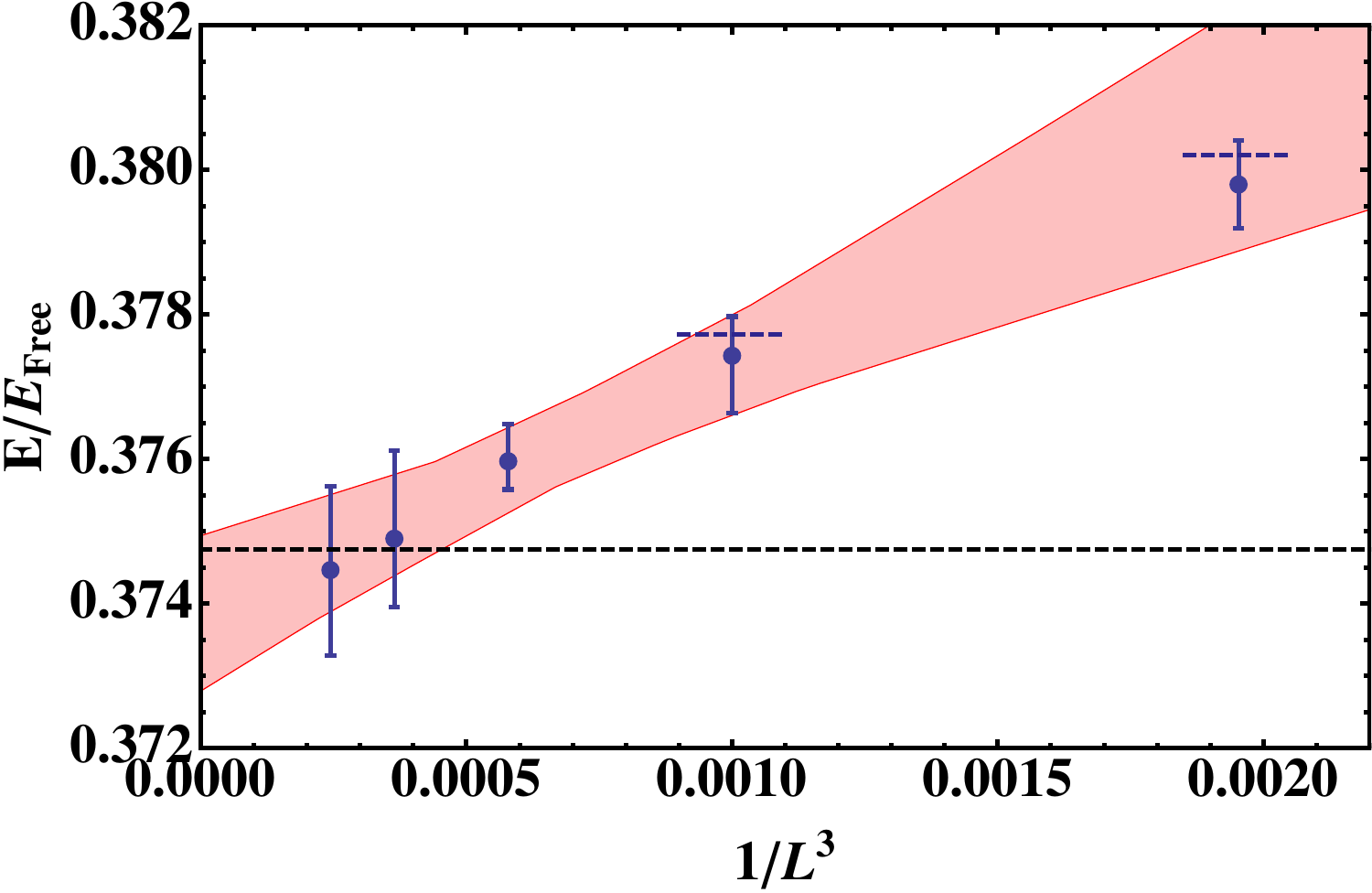}
\caption{%
\label{fig:N3_untrapped}%
Energy of $N=2+1$ unitary fermions in a zero total momentum eigenstate as a function of $1/L^3$.
Blue data points and associated error bars were obtained from numerical simulation, short blue dashed lines at $L=8$ and $L=10$ indicate results from exact diagonalization of the three fermion transfer matrix.
Red error band indicates the infinite volume extrapolation result previously reported in Ref. \cite{Endres:2011er} using simulation data.
Black dashed line indicated the exact infinite volume result of Pricoupenko and Castin reported in Ref. \cite{2007JPhA...4012863P}.
}
\end{center}
\end{figure}

In \Fig{N3_untrapped}, we plot simulation results for the energy of three unitary fermions at zero total momentum on lattice sizes up to $L=16$ and for $N_\calO=5$.
These results were originally reported in Ref. \cite{Endres:2011er}.
Exact ground state energies for the $A_1^+$ irrep obtained from diagonalizing the transfer matrix at $L=8$ and $L=10$ are indicated in the figure, and agree with our simulation results to within errors of $0.16\%$ and $0.18\%$, respectively.
As discussed in Ref. \cite{Endres:2011er}, the leading volume-dependent corrections to the energy of more than two unitary fermions with many two-body $s$-wave operators tuned is expected to be of order $L^{-3}$, coming from an untuned two-derivative two-body $p$-wave operator.
Subleading corrections are expected to be of order $L^{-4.33}$, due to the lowest dimension three-body operator, which has $\ell=0$ and scaling dimension $4.67$ \cite{2004cond.mat.12764T,Griesshammer:2005ga,Griesshammer:2005sj,2007PhRvD..76h6004N}.
Performing a fit to the data using the functional form $c_0+c_1/L^3$ yields an infinite volume extrapolation result of $E/E_{Free} = 0.3735^{+0.0014}_{-0.0007}$, and is consistent with the exact infinite volume result of Pricoupenko and Castin \cite{2007JPhA...4012863P} within $0.3\%$ uncertainties.

\begin{figure}
\begin{center}
\includegraphics[width=\figwidth]{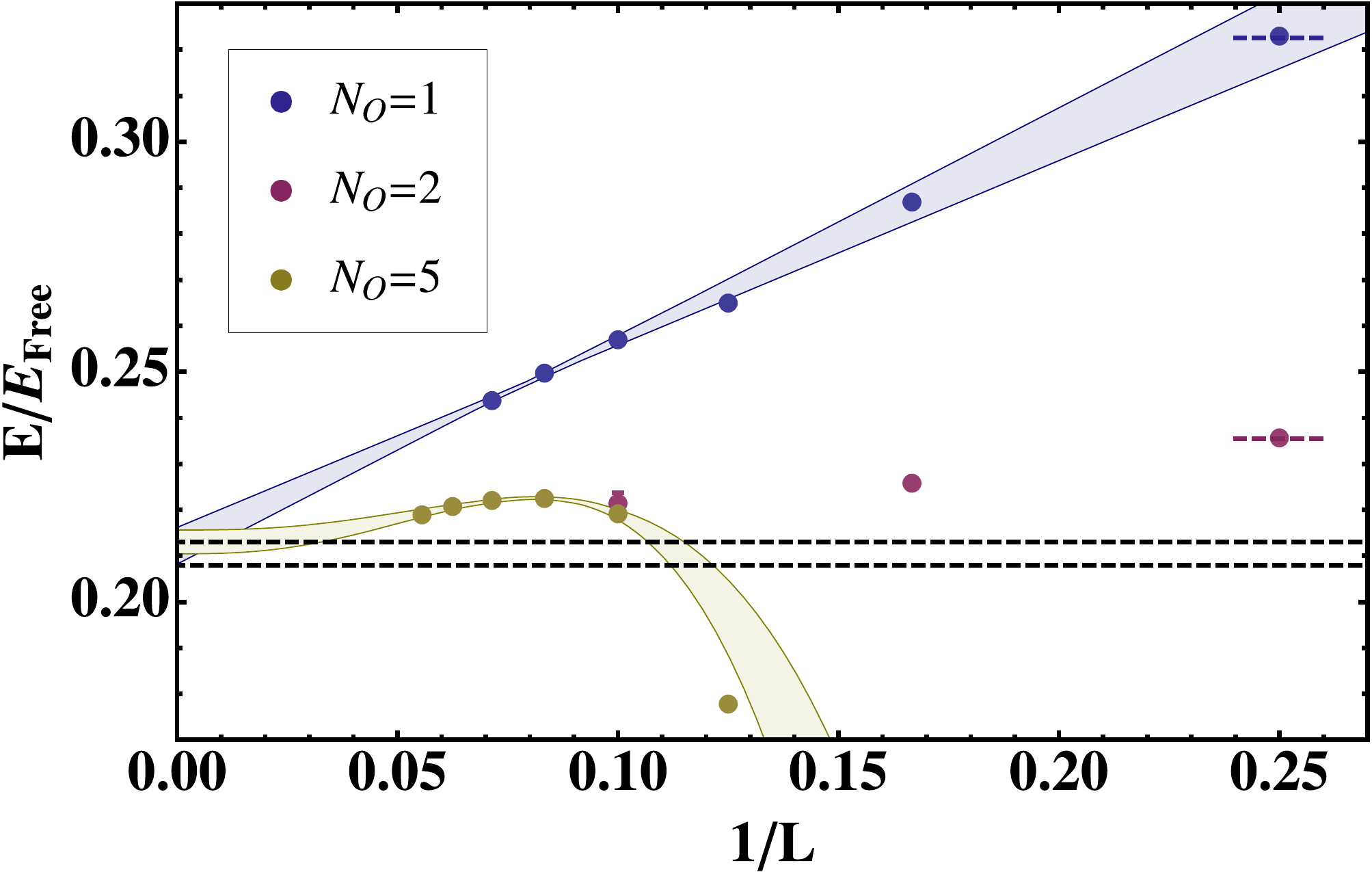}
\caption{%
\label{fig:N4_untrapped}%
Ground state energy of $N=2+2$ unitary fermions as a function of $1/L$.
Error bars include statistical and fitting systematic errors combined in quadrature.
The blue and yellow bands represent fit results to $N_\calO = 1$ and $N_\calO=5$ data as discussed in the text, with error bands reflecting both statistical and systematic errors.
Black dashed lines indicate the error band obtained from an infinite volume extrapolation of exact benchmark calculations reported in Ref. \cite{PhysRevA.83.063619}.
Short dashed lines at $L=4$ indicate energies obtained by exact diagonalizing the four-body transfer matrix.
}
\end{center}
\end{figure}

In \Fig{N4_untrapped}, we have summarized simulation results for the ground state energy of four unitary fermions for up to $N_\calO=5$ tuned couplings and lattice sizes up to $L=18$.
Exact lattice energies obtained for $L=4$ are plotted in \Fig{N4_untrapped} for $N_\calO=1$ and $2$ couplings tuned to unitarity.
In each case the exact ground state energies obtained from the transfer matrix are consistent with the simulation results within uncertainties.
In a high precision check, we found that the ground state energy of four unitary fermions at $L=4$ and $N_\calO=2$ obtained from ensembles of approximately $4B$ configurations agreed with exact results to within errors of $0.05\%$.

For $N_\calO=1$, the leading volume correction to the ground state energy for four fermions will be of order $1/L$, due to the untuned effective range operator.
To extract the ground state energy at $L=\infty$, we therefore used $c_0+c_1/L$ as our fit function for the extrapolation.
We take into account systematic errors in the infinite volume extrapolation by varying the fit interval from $L=[4,14]$ to $L=[10,14]$, and obtain $E/E_{Free}=0.2122(40)$ for the ground state energy.
For the highly tuned $N_\calO=5$ case, we expect the leading volume dependence for four fermions to be $L^{-3}$, using the same reasons as for three unitary fermions.
Unlike the case for three fermions, however, the lowest dimension three-fermion operator is expected to have $\ell=1$ and scaling dimension $4.27$ rather than $\ell=0$.
The reason is that three of the four fermions are not restricted to a specific angular momentum state. 
The subleading volume dependence is therefore expected to scale as $L^{-3.55}$.
Additional subleading terms scale as $L^{-4.33}$ and $L^{-5}$ corresponding to the $\ell=0$ three-body operator and the four-derivative $p$-wave and $d$-wave two-body operators, respectively. 
By considering the leading $L$-dependence induced by these operators, we use the fit function: $c_0+c_1/L^3+c_2/L^{3.55}$ to extrapolate the energy in the case of $N_\calO=5$.
The fit result over the interval $L=[10,16]$ is shown in \Fig{N4_untrapped}, and at infinite volume we obtain $E/E_{Free}=0.2130(26)$.
Both our $N_\calO=1$ and $N_\calO=5$ results for the ground state energy of four unitary fermions are consistent with the benchmark calculation reported \cite{PhysRevA.83.063619}, within the given uncertainties.

\subsection{Many-body Results}
\label{sec:analysis_and_results.unitary_fermions_in_a_finite_box.many-body_results}

\begin{figure}
\includegraphics[width=\figwidth]{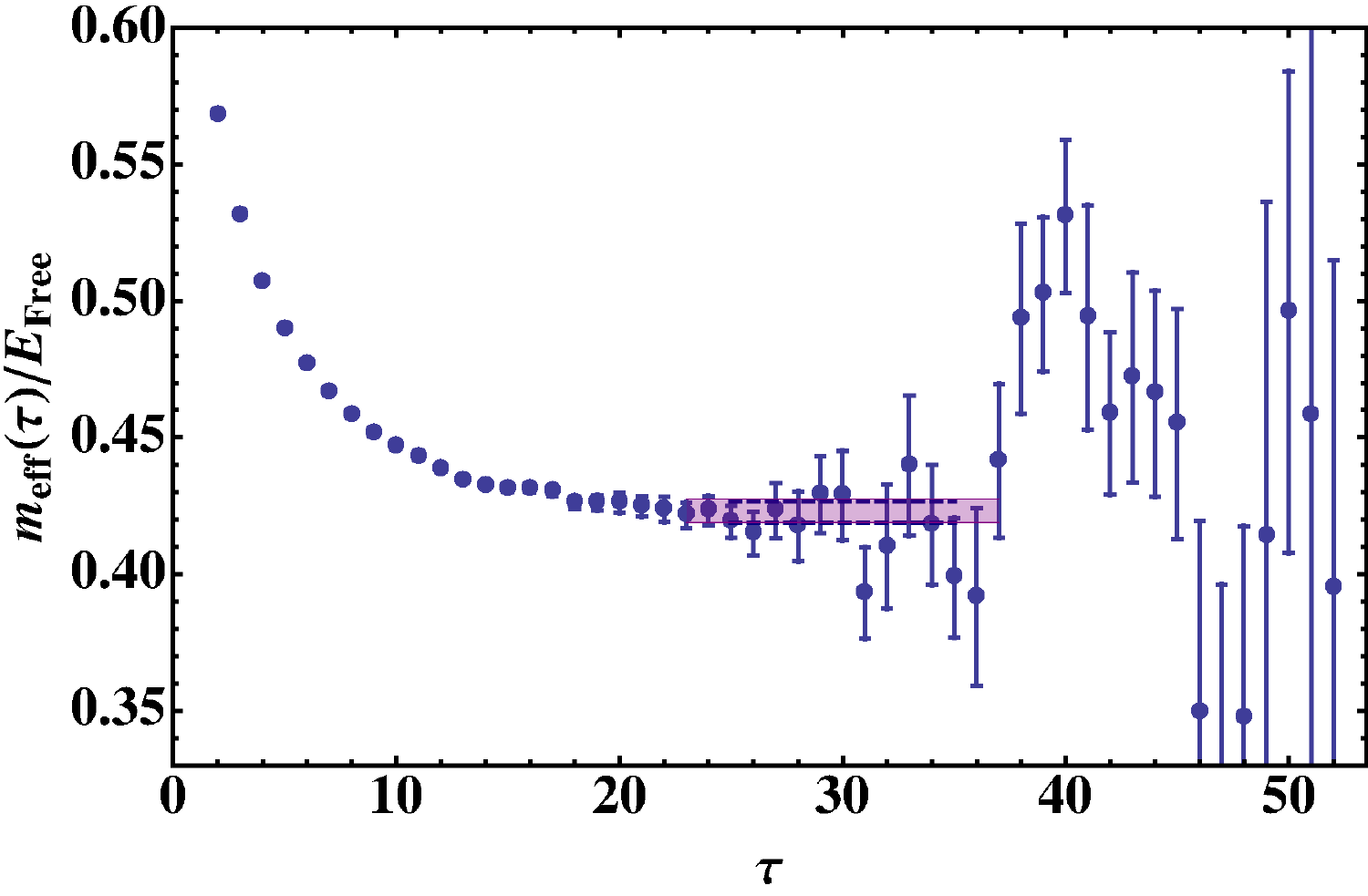}
\includegraphics[width=\figwidth]{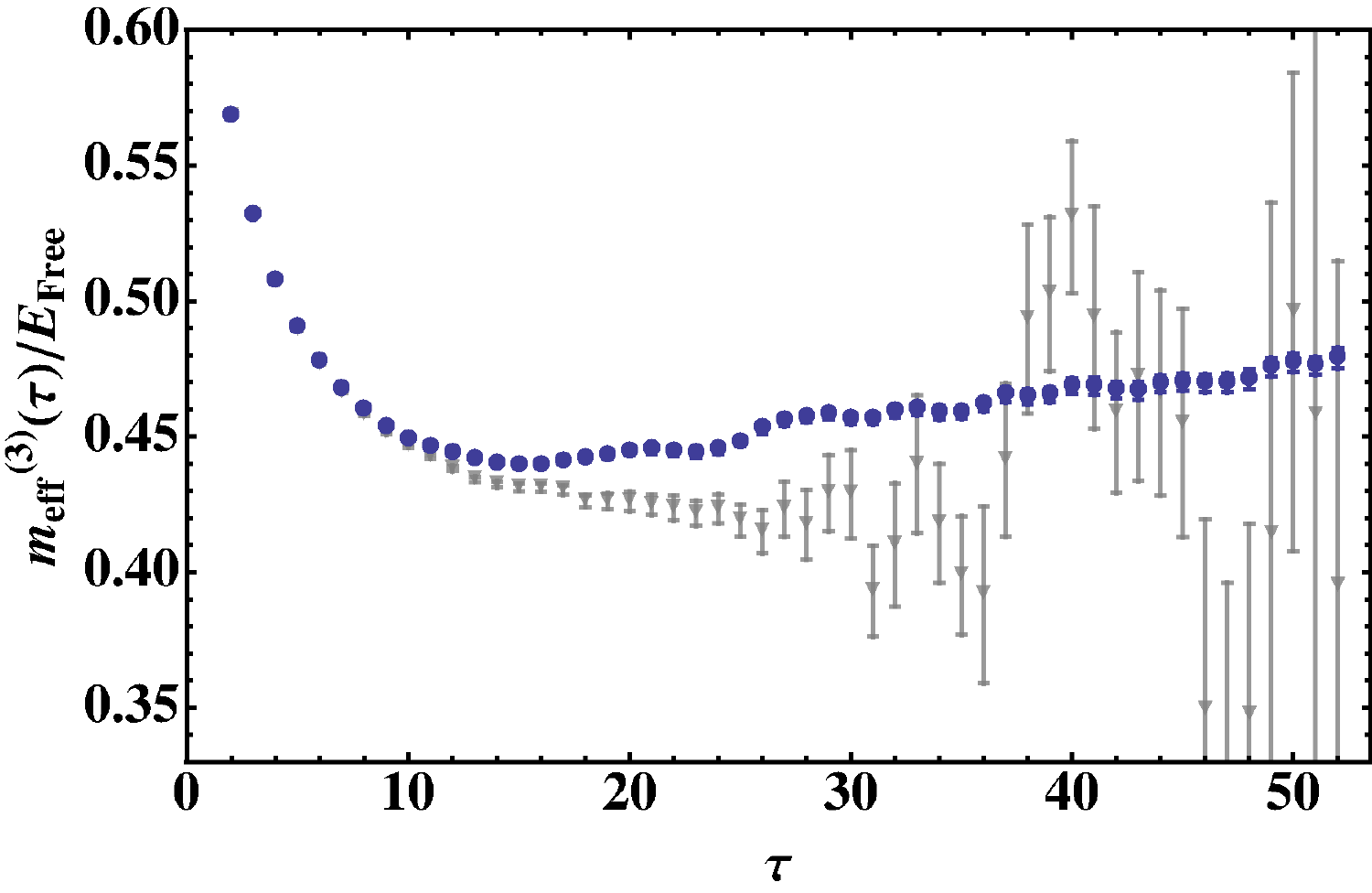}
\includegraphics[width=\figwidth]{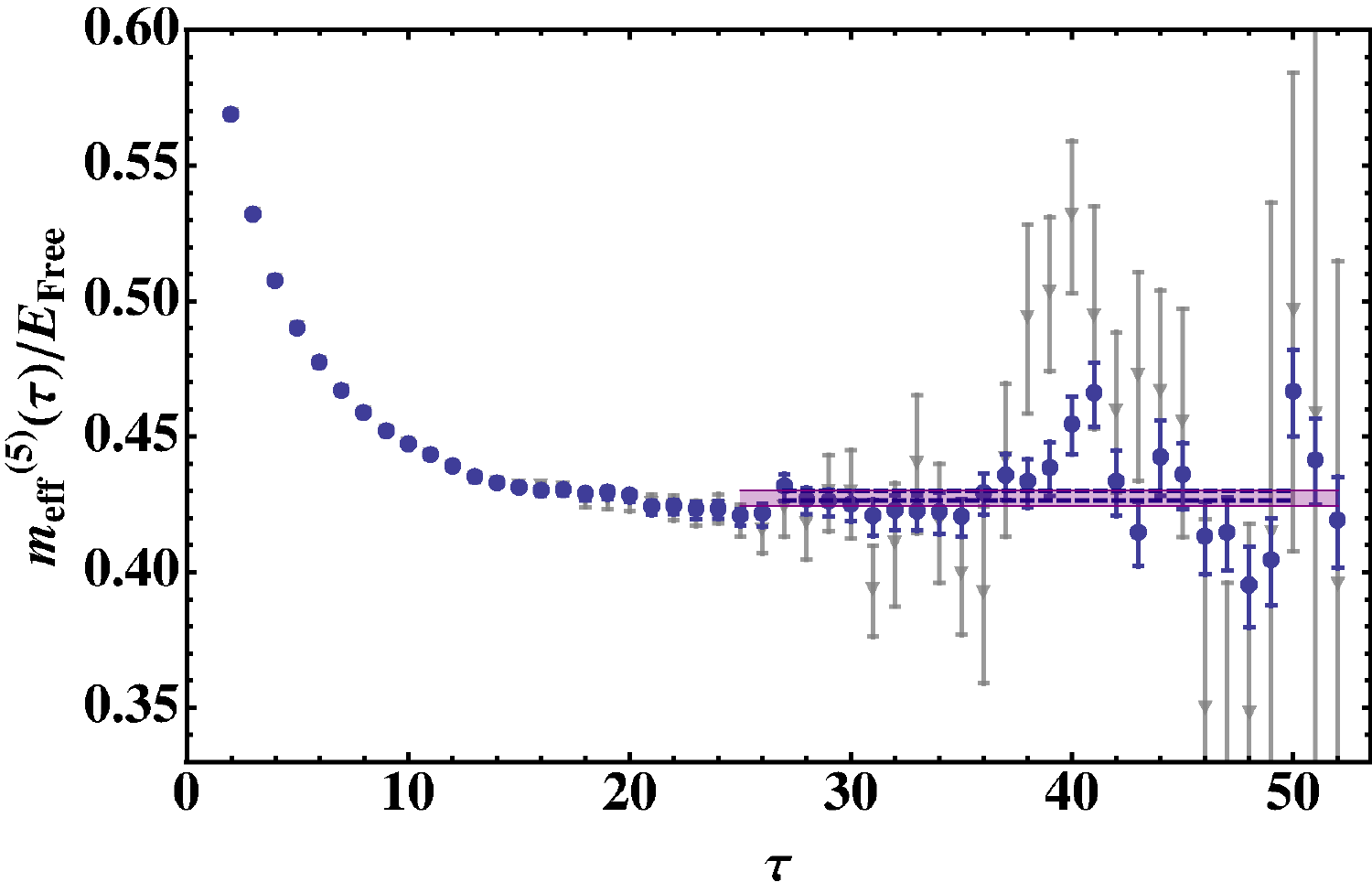}
\includegraphics[width=\figwidth]{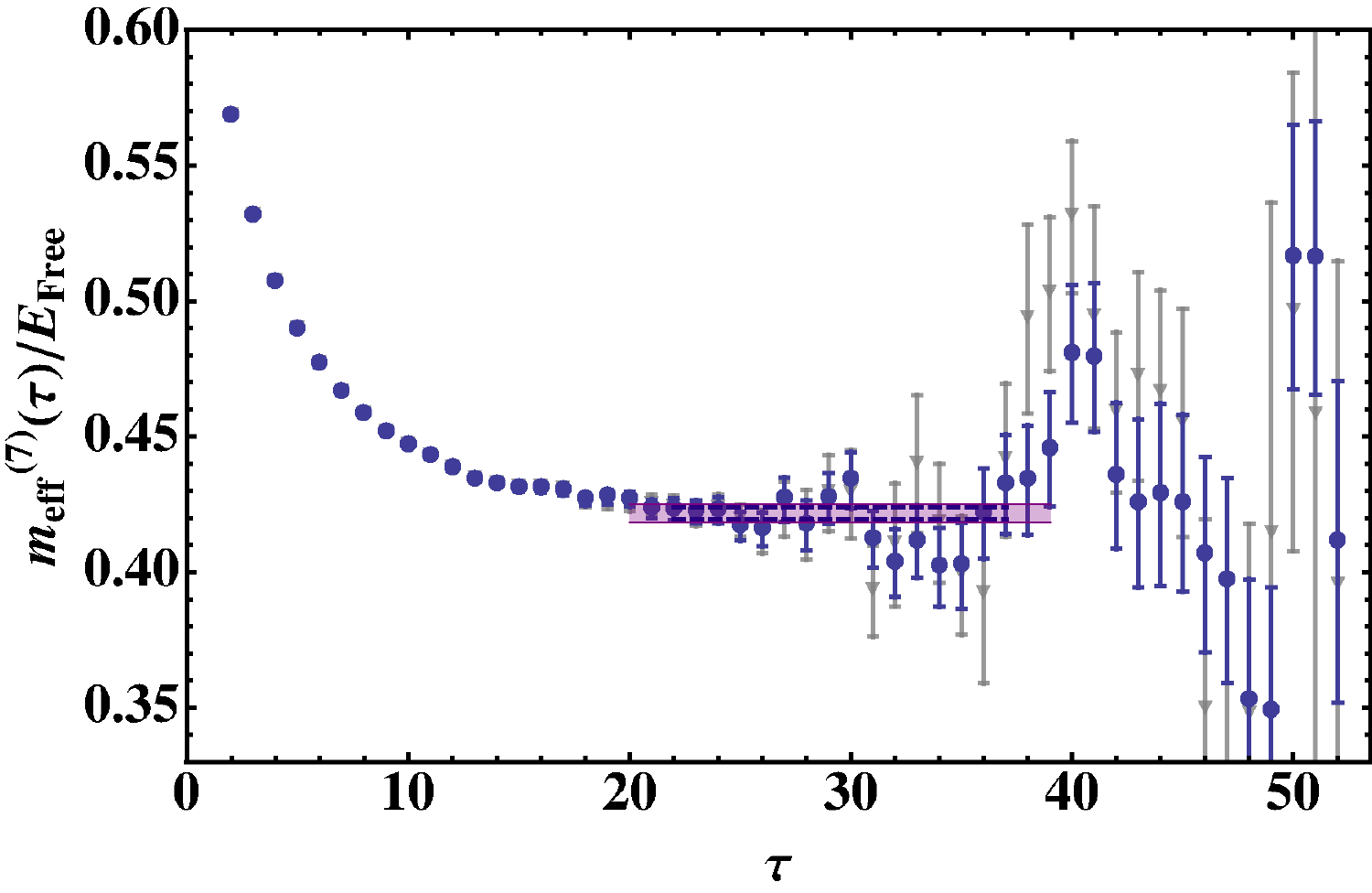}
\caption{%
\label{fig:untrapped_N10_effm}%
Conventional ($m_{eff}$) and cumulant ($m_{eff}^{(N_\kappa)}$) effective mass plots for $N=10$ unitary fermions with $\Delta \tau=2$ on a $L=12$ lattice.
Upper-left panel shows conventional, upper-right shows cumulant with $N_\kappa=3$, lower-left shows cumulant with $N_\kappa=5$, and lower-right shows cumulant with $N_\kappa=7$.
The purple band in the effective mass plots represent fits results to the plateau region when one exists; the gray data in the cumulant effective mass plots represent the effective mass obtained by using the conventional method.
}
\end{figure}

\begin{figure}
\includegraphics[width=\figwidth]{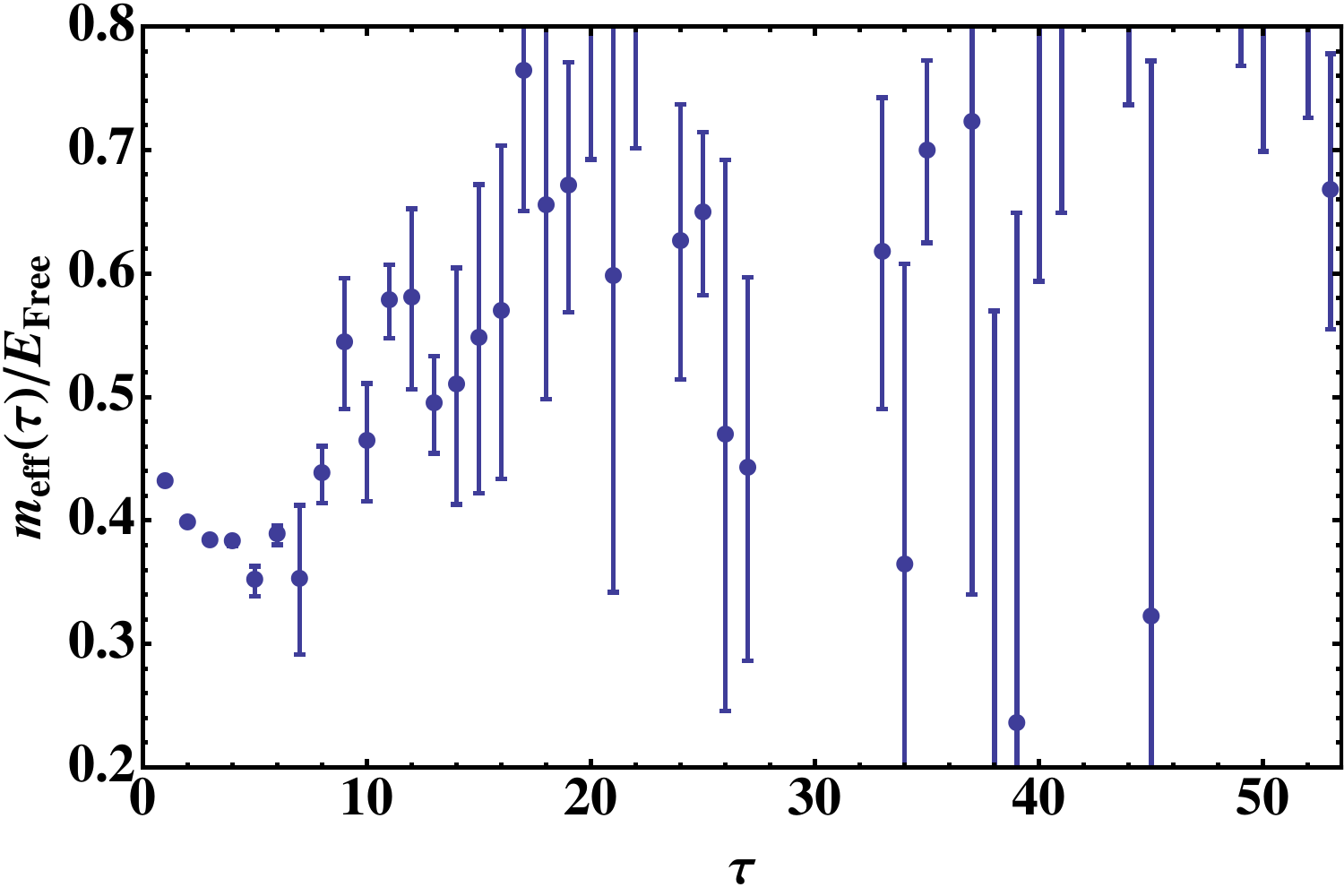}
\includegraphics[width=\figwidth]{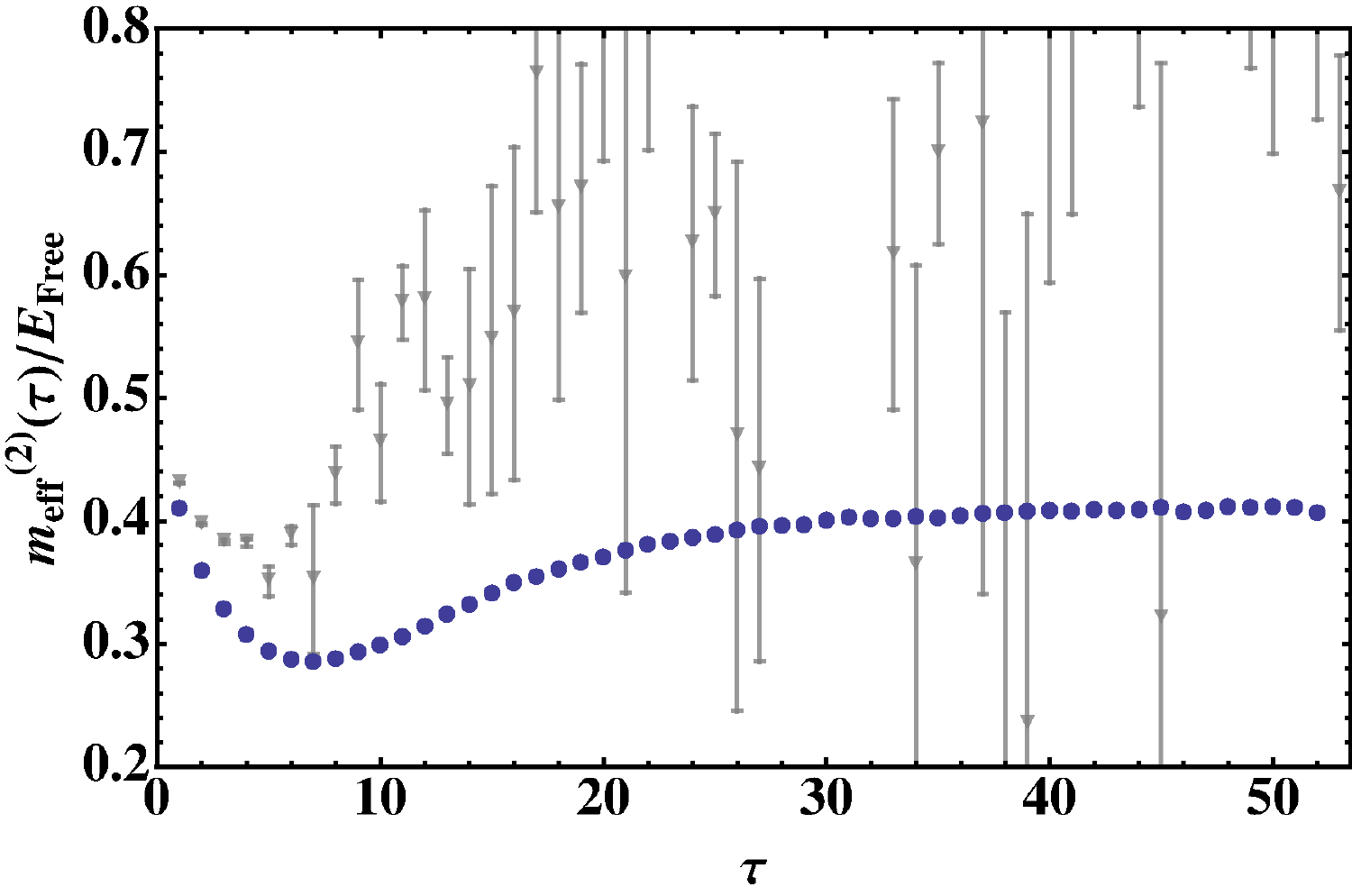}
\includegraphics[width=\figwidth]{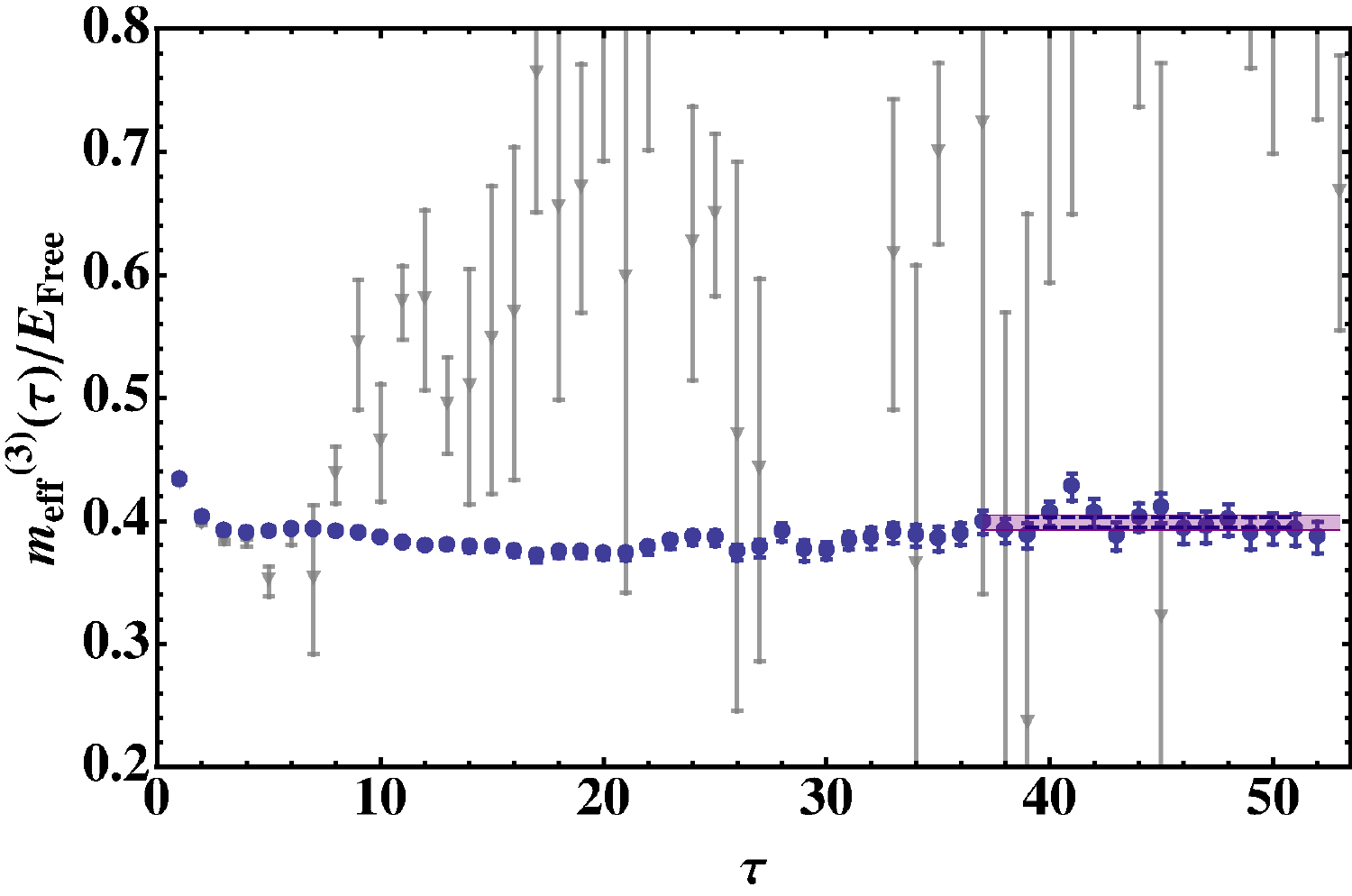}
\includegraphics[width=\figwidth]{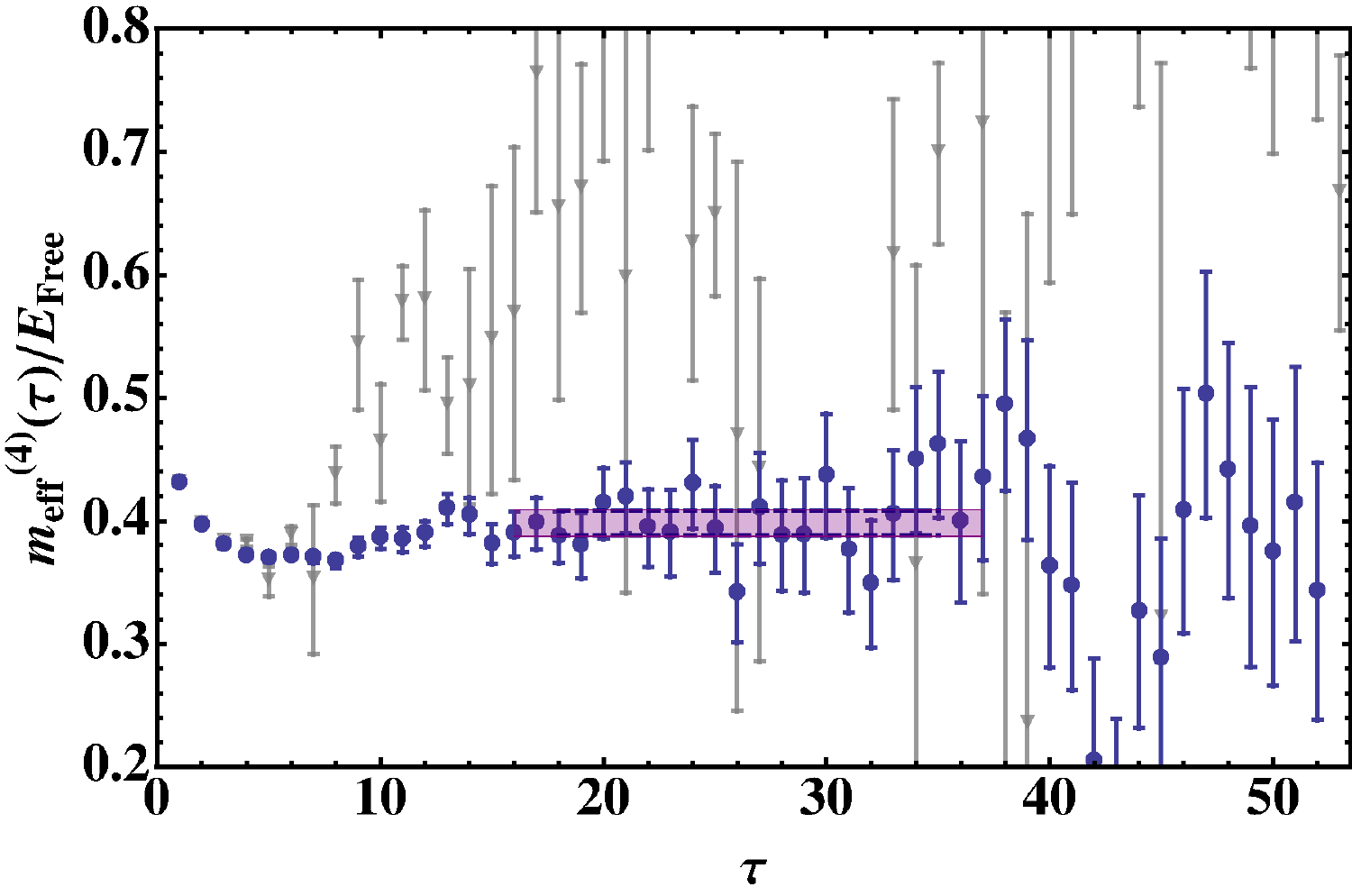}
\caption{%
\label{fig:untrapped_N50_effm}%
Conventional ($m_{eff}$) and cumulant ($m_{eff}^{(N_\kappa)}$) effective mass plots for $N=50$ unitary fermions with $\Delta \tau=2$ on a $L=12$ lattice.
Upper-left panel shows conventional, upper-right shows cumulant with $N_\kappa=2$, lower-left shows cumulant with $N_\kappa=3$ and lower-right shows cumulant with $N_\kappa=4$.
The purple band in the effective mass plots represent fit results to the plateau region when one exists (in the case of $N_\kappa=2$, a constant plus an exponential fit function was used), the gray data in the cumulant effective mass plots represent the effective mass obtained by using the conventional method.
}
\end{figure}

To determine the Bertsch parameter, we calculate the ground state energies of up to $66$ untrapped and unpolarized unitary fermions using the many-body ensembles described in \Tab{untrapped_ensembles}.
For small $N$, the distribution overlap problem is absent and the conventional effective mass defined in \Eq{eff_mass} typically shows an acceptable plateau.
On the other hand, for large $N$ the conventional effective mass exhibits a significant overlap problem and we generally fail to find plateaus.
Examples of each of these scenarios are shown in \Fig{untrapped_N10_effm} (upper-left) and \Fig{untrapped_N50_effm} (upper-left) for $N=10$ and $N=50$, respectively.
The conventional effective mass for $N=10$ in \Fig{untrapped_N10_effm} shows a plateau beginning at around $\tau\sim23$ and the ground state energy may be calculated by performing a constant fit to the plateau region before the onset of severe noise at $\tau\sim37$.
However, the conventional effective mass for $N=50$ in \Fig{untrapped_N50_effm} drifts upward beginning at $\tau\sim7$ and exhibits no plateau before the onset of an overlap problem. 

\begin{figure}
\includegraphics[width=\figwidth]{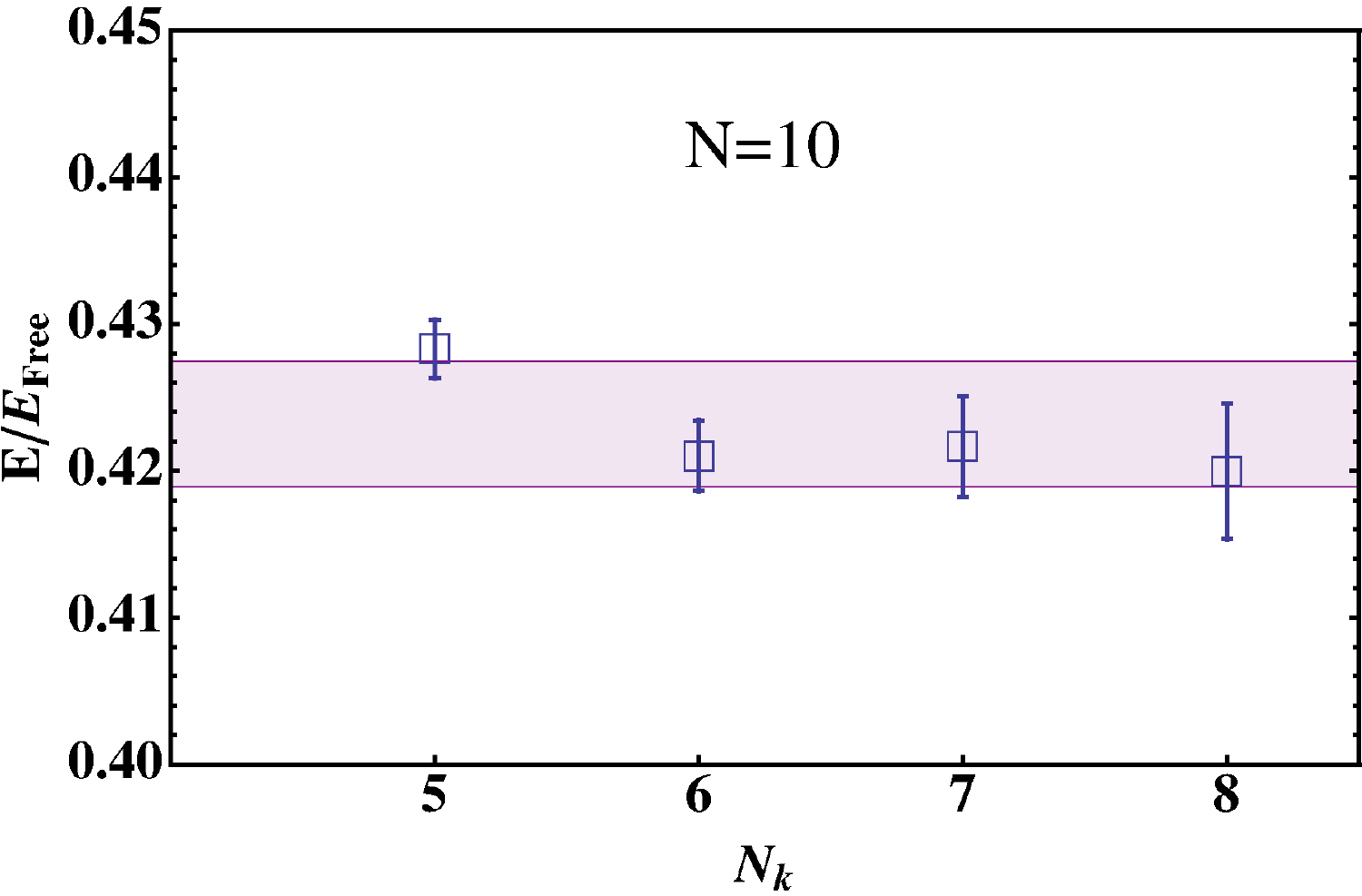}
\includegraphics[width=\figwidth]{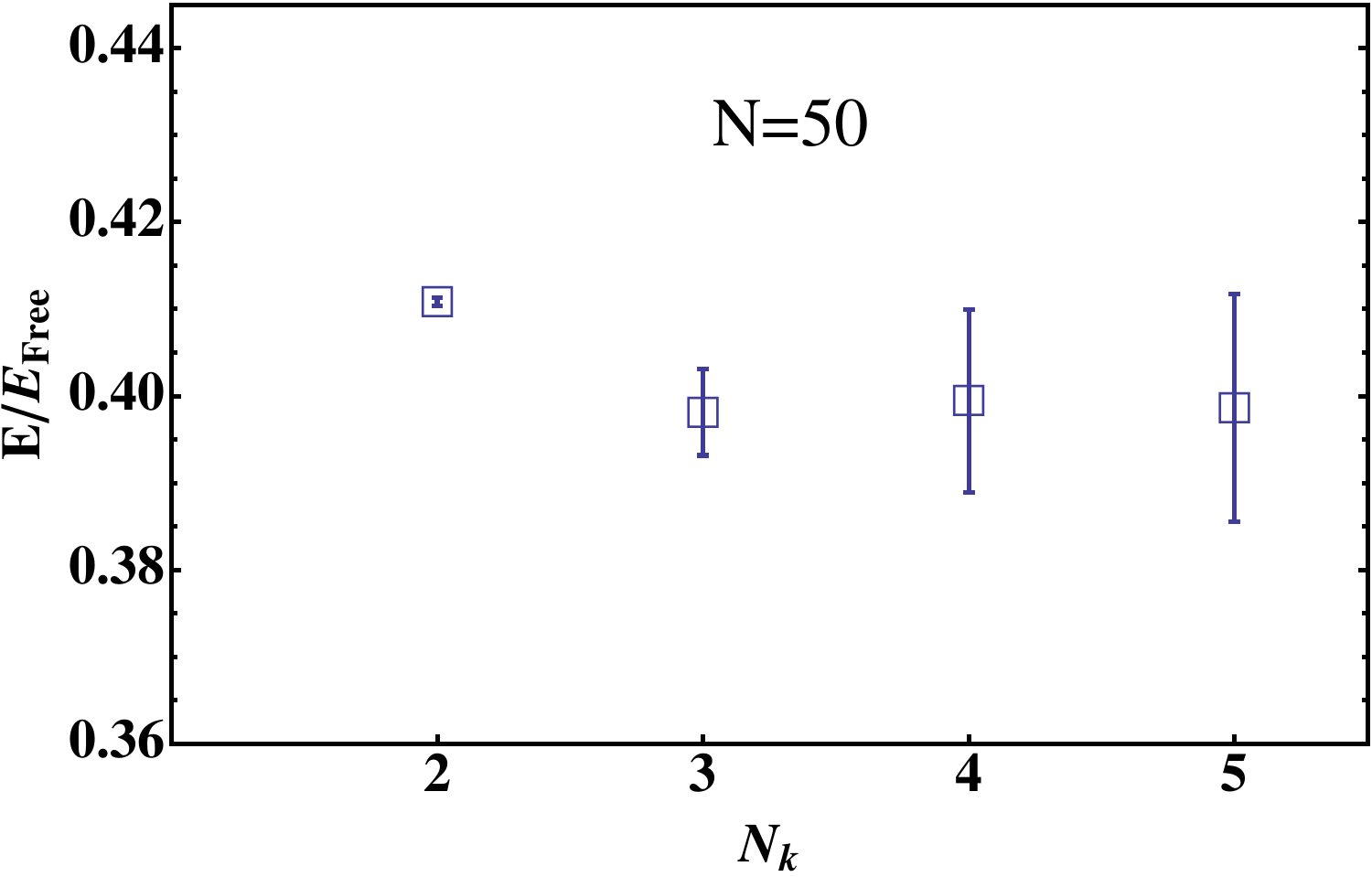}
\caption{%
\label{fig:untrapped_conv}%
Left: data points are the energies obtained using the cumulant effective mass for $N=10$ fermions, purple band is the energy obtained by using the conventional method. 
Right: data points are the energies obtained using the cumulant effective mass for $N=50$ fermions.
}
\end{figure}

As discussed in \Sec{analysis_technique}, we have used a cumulant expansion technique to overcome the distribution overlap problem for large $N$.
In particular, using the effective mass defined in \Eq{eff_mass_cum} at moderate truncation orders $N_\kappa$, we find that $m_{eff}^{(N_\kappa)}(\tau)$ exhibits clean plateaus for $N\le66$.
By performing a constant fit to the plateau region, we estimate the energy for each $N_\kappa$.
For small $N$, we may verify the cumulant method by comparing energies with those measured using the conventional analysis.
In \Fig{untrapped_N10_effm}, we plot $m_{eff}^{(N_\kappa)}$ for $N=10$ with $N_\kappa=3, 5, 7$ along with the conventional effective mass for comparison.
The cumulant effective mass with $N_\kappa=3$ shows a clean signal, but does not exhibit a plateau for any given time extent.
However, as we increase the truncation order $N_\kappa$, $m_{eff}^{(N_\kappa)}$ appears to converge, and beyond $N_\kappa=6$ the addition of higher cumulants only increases statistical noise without changing the plateau.
A plot of the energies obtained at each truncation order is presented in \Fig{untrapped_conv} (left), along with the energy measured from the conventional effective mass.
Our best estimation of the energy using the cumulant expansion is obtained for $N_\kappa=6$ using the convergence criteria outlined in Ref. \cite{PhysRevLett.107.201601}, and is consistent with the energy obtained by the conventional approach. 

For large numbers of unitary fermions, the conventional method fails to exhibit a plateau due to the onset of an overlap problem, as demonstrated in \Fig{untrapped_N50_effm} for $N=50$.
We must therefore rely entirely on the cumulant expansion to estimate energies in this case.
In \Fig{untrapped_N50_effm} we plot the effective masses from the cumulant expansion at truncation orders $N_\kappa=2,3,$ and $4$.
We find that $m_{eff}^{(N_\kappa)}$ for small $N_\kappa$ has a clean signal and the fit results to the plateau region shown in \Fig{untrapped_conv} (right) quickly converge as a function of $N_\kappa$.
Using the convergence criteria described in \cite{PhysRevLett.107.201601}, we choose $N_\kappa=3$ as the optimal truncation order for the cumulant expansion in this example.

\begin{figure}
\includegraphics[width=\figwidth]{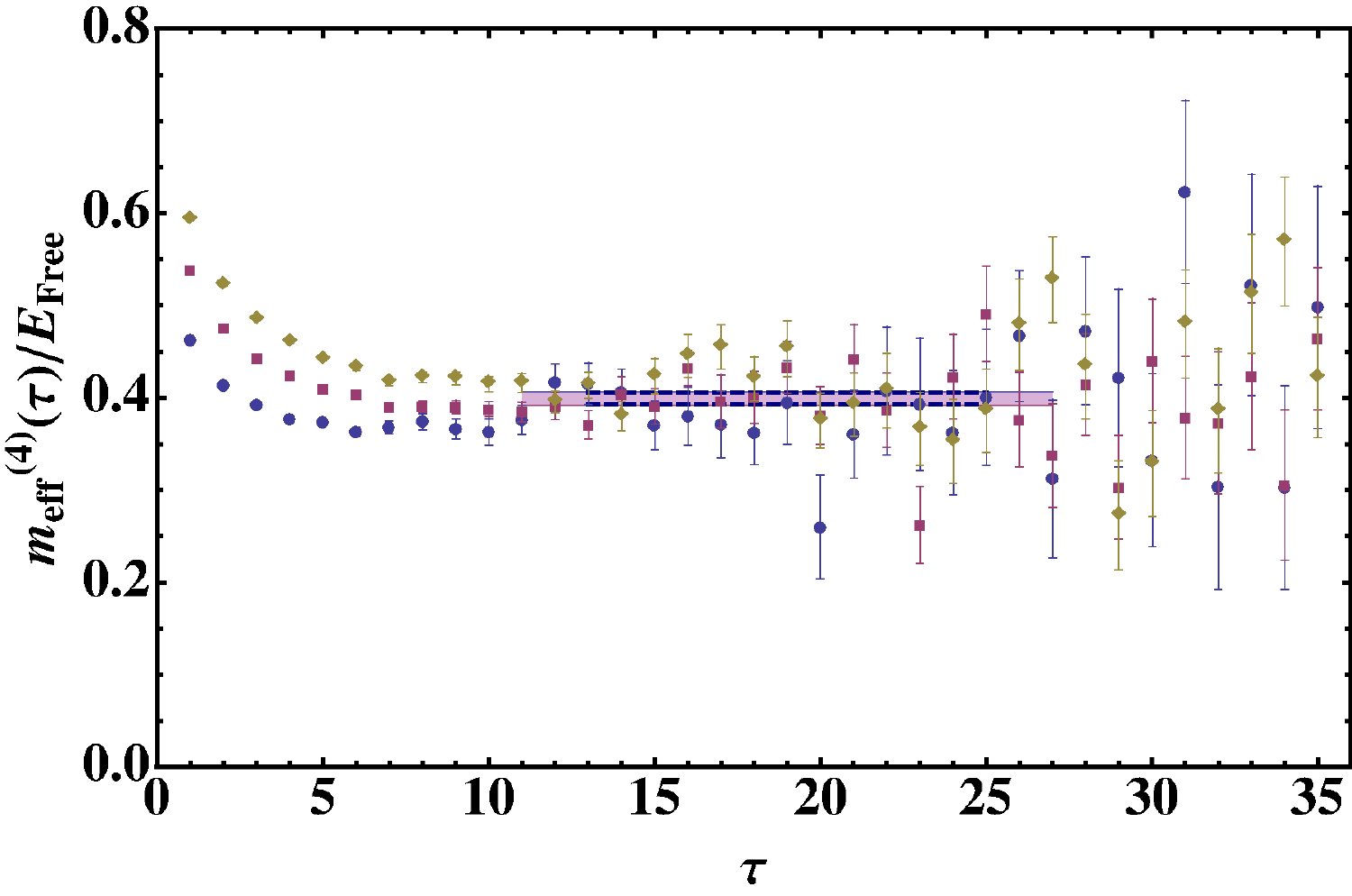}
\includegraphics[width=\figwidth]{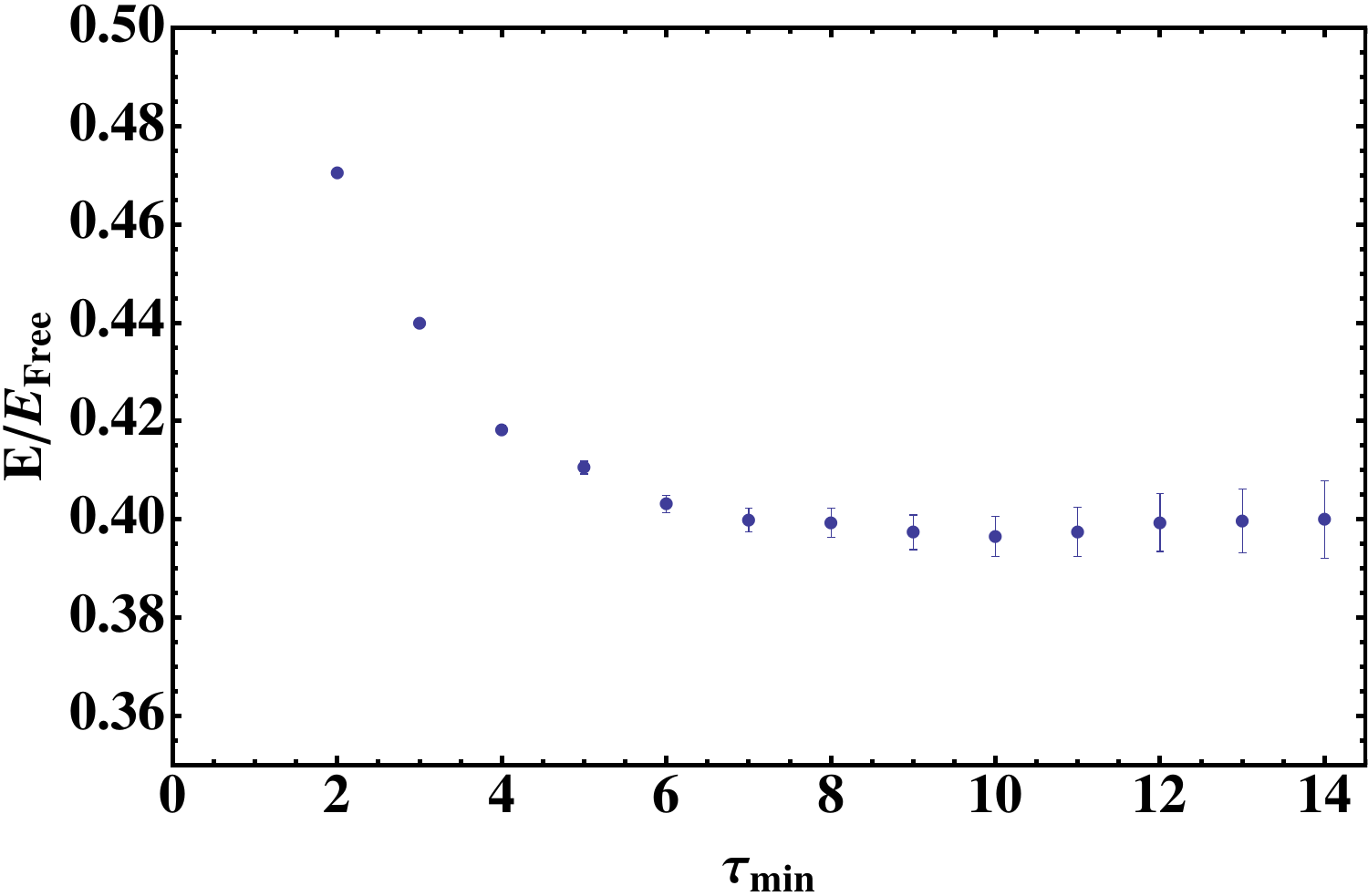}
\caption{%
\label{fig:untrapped_diff_effm}%
(Left) Effective mass plot obtained with an $N_\kappa=4$ truncation for $N=60$ untrapped unitary fermions on a $L=14$ lattice. Yellow diamonds, purple squares, and blue circles correspond to ensembles with sinks using $\beta=1.0$, $0.75$, and $0.5$, respectively. The dashed line represents the statistical uncertainty from a simultaneous fit, while the purple band represents the combined fitting statistical and systematic uncertainties. (Right) Results for simultaneous fits to the data in the left plot as a function of the beginning of the time interval used for fitting. The endpoint was held fixed at $\tau_{\mbox{max}}=25$.
}
\end{figure}

To improve our estimates of the ground state energy, we have considered several different choices for the free parameter $\beta$ appearing in the sink wavefunction defined in \Eq{sink_wavefunc}.
The optimal values which we considered are provided in \Tab{untrapped_ensembles}.
By performing simultaneous fits to the correlators at different values of $\beta$, we are able to obtain greater reliability in our ground state energies.
As an example, we plot the effective masses for $N=60$ unitary fermions on an $L=14$ lattice using the cumulant expansion method truncated at $N_k=4$ for three values of $\beta$ in \Fig{untrapped_diff_effm} (left).
In the same figure (right) is a plot of the results from a simultaneous fit to all three correlation functions as a function of the starting time of the fitting region (the endpoint defined as $\tau_{\mbox{max}}=25$ is held fixed).
Using such simultaneous fits, we are able to extract reliable energies from the plateau region where the three effective masses are statistically indistinguishable.

\begingroup
\squeezetable
\begin{table}[h!]
\caption{%
\label{tab:untrapped_paired_energies}%
Energies in units of the free gas energy for $N=N^\uparrow+N^\downarrow$ paired unitary fermions in a finite box. 
Extrapolated results for $L=\infty$ reflect the fit parameter $c_0(N)$ obtained using a three-parameter fit described in the text. 
The uncertainties represent the fitting statistical and systematic uncertainties combined in a quadrature.
}
\begin{ruledtabular}
\begin{tabular}{c|ccccc}
N & $L=10~~(N_k)$ & $L=12~~(N_k)$ & $L=14~~(N_k)$ & $L=16~~(N_k)$ & $L=\infty$\\
\hline
4\footnote{The ground state energies for $N=4$ are calculated from the ensembles used in \Sec{analysis_and_results.unitary_fermions_in_a_finite_box.few-body_results}}   & $0.2989^{+0.0012}_{-0.0011}$ & $0.3035^{+0.0032}_{-0.0032}$ & $0.3029^{+0.0055}_{-0.0047}$ & $0.3011^{+0.0013}_{-0.0010}$ & $0.2906^{+0.0035}_{-0.0035}$\\
6  & $0.403^{+0.007}_{-0.007}~~(8)$ & $0.401^{+0.004}_{-0.003}~~(8)$ & $0.408^{+0.001}_{-0.001}~~(8)$ & $0.406^{+0.002}_{-0.001}~~(8)$ & $0.401^{+0.010}_{-0.009}$\\
8  & $0.410^{+0.013}_{-0.009}~~(7)$ & $0.424^{+0.004}_{-0.004}~~(8)$ & $0.429^{+0.005}_{-0.004}~~(8)$ & $0.427^{+0.003}_{-0.003}~~(8)$ & $0.421^{+0.016}_{-0.017}$\\
10 & $0.420^{+0.009}_{-0.008}~~(6)$ & $0.421^{+0.005}_{-0.004}~~(8)$ & $0.427^{+0.004}_{-0.003}~~(8)$ & $0.422^{+0.004}_{-0.005}~~(8)$ & $0.424^{+0.019}_{-0.024}$\\
12 & $0.417^{+0.009}_{-0.012}~~(6)$ & $0.410^{+0.006}_{-0.006}~~(7)$ & $0.417^{+0.003}_{-0.002}~~(6)$ & $0.410^{+0.003}_{-0.003}~~(8)$ & $0.408^{+0.017}_{-0.025}$\\
14 & $0.406^{+0.004}_{-0.005}~~(5)$ & $0.399^{+0.006}_{-0.006}~~(6)$ & $0.404^{+0.003}_{-0.002}~~(7)$ & $0.395^{+0.004}_{-0.004}~~(8)$ & $0.392^{+0.018}_{-0.016}$\\
16 & $0.407^{+0.006}_{-0.007}~~(5)$ & $0.409^{+0.009}_{-0.010}~~(6)$ & $0.411^{+0.005}_{-0.003}~~(5)$ & $0.394^{+0.005}_{-0.004}~~(6)$ & $0.340^{+0.014}_{-0.018}$\\
18 & $0.404^{+0.014}_{-0.007}~~(5)$ & $0.411^{+0.012}_{-0.010}~~(6)$ & $0.412^{+0.008}_{-0.004}~~(7)$ & $0.400^{+0.008}_{-0.004}~~(6)$ & $0.359^{+0.025}_{-0.024}$\\
20 & $0.403^{+0.013}_{-0.009}~~(5)$ & $0.416^{+0.008}_{-0.009}~~(6)$ & $0.415^{+0.008}_{-0.006}~~(7)$ & $0.406^{+0.006}_{-0.007}~~(6)$ & $0.368^{+0.028}_{-0.028}$\\
22 & $0.411^{+0.009}_{-0.021}~~(5)$ & $0.416^{+0.007}_{-0.006}~~(6)$ & $0.413^{+0.010}_{-0.008}~~(6)$ & $0.406^{+0.008}_{-0.009}~~(6)$ & $0.380^{+0.026}_{-0.055}$\\
24 & $0.404^{+0.003}_{-0.003}~~(3)$ & $0.415^{+0.010}_{-0.009}~~(6)$ & $0.415^{+0.008}_{-0.012}~~(6)$ & $0.404^{+0.009}_{-0.010}~~(6)$ & $0.367^{+0.031}_{-0.048}$\\
26 & $0.400^{+0.003}_{-0.004}~~(3)$ & $0.413^{+0.014}_{-0.012}~~(6)$ & $0.413^{+0.012}_{-0.008}~~(6)$ & $0.404^{+0.005}_{-0.002}~~(6)$ & $0.377^{+0.016}_{-0.017}$\\
28 & $0.398^{+0.003}_{-0.004}~~(3)$ & $0.410^{+0.010}_{-0.010}~~(6)$ & $0.409^{+0.005}_{-0.004}~~(6)$ & $0.401^{+0.005}_{-0.003}~~(6)$ & $0.367^{+0.017}_{-0.023}$\\
30 & $0.394^{+0.003}_{-0.004}~~(3)$ & $0.407^{+0.010}_{-0.011}~~(6)$ & $0.405^{+0.006}_{-0.006}~~(6)$ & $0.397^{+0.006}_{-0.004}~~(6)$ & $0.367^{+0.022}_{-0.024}$\\
32 & $0.392^{+0.003}_{-0.005}~~(3)$ & $0.401^{+0.014}_{-0.012}~~(6)$ & $0.402^{+0.009}_{-0.008}~~(6)$ & $0.393^{+0.007}_{-0.004}~~(6)$ & $0.359^{+0.030}_{-0.027}$\\
34 & $0.391^{+0.003}_{-0.006}~~(3)$ & $0.399^{+0.008}_{-0.008}~~(4)$ & $0.395^{+0.011}_{-0.008}~~(6)$ & $0.393^{+0.003}_{-0.002}~~(5)$ & $0.373^{+0.021}_{-0.029}$\\
36 & $0.389^{+0.003}_{-0.005}~~(3)$ & $0.398^{+0.008}_{-0.009}~~(4)$ & $0.396^{+0.007}_{-0.006}~~(6)$ & $0.388^{+0.004}_{-0.003}~~(5)$ & $0.356^{+0.032}_{-0.029}$\\
38 & $0.388^{+0.004}_{-0.006}~~(3)$ & $0.398^{+0.008}_{-0.010}~~(4)$ & $0.397^{+0.006}_{-0.009}~~(4)$ & $0.384^{+0.003}_{-0.003}~~(5)$ & $0.334^{+0.032}_{-0.024}$\\
40 & $0.389^{+0.004}_{-0.005}~~(3)$ & $0.396^{+0.009}_{-0.008}~~(4)$ & $0.394^{+0.008}_{-0.008}~~(4)$ & $0.387^{+0.004}_{-0.004}~~(5)$ & $0.356^{+0.023}_{-0.030}$\\
42 & $0.389^{+0.004}_{-0.005}~~(3)$ & $0.394^{+0.007}_{-0.008}~~(4)$ & $0.395^{+0.010}_{-0.009}~~(4)$ & $0.388^{+0.004}_{-0.004}~~(5)$ & $0.365^{+0.024}_{-0.025}$\\
44 & $0.389^{+0.004}_{-0.005}~~(3)$ & $0.395^{+0.007}_{-0.009}~~(4)$ & $0.394^{+0.010}_{-0.008}~~(4)$ & $0.388^{+0.002}_{-0.003}~~(4)$ & $0.365^{+0.022}_{-0.025}$\\
46 & $0.389^{+0.004}_{-0.005}~~(3)$ & $0.396^{+0.008}_{-0.009}~~(4)$ & $0.392^{+0.006}_{-0.006}~~(4)$ & $0.389^{+0.002}_{-0.003}~~(4)$ & $0.370^{+0.023}_{-0.022}$\\
48 & $0.390^{+0.005}_{-0.005}~~(3)$ & $0.399^{+0.003}_{-0.003}~~(3)$ & $0.392^{+0.007}_{-0.007}~~(4)$ & $0.389^{+0.002}_{-0.003}~~(4)$ & $0.361^{+0.012}_{-0.013}$\\
50 & $0.387^{+0.006}_{-0.004}~~(3)$ & $0.399^{+0.003}_{-0.003}~~(3)$ & $0.392^{+0.005}_{-0.006}~~(4)$ & $0.389^{+0.002}_{-0.003}~~(4)$ & $0.359^{+0.012}_{-0.011}$\\
52 & $0.388^{+0.006}_{-0.004}~~(3)$ & $0.399^{+0.003}_{-0.003}~~(3)$ & $0.392^{+0.006}_{-0.006}~~(4)$ & $0.389^{+0.002}_{-0.003}~~(4)$ & $0.359^{+0.012}_{-0.012}$\\
54 & $0.388^{+0.006}_{-0.004}~~(3)$ & $0.399^{+0.003}_{-0.004}~~(3)$ & $0.393^{+0.007}_{-0.008}~~(4)$ & $0.388^{+0.003}_{-0.003}~~(4)$ & $0.357^{+0.019}_{-0.012}$\\
56 & $0.388^{+0.004}_{-0.003}~~(3)$ & $0.399^{+0.004}_{-0.004}~~(3)$ & $0.396^{+0.005}_{-0.007}~~(4)$ & $0.391^{+0.003}_{-0.003}~~(4)$ & $0.364^{+0.020}_{-0.011}$\\
58 & $0.389^{+0.004}_{-0.004}~~(3)$ & $0.399^{+0.004}_{-0.004}~~(3)$ & $0.397^{+0.006}_{-0.008}~~(4)$ & $0.393^{+0.004}_{-0.003}~~(4)$ & $0.370^{+0.019}_{-0.012}$\\
60 & $0.389^{+0.005}_{-0.004}~~(3)$ & $0.399^{+0.004}_{-0.004}~~(3)$ & $0.400^{+0.004}_{-0.004}~~(4)$ & $0.395^{+0.004}_{-0.003}~~(4)$ & $0.374^{+0.019}_{-0.013}$\\
62 & $0.389^{+0.004}_{-0.004}~~(3)$ & $0.399^{+0.004}_{-0.004}~~(3)$ & $0.402^{+0.005}_{-0.004}~~(4)$ & $0.396^{+0.004}_{-0.004}~~(4)$ & $0.377^{+0.018}_{-0.013}$\\
64 & $0.389^{+0.004}_{-0.004}~~(3)$ & $0.400^{+0.003}_{-0.004}~~(3)$ & $0.402^{+0.005}_{-0.005}~~(4)$ & $0.397^{+0.004}_{-0.004}~~(4)$ & $0.380^{+0.018}_{-0.013}$\\
66 & $0.390^{+0.005}_{-0.004}~~(3)$ & $0.400^{+0.003}_{-0.004}~~(3)$ & $0.404^{+0.005}_{-0.005}~~(4)$ & $0.398^{+0.004}_{-0.004}~~(4)$ & $0.382^{+0.019}_{-0.014}$\\
\end{tabular}
\end{ruledtabular}
\end{table}
\endgroup

\begin{figure}
\includegraphics[width=\figwidth]{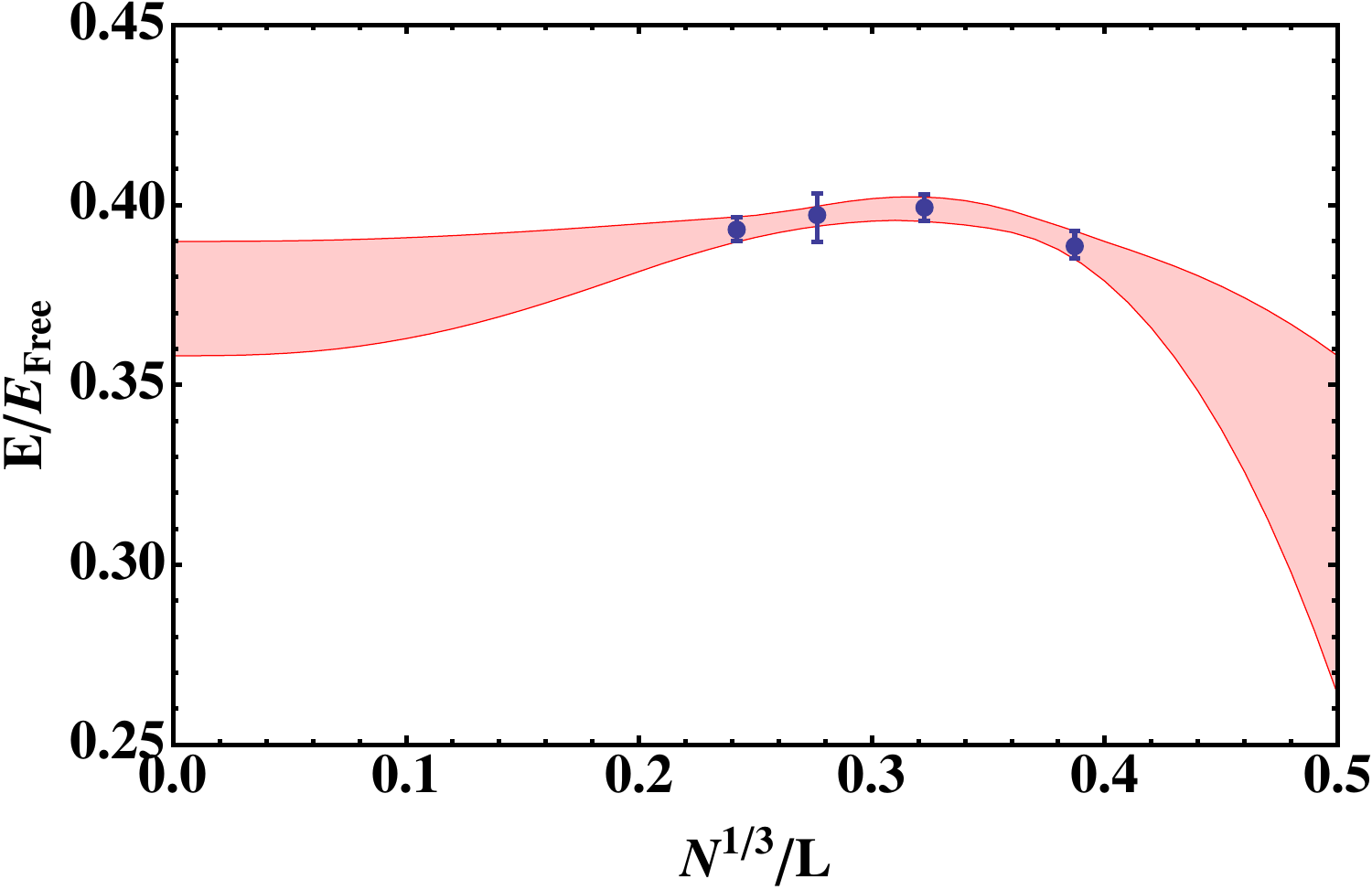}
\includegraphics[width=\figwidth]{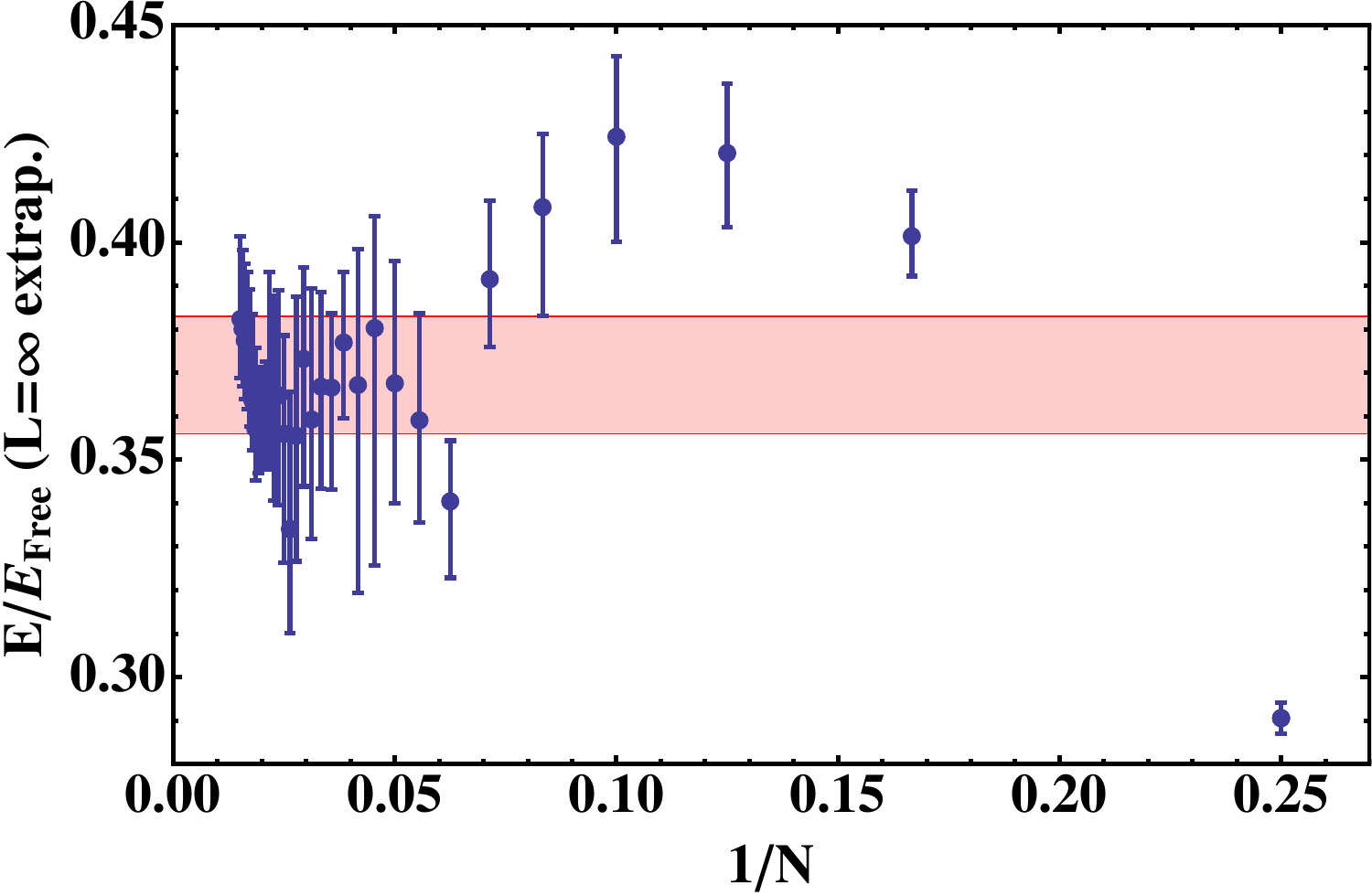}
\caption{%
\label{fig:untrapped_bertsch}%
(Left) An infinite volume extrapolation of the ground state energy for $N=58$ unitary fermions.
(Right) Ground state energy extrapolated to zero density as a function of $1/N$. The red band represents a constant fit to the energies for $40\leq N \leq 66$.
}
\end{figure}

The ground state energies for up to $66$ unpolarized unitary fermions at finite volume were estimated using the analysis techniques described above and reported in \Tab{untrapped_paired_energies} along with the truncation order $N_\kappa$ used to obtain the result at each $N$.
Energies are quoted in units of the thermodynamic limit definition of the free gas energy, $E_{Free}(N/V)$, for each value of $N$ and $L$.
The quoted errors represent both fitting statistical and systematic errors combined in quadrature as discussed in \Sec{analysis_technique}.
For each fixed value of  $N$, we performed an infinite volume (or equivalently a zero density) extrapolation of the energy using data obtained at different volumes. 
As was the case for $N=4$ fermions, we expect the leading and subleading volume dependence of the energies obtained using $N_{\mathcal{O}}=5$ tuned couplings to scale as $L^{-3}$ and $L^{-3.55}$, respectively, corresponding to effects induced by $\ell=1$ two- and three-body operators.
We therefore used the fit function $c_0(N)+c_1(N)/L^3+c_2(N)/L^{3.55}$ to perform an infinite volume the extrapolation.
An example of such a fit for $N=58$ untrapped fermions is shown in \Fig{untrapped_bertsch} (left).

The infinite volume extrapolated energies $E(N)/E_{Free}(N)\equiv c_0(N)$ are tabulated in \Tab{untrapped_paired_energies} and plotted in \Fig{untrapped_bertsch} (right) as a function of the inverse of fermion number.
Our results show that the shell structure is present in the first and second shells ($4\leq N \leq 38$), which is much more evident in the energies at finite volume. 
On the other hand, we find little evidence for shell effects within the last two shells (i.e., $40\leq N \leq 66$), suggesting that within the numerical uncertainty of our measurements, we are sufficiently near the thermodynamic limit to perform a thermodynamic limit extrapolation of the Bertsch parameter, given by $\xi = \lim_{N\to\infty} c_0(N)$. 
Note that the $N$-dependence is expected to be correlated since  the energies at different $N$ were determined from the same ensemble. 
To estimate the Bertsch parameter, we have performed a correlated constant fit to the infinite volume extrapolated energies over the fit range $N\in[40,66]$, obtaining the estimated value: $\xi = 0.366^{+0.016}_{-0.011}$.

\begin{figure}
\includegraphics[width= 0.8\textwidth]{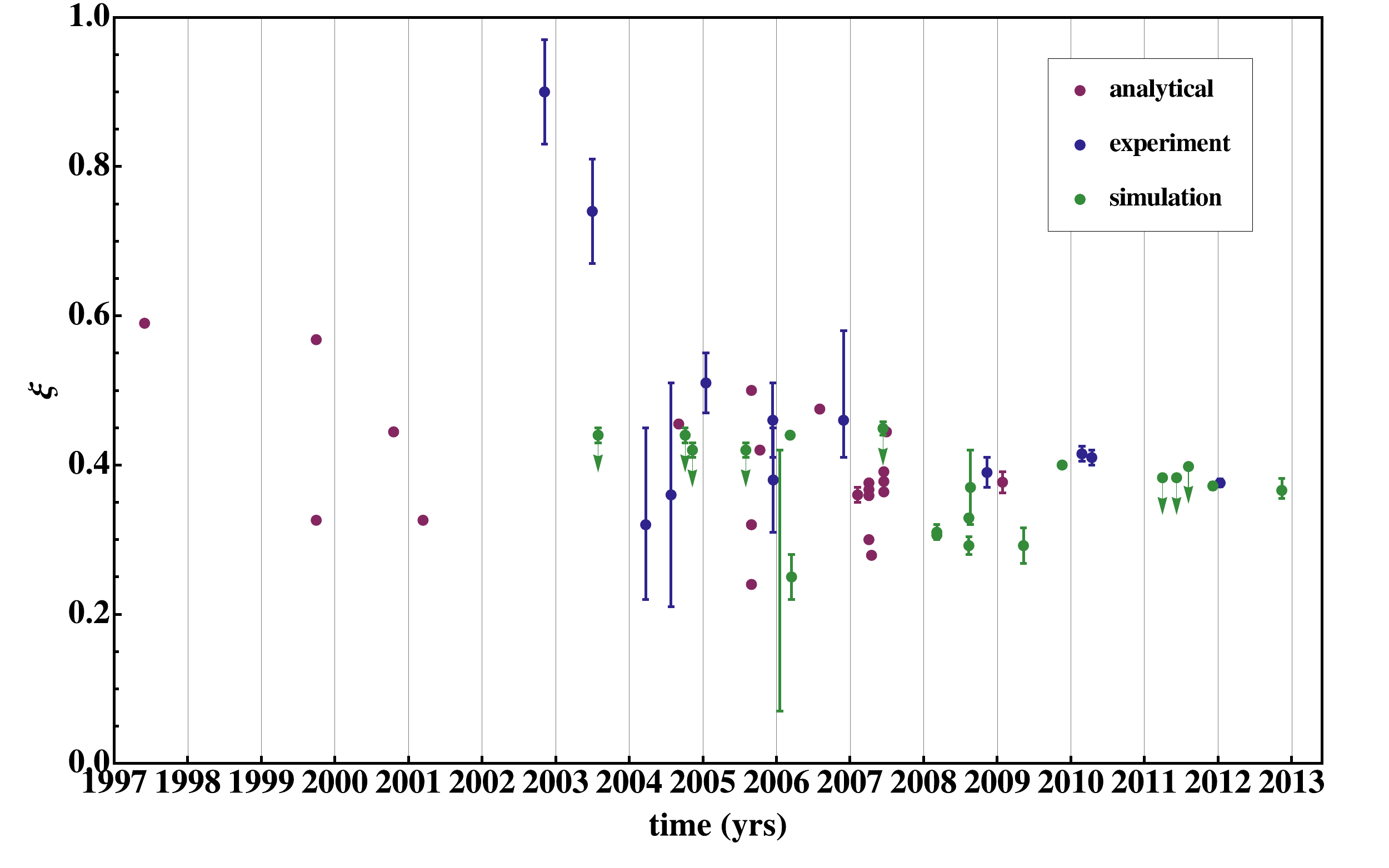}
\caption{%
\label{fig:bertsch_history}%
Historical results for the Bertsch parameter determined experimentally, by analytic calculation, and by numerical simulation.
Numerical values and citations are tabulated in \Tab{bertsch_history}; our value is indicated as the latest simulation data point.
}
\end{figure}

The Bertsch parameter has been extensively studied in the past using quantum Monte-Carlo (QMC) simulations.
The earliest works based on a variational approach found an upper bound of $\xi \leq 0.42(1)$ \cite{PhysRevA.70.043602,PhysRevLett.93.200404}, while a more recent QMC calculation for $N=66$ with an extrapolation to zero range reported an upper bound of $\xi \leq 0.383(3)$ \cite{PhysRevLett.106.235303,PhysRevA.83.041601}. 
Numerous lattice simulations of two-component fermions in the  unitary limit have been reported at both zero and nonzero temperature.
References \cite{PhysRevC.79.054003} and \cite{PhysRevA.78.023625} quoted the Bertsch parameter values $0.292(24)$ and $0.37(5)$, respectively, from finite temperature lattice simulations extrapolated to zero temperature.
A different zero temperature lattice calculation with an infinite volume extrapolation for $N=10$ and $N=14$ yielded $\xi=0.292(12)$ and $0.329(5)$, respectively \cite{PhysRevC.78.024001}. 
The Bertsch parameter has also been measured in several atomic experiments by studying pair correlation and absorption rates of $^{6}{\textrm{Li}}$ and $^{40}{\textrm{K}}$ in a harmonic trap.
Some recent experimental measurements reported $\xi=0.39(2)$ \cite{springerlink:10.1007/s10909-008-9850-2} by Duke and $0.41(1)$ \cite{Navon07052010} by the Paris group.
The most recent experimental determination by the group from the
Massachusetts Institute of Technology (MIT) found $0.376(4)$ \cite{Ku12012012}.
In \Fig{bertsch_history}, we summarize all analytical, numerical and experimental estimates of $\xi$ to date along with our value of the Bertsch parameter obtained from the simulations of up to $N=66$ untrapped unitary fermions.
References for the historical results are provided in \Tab{bertsch_history}.
Our determination of the Bertsch parameter appears as the latest data point in \Fig{bertsch_history} and is statistically consistent with other recent findings.

\begingroup
\squeezetable
\begin{table}
\caption{%
\label{tab:bertsch_history}%
Historical results for the Bertsch parameter $\xi$ determined experimentally (exp.), by numerical simulation (sim.) and by analytic calculation (anal.), along with publication (pub.) date.
Values obtained variationally are upper bounds, and are indicated with an asterisk; simulation results without a quoted error bar should be regarded as approximate.
}
\begin{ruledtabular}
\begin{tabular}{ccl|ccl|ccl}
pub. date & $\xi$ (exp.) & ref. & pub. date &  $\xi$ (sim.) & ref. & pub. date &  $\xi$ (anal.) & ref. \\ 
\hline
2002-11-07 & 0.90(7)            & \cite{O'Hara13122002}                         &  2003-07-31 & $0.44(1)^*$    & \cite{PhysRevLett.91.050401}                   & 1997-06-01 & 0.59      & \cite{PhysRevB.55.15153}      \\
2003-07-02 & 0.74(7)            & \cite{PhysRevA.68.011401}                     &  2004-10-05 & $0.44(1)^*$    & \cite{PhysRevA.70.043602}                      & 1999-10-01 & 0.326     & \cite{PhysRevC.60.054311}     \\
2004-07-27 & 0.36(15)           & \cite{PhysRevLett.93.050401}                  &  2004-11-10 & $0.42(1)^*$    & \cite{PhysRevLett.93.200404}                   & 1999-10-01 & 0.568     & \cite{PhysRevC.60.054311}     \\
2004-03-23 & $0.32^{+13}_{-10}$ & \cite{PhysRevLett.92.120401}                  &  2005-08-02 & $0.42(1)^*$    & \cite{PhysRevLett.95.060401}                   & 2000-10-19 & 4/9       & \cite{2000nucl.th..10066S}    \\
2005-01-16 & 0.51(4)            & \cite{Kinast25022005}                         &  2006-01-18 & 0.07-0.42      & \cite{PhysRevC.73.015202}                      & 2001-03-14 & 0.326     & \cite{PhysRevA.63.043606}     \\
2005-12-14 & 0.46(5)            & \cite{Partridge27012006}                      &  2006-03-10 & 0.44           & \cite{PhysRevLett.96.090404}                   & 2004-09-03 & 0.455     & \cite{PhysRevLett.93.100404}  \\
2005-12-16 & 0.38(7)            & \cite{PhysRevLett.95.250404}                  &  2006-03-17 & 0.25(3)        & \cite{PhysRevB.73.115112}                      & 2005-08-30 & 0.32      & \cite{2005NuPhA.762...82S}    \\
2006-11-30 & $0.46^{+12}_{-5}$  & \cite{PhysRevLett.97.220406}                  &  2007-06-14 & $0.449(9)^*$   & \cite{1367-2630-9-6-163}                       & 2005-08-30 & 0.24      & \cite{2005NuPhA.762...82S}    \\
2008-11-11 & 0.39(2)            & \cite{springerlink:10.1007/s10909-008-9850-2} &  2008-03-07 & 0.31(1)        & \cite{springerlink:10.1140/epja/i2008-10537-2} & 2005-08-30 & 0.5       & \cite{2005NuPhA.762...82S}    \\
2010-04-15 & 0.41(1)            & \cite{Navon07052010}                          &  2008-03-07 & 0.306(1)       & \cite{springerlink:10.1140/epja/i2008-10537-2} & 2005-10-11 & 0.42      & \cite{PhysRevA.72.041603}     \\
2010-02-25 & 0.415(10)          & \cite{2010Natur.463.1057N}                    &  2008-08-13 & 0.292(12)      & \cite{PhysRevC.78.024001}                      & 2006-08-04 & 0.475     & \cite{PhysRevLett.97.050403}  \\
2012-01-12 & 0.376(4)           & \cite{Ku12012012}                             &  2008-08-13 & 0.329(5)       & \cite{PhysRevC.78.024001}                      & 2007-02-08 & 0.36(1)   & \cite{PhysRevA.75.023610}     \\
           &                    &                                               &  2008-08-21 & 0.37 (5)       & \cite{PhysRevA.78.023625}                      & 2007-04-18 & 0.279     & \cite{PhysRevA.75.043614}     \\
           &                    &                                               &  2009-05-11 & 0.292(24)      & \cite{PhysRevC.79.054003}                      & 2007-04-05 & 0.300     & \cite{PhysRevA.75.043605}     \\
           &                    &                                               &  2009-11-19 & 0.4            & \cite{PhysRevLett.103.210403}                  & 2007-04-05 & 0.367     & \cite{PhysRevA.75.043605}     \\
           &                    &                                               &  2011-04-01 & $0.383(1)^*$   & \cite{PhysRevA.83.041601}                      & 2007-04-05 & 0.359     & \cite{PhysRevA.75.043605}     \\
           &                    &                                               &  2011-06-10 & $0.383(1)^*$   & \cite{PhysRevLett.106.235303}                  & 2007-04-05 & 0.376     & \cite{PhysRevA.75.043605}     \\ 
           &                    &                                               &  2011-08-08 & $0.398(3)^*$   & \cite{PhysRevA.84.023615}                      & 2007-06-18 & 0.391     & \cite{PhysRevA.75.063617}     \\
           &                    &                                               &  2011-12-07 & 0.372(5)       & \cite{PhysRevA.84.061602}                      & 2007-06-18 & 0.364     & \cite{PhysRevA.75.063617}     \\
           &                    &                                               &             &                &                                                & 2007-06-18 & 0.378     & \cite{PhysRevA.75.063617}     \\
           &                    &                                               &             &                &                                                & 2007-07-01 & 4/9       & \cite{0256-307X-24-7-011}     \\
           &                    &                                               &             &                &                                                & 2009-01-27 & 0.377(14) & \cite{PhysRevA.79.013627}     \\
\end{tabular}
\end{ruledtabular}
\end{table}
\endgroup

\section{Conclusion}
\label{sec:conclusion}

We have studied up to $66$ unpolarized unitary fermions in a periodic box by applying a lattice Monte Carlo method developed for studying large numbers of strongly interacting nonrelativistic spin-$1/2$ fermions \cite{Endres:2011er}.
Our method differs from methods used in the past in that it does not make use of importance sampling, nor is it variational in nature.
As such, our approach not only allows us to study unpolarized Fermi systems, but also systems with unequal numbers of spin up and spin down fermions.
One of the main obstacles in calculating ground state energies of large numbers of fermions using our method is that it exhibits a severe distribution overlap problem, resulting in unreliable estimates of correlation functions.
To solve this problem, we use a cumulant expansion technique for the logarithm of correlators \cite{PhysRevLett.107.201601}, which allows us to determine energies in a reliable manner with controlled systematic errors.
The successful application of our method to unitary fermions gives us confidence that these techniques may prove useful in other situations where importance sampling is difficult.
Conventional importance sampling schemes for Fermi gas calculations often use the $N$-body correlator itself as an importance measure and so the ensemble generated is only of use for estimating a single observable for which it was designed.
Our approach offers an advantage over such importance sampling schemes in that one may use the ensemble generated to reliably estimate all desired observables.
Thus our approach avoids the multiplicative enhancement in computational cost by the number of measured observables which is inherent in calculations based on importance sampling.

Our main findings for this study are summarized as follows:
\begin{enumerate}
\item The exact diagonalization of the transfer matrix for two, three, and four fermions enables us to verify our simulation results, and to study systematic errors from spatial and temporal discretization.
Our results for the spectrum of three and four fermions are in good agreement with benchmark calculations from other groups.
While few-body systems were used in this paper as a way to test our methodology, they are interesting in their own right, and it looks feasible to use our methods in the future to measure the fascinating anomalous scaling behavior expected of three-body interactions.
\item As part of our study of systematic effects, we have calculated the two-body $s$- and $p$-wave scattering phase phase shifts and few-body excited state energies in the lattice theory for various choices of $L$ and $N_\calO$.
\item Due to the highly improved, Galilean invariant action for which the first few terms in the effective range expansion for $s$-wave scattering have been systematically eliminated, we find mild volume dependence in the energies for the four volumes considered.
The remaining finite-volume or discretization effects, where the leading contributions come from the $p$-wave and three-body operators, are eliminated by performing an infinite volume or equivalently a zero density extrapolation.
\item The many-body ground state energies (measured in units of the noninteracting energies) show no discernible shell effects for $N\ge 40$, implying that the system is near the thermodynamic limit and therefore a reliable determination of the Bertsch parameter is possible.
We determined the Bertsch parameter to approximately 4\% statistical and systematic uncertainties and find agreement with the most recent experimental and numerical determinations by other research groups, thus demonstrating the success of our lattice construction and novel analysis methods as applied to many-body calculations.
\end{enumerate}

Our work shows that by combining novel statistical techniques and by perfecting the action it is possible to perform lattice Monte Carlo calculations that are quite competitive with methods that employ costly importance sampling.
It is possible that the operator basis in \Eq{Ofunc} that we chose for perfecting the interactions is not the optimal set, and that our method for fixing the operator coefficients is not the optimal strategy.
An interesting direction for future research would be to understand better whether there exists such an optimal strategy for perfecting the action, and whether there are benefits in combining both the perfect action and importance sampling techniques to further extend the computational reach of simulations for trapped atoms.

Note added: After this work was completed, Ref. \cite{2012arXiv1203.2565D} appeared using similar methods as described here, with compatible results for the unextrapolated Bertsch parameter at similar volumes.

\acknowledgments
Computations for this work were carried out in part on facilities of the USQCD Collaboration, which are funded by the Office of Science of the U.S. Department of Energy.
This research was also supported in part by the National Science Foundation through TeraGrid resources provided by the TeraGrid Science Gateways program.
This work was supported by U. S. Department of Energy Grant No. DE-FG02-00ER41132 (to D. B. K., J.-W. L., and A. N. N.) and  by MEXT Grant-in-Aid for Young Scientists (B) (23740227) (to M. G. E.).

\appendix
\section{Transfer matrices}
\label{sec:transfer_matrices}

The ensemble average of a direct product of $N$ propagators $K^{-1}(\tau;0)$ yields a $V^N$-dimensional matrix:
\begin{eqnarray}
\hat\calU_N(\tau) = \langle \underbrace{ K^{-1}(\tau;0) \otimes \ldots \otimes K^{-1}(\tau;0) }_N \rangle\ ,
\end{eqnarray}
which may be related to the Euclidean time-evolution operator for a system of $N$ particles.
Specifically one may define the time evolution operator as
\begin{eqnarray}
\calU_N(\tau) = \hat\calU_N(0)^{-1/2}\hat\calU_N(\tau)\hat\calU_N(0)^{-1/2}\ ,
\end{eqnarray}
which satisfies the properties:
\begin{eqnarray}
\calU_N(\tau)^\dagger &=& \calU_N(\tau)\ ,\cr
\calU_N(\tau)\calU_N(\tau') &=& \calU_N(\tau+\tau') + \calO(b_\tau)\ , \cr
\calU_N(0) &=& 1\ .
\end{eqnarray}
One may then derive an analytic expression for the $N$-particle transfer matrix, given by $\calT_N = \calU_N(1)$, the eigenvalues of which yield the exponentiated energies of the $N$-body system.

Explicitly, the matrix elements of the $N$-body transfer matrix in momentum space is given by:
\begin{eqnarray}
\langle \bfq_1',\ldots,\bfq_N' | \calT_N | \bfq_1, \ldots, \bfq_N\rangle
&=& \prod_{i=1}^N \frac{1}{\sqrt{\xi(\bfq_i')\xi(\bfq_i)}} \left[ \prod_{i=1}^N \delta_{\bfq_i',\bfq_i} \right. \cr
&+& \left. \frac{1}{2V} \sum_{i\neq j}^N C(\bfq_i'-\bfq_i) \delta_{\bfq_i'-\bfq_i,\bfq_j'-\bfq_j} \prod_{k\neq i,j}^N \delta_{\bfq_k',\bfq_k} +  \ldots  \right]\ ,
\end{eqnarray}
where the ellipses represents higher order terms involving two or more contributions from the interaction $C$.
In addition, the ellipses include contact terms which come from a slight modification of Wick's theorem in the case of $Z_2$ fields.
\footnote{For example, $\langle \phi_\bfx \phi_\bfy \phi_\bfz \phi_\bfw \rangle = \delta_{\bfx\bfy}\delta_{\bfz\bfw} + \delta_{\bfx\bfz}\delta_{\bfy\bfw} + \delta_{\bfx\bfw}\delta_{\bfy\bfz} + w \delta_{\bfx\bfy}\delta_{\bfy\bfz}\delta_{\bfz\bfw}$, where $w=0$ for Gaussian and $w=1$ for $Z_2$ fields.} 
Such contributions only appear in the case of four or more particles, however.

In the case of $N=N^\uparrow + N^\downarrow$ fermions, the above transfer matrix must be antisymmetrized with respect to momenta corresponding to each species.
Although the transfer matrix is $V^N$ dimensional, we may reduce the dimensionality by a power of volume by projecting onto the center-of-mass (c.m.) frame using the projection operator:
\begin{eqnarray}
\langle \bfq_1',\ldots,\bfq_N' |\calP_{\mbox{c.m.}} |\bfq_1,\ldots,\bfq_N \rangle = \delta_{\bfq_1'+\ldots+\bfq_N',0} \prod_{i=1}^N \delta_{\bfq_i',\bfq_i}\ .
\end{eqnarray}
A further reduction of the dimensionality may be performed by projecting the transfer matrix onto the positive and negative parity irreducible representations (irreps) $r = A_1^\pm, A_2^\pm, E^\pm, T_1^\pm, T_2^\pm$ of the octahedral group $O_h$, using the projection operator:
\begin{eqnarray}
\langle \bfq_1',\ldots,\bfq_N'| \calP_{r}| \bfq_1,\ldots,\bfq_N \rangle = \frac{1}{48} \sum_{g} \chi_r(g) \prod_{i=1}^N \delta_{R(g)\bfq_i',\bfq_i}\ .
\end{eqnarray}
In this expression, the sum is over the 48 group elements $g$ of $O_h$, $\chi_r(g)$ are the characters of the representation $r$, and $R(g)$ are the three-dimensional rotation matrices corresponding to each group element $g$.
The irreps $r$ have dimensionality $d_r = 1, 1, 2, 3$, and $3$, respectively.

\bibliography{unitary}

\begin{thebibliography}{70}
\expandafter\ifx\csname natexlab\endcsname\relax\def\natexlab#1{#1}\fi
\expandafter\ifx\csname bibnamefont\endcsname\relax
  \def\bibnamefont#1{#1}\fi
\expandafter\ifx\csname bibfnamefont\endcsname\relax
  \def\bibfnamefont#1{#1}\fi
\expandafter\ifx\csname citenamefont\endcsname\relax
  \def\citenamefont#1{#1}\fi
\expandafter\ifx\csname url\endcsname\relax
  \def\url#1{\texttt{#1}}\fi
\expandafter\ifx\csname urlprefix\endcsname\relax\def\urlprefix{URL }\fi
\providecommand{\bibinfo}[2]{#2}
\providecommand{\eprint}[2][]{\url{#2}}

\bibitem[{\citenamefont{Baker}(2000)}]{baker2000mbx}
\bibinfo{author}{\bibfnamefont{G.}~\bibnamefont{Baker}}, in
  \emph{\bibinfo{booktitle}{Recent progress in many-body theories: the
  proceedings of the 10th international conference, Seattle, USA, September
  10-15, 1999}} (\bibinfo{organization}{World Scientific Pub Co Inc},
  \bibinfo{year}{2000}), vol.~\bibinfo{volume}{3}, p.~\bibinfo{pages}{15}.

\bibitem[{\citenamefont{O'Hara et~al.}(2002)\citenamefont{O'Hara, Hemmer, Gehm,
  Granade, and Thomas}}]{O'Hara13122002}
\bibinfo{author}{\bibfnamefont{K.~M.} \bibnamefont{O'Hara}},
  \bibinfo{author}{\bibfnamefont{S.~L.} \bibnamefont{Hemmer}},
  \bibinfo{author}{\bibfnamefont{M.~E.} \bibnamefont{Gehm}},
  \bibinfo{author}{\bibfnamefont{S.~R.} \bibnamefont{Granade}},
  \bibnamefont{and} \bibinfo{author}{\bibfnamefont{J.~E.}
  \bibnamefont{Thomas}}, \bibinfo{journal}{Science}
  \textbf{\bibinfo{volume}{298}}, \bibinfo{pages}{2179} (\bibinfo{year}{2002}),
  \eprint{http://www.sciencemag.org/content/298/5601/2179.full.pdf},
  \urlprefix\url{http://www.sciencemag.org/content/298/5601/2179.abstract}.

\bibitem[{\citenamefont{Gehm et~al.}(2003)\citenamefont{Gehm, Hemmer, Granade,
  O'Hara, and Thomas}}]{PhysRevA.68.011401}
\bibinfo{author}{\bibfnamefont{M.~E.} \bibnamefont{Gehm}},
  \bibinfo{author}{\bibfnamefont{S.~L.} \bibnamefont{Hemmer}},
  \bibinfo{author}{\bibfnamefont{S.~R.} \bibnamefont{Granade}},
  \bibinfo{author}{\bibfnamefont{K.~M.} \bibnamefont{O'Hara}},
  \bibnamefont{and} \bibinfo{author}{\bibfnamefont{J.~E.}
  \bibnamefont{Thomas}}, \bibinfo{journal}{Phys. Rev. A}
  \textbf{\bibinfo{volume}{68}}, \bibinfo{pages}{011401}
  (\bibinfo{year}{2003}),
  \urlprefix\url{http://link.aps.org/doi/10.1103/PhysRevA.68.011401}.

\bibitem[{\citenamefont{Bourdel et~al.}(2004)\citenamefont{Bourdel, Khaykovich,
  Cubizolles, Zhang, Chevy, Teichmann, Tarruell, Kokkelmans, and
  Salomon}}]{PhysRevLett.93.050401}
\bibinfo{author}{\bibfnamefont{T.}~\bibnamefont{Bourdel}},
  \bibinfo{author}{\bibfnamefont{L.}~\bibnamefont{Khaykovich}},
  \bibinfo{author}{\bibfnamefont{J.}~\bibnamefont{Cubizolles}},
  \bibinfo{author}{\bibfnamefont{J.}~\bibnamefont{Zhang}},
  \bibinfo{author}{\bibfnamefont{F.}~\bibnamefont{Chevy}},
  \bibinfo{author}{\bibfnamefont{M.}~\bibnamefont{Teichmann}},
  \bibinfo{author}{\bibfnamefont{L.}~\bibnamefont{Tarruell}},
  \bibinfo{author}{\bibfnamefont{S.~J. J. M.~F.} \bibnamefont{Kokkelmans}},
  \bibnamefont{and} \bibinfo{author}{\bibfnamefont{C.}~\bibnamefont{Salomon}},
  \bibinfo{journal}{Phys. Rev. Lett.} \textbf{\bibinfo{volume}{93}},
  \bibinfo{pages}{050401} (\bibinfo{year}{2004}),
  \urlprefix\url{http://link.aps.org/doi/10.1103/PhysRevLett.93.050401}.

\bibitem[{\citenamefont{Bartenstein et~al.}(2004)\citenamefont{Bartenstein,
  Altmeyer, Riedl, Jochim, Chin, Denschlag, and Grimm}}]{PhysRevLett.92.120401}
\bibinfo{author}{\bibfnamefont{M.}~\bibnamefont{Bartenstein}},
  \bibinfo{author}{\bibfnamefont{A.}~\bibnamefont{Altmeyer}},
  \bibinfo{author}{\bibfnamefont{S.}~\bibnamefont{Riedl}},
  \bibinfo{author}{\bibfnamefont{S.}~\bibnamefont{Jochim}},
  \bibinfo{author}{\bibfnamefont{C.}~\bibnamefont{Chin}},
  \bibinfo{author}{\bibfnamefont{J.~H.} \bibnamefont{Denschlag}},
  \bibnamefont{and} \bibinfo{author}{\bibfnamefont{R.}~\bibnamefont{Grimm}},
  \bibinfo{journal}{Phys. Rev. Lett.} \textbf{\bibinfo{volume}{92}},
  \bibinfo{pages}{120401} (\bibinfo{year}{2004}),
  \urlprefix\url{http://link.aps.org/doi/10.1103/PhysRevLett.92.120401}.

\bibitem[{\citenamefont{Kinast et~al.}(2005)\citenamefont{Kinast, Turlapov,
  Thomas, Chen, Stajic, and Levin}}]{Kinast25022005}
\bibinfo{author}{\bibfnamefont{J.}~\bibnamefont{Kinast}},
  \bibinfo{author}{\bibfnamefont{A.}~\bibnamefont{Turlapov}},
  \bibinfo{author}{\bibfnamefont{J.~E.} \bibnamefont{Thomas}},
  \bibinfo{author}{\bibfnamefont{Q.}~\bibnamefont{Chen}},
  \bibinfo{author}{\bibfnamefont{J.}~\bibnamefont{Stajic}}, \bibnamefont{and}
  \bibinfo{author}{\bibfnamefont{K.}~\bibnamefont{Levin}},
  \bibinfo{journal}{Science} \textbf{\bibinfo{volume}{307}},
  \bibinfo{pages}{1296} (\bibinfo{year}{2005}),
  \eprint{http://www.sciencemag.org/content/307/5713/1296.full.pdf},
  \urlprefix\url{http://www.sciencemag.org/content/307/5713/1296.abstract}.

\bibitem[{\citenamefont{Partridge et~al.}(2006)\citenamefont{Partridge, Li,
  Kamar, Liao, and Hulet}}]{Partridge27012006}
\bibinfo{author}{\bibfnamefont{G.~B.} \bibnamefont{Partridge}},
  \bibinfo{author}{\bibfnamefont{W.}~\bibnamefont{Li}},
  \bibinfo{author}{\bibfnamefont{R.~I.} \bibnamefont{Kamar}},
  \bibinfo{author}{\bibfnamefont{Y.-a.} \bibnamefont{Liao}}, \bibnamefont{and}
  \bibinfo{author}{\bibfnamefont{R.~G.} \bibnamefont{Hulet}},
  \bibinfo{journal}{Science} \textbf{\bibinfo{volume}{311}},
  \bibinfo{pages}{503} (\bibinfo{year}{2006}),
  \eprint{http://www.sciencemag.org/content/311/5760/503.full.pdf},
  \urlprefix\url{http://www.sciencemag.org/content/311/5760/503.abstract}.

\bibitem[{\citenamefont{Regal et~al.}(2005)\citenamefont{Regal, Greiner,
  Giorgini, Holland, and Jin}}]{PhysRevLett.95.250404}
\bibinfo{author}{\bibfnamefont{C.~A.} \bibnamefont{Regal}},
  \bibinfo{author}{\bibfnamefont{M.}~\bibnamefont{Greiner}},
  \bibinfo{author}{\bibfnamefont{S.}~\bibnamefont{Giorgini}},
  \bibinfo{author}{\bibfnamefont{M.}~\bibnamefont{Holland}}, \bibnamefont{and}
  \bibinfo{author}{\bibfnamefont{D.~S.} \bibnamefont{Jin}},
  \bibinfo{journal}{Phys. Rev. Lett.} \textbf{\bibinfo{volume}{95}},
  \bibinfo{pages}{250404} (\bibinfo{year}{2005}),
  \urlprefix\url{http://link.aps.org/doi/10.1103/PhysRevLett.95.250404}.

\bibitem[{\citenamefont{Stewart et~al.}(2006)\citenamefont{Stewart, Gaebler,
  Regal, and Jin}}]{PhysRevLett.97.220406}
\bibinfo{author}{\bibfnamefont{J.~T.} \bibnamefont{Stewart}},
  \bibinfo{author}{\bibfnamefont{J.~P.} \bibnamefont{Gaebler}},
  \bibinfo{author}{\bibfnamefont{C.~A.} \bibnamefont{Regal}}, \bibnamefont{and}
  \bibinfo{author}{\bibfnamefont{D.~S.} \bibnamefont{Jin}},
  \bibinfo{journal}{Phys. Rev. Lett.} \textbf{\bibinfo{volume}{97}},
  \bibinfo{pages}{220406} (\bibinfo{year}{2006}),
  \urlprefix\url{http://link.aps.org/doi/10.1103/PhysRevLett.97.220406}.

\bibitem[{\citenamefont{Luo and
  Thomas}(2009)}]{springerlink:10.1007/s10909-008-9850-2}
\bibinfo{author}{\bibfnamefont{L.}~\bibnamefont{Luo}} \bibnamefont{and}
  \bibinfo{author}{\bibfnamefont{J.}~\bibnamefont{Thomas}},
  \bibinfo{journal}{Journal of Low Temperature Physics}
  \textbf{\bibinfo{volume}{154}}, \bibinfo{pages}{1} (\bibinfo{year}{2009}),
  ISSN \bibinfo{issn}{0022-2291}, \bibinfo{note}{10.1007/s10909-008-9850-2},
  \urlprefix\url{http://dx.doi.org/10.1007/s10909-008-9850-2}.

\bibitem[{\citenamefont{Navon et~al.}(2010)\citenamefont{Navon, Nascimb√®ne,
  Chevy, and Salomon}}]{Navon07052010}
\bibinfo{author}{\bibfnamefont{N.}~\bibnamefont{Navon}},
  \bibinfo{author}{\bibfnamefont{S.}~\bibnamefont{Nascimb√®ne}},
  \bibinfo{author}{\bibfnamefont{F.}~\bibnamefont{Chevy}}, \bibnamefont{and}
  \bibinfo{author}{\bibfnamefont{C.}~\bibnamefont{Salomon}},
  \bibinfo{journal}{Science} \textbf{\bibinfo{volume}{328}},
  \bibinfo{pages}{729} (\bibinfo{year}{2010}),
  \eprint{http://www.sciencemag.org/content/328/5979/729.full.pdf},
  \urlprefix\url{http://www.sciencemag.org/content/328/5979/729.abstract}.

\bibitem[{\citenamefont{{Nascimb{\`e}ne}
  et~al.}(2010)\citenamefont{{Nascimb{\`e}ne}, {Navon}, {Jiang}, {Chevy}, and
  {Salomon}}}]{2010Natur.463.1057N}
\bibinfo{author}{\bibfnamefont{S.}~\bibnamefont{{Nascimb{\`e}ne}}},
  \bibinfo{author}{\bibfnamefont{N.}~\bibnamefont{{Navon}}},
  \bibinfo{author}{\bibfnamefont{K.~J.} \bibnamefont{{Jiang}}},
  \bibinfo{author}{\bibfnamefont{F.}~\bibnamefont{{Chevy}}}, \bibnamefont{and}
  \bibinfo{author}{\bibfnamefont{C.}~\bibnamefont{{Salomon}}},
  \bibinfo{journal}{\nat} \textbf{\bibinfo{volume}{463}}, \bibinfo{pages}{1057}
  (\bibinfo{year}{2010}), \eprint{0911.0747}.

\bibitem[{\citenamefont{Ku et~al.}(2012)\citenamefont{Ku, Sommer, Cheuk, and
  Zwierlein}}]{Ku12012012}
\bibinfo{author}{\bibfnamefont{M.~J.~H.} \bibnamefont{Ku}},
  \bibinfo{author}{\bibfnamefont{A.~T.} \bibnamefont{Sommer}},
  \bibinfo{author}{\bibfnamefont{L.~W.} \bibnamefont{Cheuk}}, \bibnamefont{and}
  \bibinfo{author}{\bibfnamefont{M.~W.} \bibnamefont{Zwierlein}},
  \bibinfo{journal}{Science}  (\bibinfo{year}{2012}),
  \eprint{http://www.sciencemag.org/content/early/2012/01/11/science.1214987.f%
ull.pdf},
  \urlprefix\url{http://www.sciencemag.org/content/early/2012/01/11/science.12%
14987.abstract}.

\bibitem[{\citenamefont{Engelbrecht et~al.}(1997)\citenamefont{Engelbrecht,
  Randeria, and S\'ade~Melo}}]{PhysRevB.55.15153}
\bibinfo{author}{\bibfnamefont{J.~R.} \bibnamefont{Engelbrecht}},
  \bibinfo{author}{\bibfnamefont{M.}~\bibnamefont{Randeria}}, \bibnamefont{and}
  \bibinfo{author}{\bibfnamefont{C.~A.~R.} \bibnamefont{S\'ade~Melo}},
  \bibinfo{journal}{Phys. Rev. B} \textbf{\bibinfo{volume}{55}},
  \bibinfo{pages}{15153} (\bibinfo{year}{1997}),
  \urlprefix\url{http://link.aps.org/doi/10.1103/PhysRevB.55.15153}.

\bibitem[{\citenamefont{Baker}(1999)}]{PhysRevC.60.054311}
\bibinfo{author}{\bibfnamefont{G.~A.} \bibnamefont{Baker}},
  \bibinfo{journal}{Phys. Rev. C} \textbf{\bibinfo{volume}{60}},
  \bibinfo{pages}{054311} (\bibinfo{year}{1999}),
  \urlprefix\url{http://link.aps.org/doi/10.1103/PhysRevC.60.054311}.

\bibitem[{\citenamefont{{Steele}}(2000)}]{2000nucl.th..10066S}
\bibinfo{author}{\bibfnamefont{J.~V.} \bibnamefont{{Steele}}},
  \bibinfo{journal}{ArXiv Nuclear Theory e-prints}  (\bibinfo{year}{2000}),
  \eprint{arXiv:nucl-th/0010066}.

\bibitem[{\citenamefont{Heiselberg}(2001)}]{PhysRevA.63.043606}
\bibinfo{author}{\bibfnamefont{H.}~\bibnamefont{Heiselberg}},
  \bibinfo{journal}{Phys. Rev. A} \textbf{\bibinfo{volume}{63}},
  \bibinfo{pages}{043606} (\bibinfo{year}{2001}),
  \urlprefix\url{http://link.aps.org/doi/10.1103/PhysRevA.63.043606}.

\bibitem[{\citenamefont{Perali et~al.}(2004)\citenamefont{Perali, Pieri, and
  Strinati}}]{PhysRevLett.93.100404}
\bibinfo{author}{\bibfnamefont{A.}~\bibnamefont{Perali}},
  \bibinfo{author}{\bibfnamefont{P.}~\bibnamefont{Pieri}}, \bibnamefont{and}
  \bibinfo{author}{\bibfnamefont{G.~C.} \bibnamefont{Strinati}},
  \bibinfo{journal}{Phys. Rev. Lett.} \textbf{\bibinfo{volume}{93}},
  \bibinfo{pages}{100404} (\bibinfo{year}{2004}),
  \urlprefix\url{http://link.aps.org/doi/10.1103/PhysRevLett.93.100404}.

\bibitem[{\citenamefont{{Sch{\"a}fer} et~al.}(2005)\citenamefont{{Sch{\"a}fer},
  {Kao}, and {Cotanch}}}]{2005NuPhA.762...82S}
\bibinfo{author}{\bibfnamefont{T.}~\bibnamefont{{Sch{\"a}fer}}},
  \bibinfo{author}{\bibfnamefont{C.-W.} \bibnamefont{{Kao}}}, \bibnamefont{and}
  \bibinfo{author}{\bibfnamefont{S.~R.} \bibnamefont{{Cotanch}}},
  \bibinfo{journal}{Nuclear Physics A} \textbf{\bibinfo{volume}{762}},
  \bibinfo{pages}{82} (\bibinfo{year}{2005}), \eprint{arXiv:nucl-th/0504088}.

\bibitem[{\citenamefont{Papenbrock}(2005)}]{PhysRevA.72.041603}
\bibinfo{author}{\bibfnamefont{T.}~\bibnamefont{Papenbrock}},
  \bibinfo{journal}{Phys. Rev. A} \textbf{\bibinfo{volume}{72}},
  \bibinfo{pages}{041603} (\bibinfo{year}{2005}),
  \urlprefix\url{http://link.aps.org/doi/10.1103/PhysRevA.72.041603}.

\bibitem[{\citenamefont{Nishida and Son}(2006)}]{PhysRevLett.97.050403}
\bibinfo{author}{\bibfnamefont{Y.}~\bibnamefont{Nishida}} \bibnamefont{and}
  \bibinfo{author}{\bibfnamefont{D.~T.} \bibnamefont{Son}},
  \bibinfo{journal}{Phys. Rev. Lett.} \textbf{\bibinfo{volume}{97}},
  \bibinfo{pages}{050403} (\bibinfo{year}{2006}),
  \urlprefix\url{http://link.aps.org/doi/10.1103/PhysRevLett.97.050403}.

\bibitem[{\citenamefont{Haussmann et~al.}(2007)\citenamefont{Haussmann,
  Rantner, Cerrito, and Zwerger}}]{PhysRevA.75.023610}
\bibinfo{author}{\bibfnamefont{R.}~\bibnamefont{Haussmann}},
  \bibinfo{author}{\bibfnamefont{W.}~\bibnamefont{Rantner}},
  \bibinfo{author}{\bibfnamefont{S.}~\bibnamefont{Cerrito}}, \bibnamefont{and}
  \bibinfo{author}{\bibfnamefont{W.}~\bibnamefont{Zwerger}},
  \bibinfo{journal}{Phys. Rev. A} \textbf{\bibinfo{volume}{75}},
  \bibinfo{pages}{023610} (\bibinfo{year}{2007}),
  \urlprefix\url{http://link.aps.org/doi/10.1103/PhysRevA.75.023610}.

\bibitem[{\citenamefont{Veillette et~al.}(2007)\citenamefont{Veillette, Sheehy,
  and Radzihovsky}}]{PhysRevA.75.043614}
\bibinfo{author}{\bibfnamefont{M.~Y.} \bibnamefont{Veillette}},
  \bibinfo{author}{\bibfnamefont{D.~E.} \bibnamefont{Sheehy}},
  \bibnamefont{and}
  \bibinfo{author}{\bibfnamefont{L.}~\bibnamefont{Radzihovsky}},
  \bibinfo{journal}{Phys. Rev. A} \textbf{\bibinfo{volume}{75}},
  \bibinfo{pages}{043614} (\bibinfo{year}{2007}),
  \urlprefix\url{http://link.aps.org/doi/10.1103/PhysRevA.75.043614}.

\bibitem[{\citenamefont{Arnold et~al.}(2007)\citenamefont{Arnold, Drut, and
  Son}}]{PhysRevA.75.043605}
\bibinfo{author}{\bibfnamefont{P.}~\bibnamefont{Arnold}},
  \bibinfo{author}{\bibfnamefont{J.~E.} \bibnamefont{Drut}}, \bibnamefont{and}
  \bibinfo{author}{\bibfnamefont{D.~T.} \bibnamefont{Son}},
  \bibinfo{journal}{Phys. Rev. A} \textbf{\bibinfo{volume}{75}},
  \bibinfo{pages}{043605} (\bibinfo{year}{2007}),
  \urlprefix\url{http://link.aps.org/doi/10.1103/PhysRevA.75.043605}.

\bibitem[{\citenamefont{Nishida and Son}(2007)}]{PhysRevA.75.063617}
\bibinfo{author}{\bibfnamefont{Y.}~\bibnamefont{Nishida}} \bibnamefont{and}
  \bibinfo{author}{\bibfnamefont{D.~T.} \bibnamefont{Son}},
  \bibinfo{journal}{Phys. Rev. A} \textbf{\bibinfo{volume}{75}},
  \bibinfo{pages}{063617} (\bibinfo{year}{2007}),
  \urlprefix\url{http://link.aps.org/doi/10.1103/PhysRevA.75.063617}.

\bibitem[{\citenamefont{Ji-Sheng}(2007)}]{0256-307X-24-7-011}
\bibinfo{author}{\bibfnamefont{C.}~\bibnamefont{Ji-Sheng}},
  \bibinfo{journal}{Chinese Physics Letters} \textbf{\bibinfo{volume}{24}},
  \bibinfo{pages}{1825} (\bibinfo{year}{2007}),
  \urlprefix\url{http://stacks.iop.org/0256-307X/24/i=7/a=011}.

\bibitem[{\citenamefont{Nishida}(2009)}]{PhysRevA.79.013627}
\bibinfo{author}{\bibfnamefont{Y.}~\bibnamefont{Nishida}},
  \bibinfo{journal}{Phys. Rev. A} \textbf{\bibinfo{volume}{79}},
  \bibinfo{pages}{013627} (\bibinfo{year}{2009}),
  \urlprefix\url{http://link.aps.org/doi/10.1103/PhysRevA.79.013627}.

\bibitem[{\citenamefont{Carlson et~al.}(2003)\citenamefont{Carlson, Chang,
  Pandharipande, and Schmidt}}]{PhysRevLett.91.050401}
\bibinfo{author}{\bibfnamefont{J.}~\bibnamefont{Carlson}},
  \bibinfo{author}{\bibfnamefont{S.-Y.} \bibnamefont{Chang}},
  \bibinfo{author}{\bibfnamefont{V.~R.} \bibnamefont{Pandharipande}},
  \bibnamefont{and} \bibinfo{author}{\bibfnamefont{K.~E.}
  \bibnamefont{Schmidt}}, \bibinfo{journal}{Phys. Rev. Lett.}
  \textbf{\bibinfo{volume}{91}}, \bibinfo{pages}{050401}
  (\bibinfo{year}{2003}),
  \urlprefix\url{http://link.aps.org/doi/10.1103/PhysRevLett.91.050401}.

\bibitem[{\citenamefont{Chang et~al.}(2004)\citenamefont{Chang, Pandharipande,
  Carlson, and Schmidt}}]{PhysRevA.70.043602}
\bibinfo{author}{\bibfnamefont{S.~Y.} \bibnamefont{Chang}},
  \bibinfo{author}{\bibfnamefont{V.~R.} \bibnamefont{Pandharipande}},
  \bibinfo{author}{\bibfnamefont{J.}~\bibnamefont{Carlson}}, \bibnamefont{and}
  \bibinfo{author}{\bibfnamefont{K.~E.} \bibnamefont{Schmidt}},
  \bibinfo{journal}{Phys. Rev. A} \textbf{\bibinfo{volume}{70}},
  \bibinfo{pages}{043602} (\bibinfo{year}{2004}),
  \urlprefix\url{http://link.aps.org/doi/10.1103/PhysRevA.70.043602}.

\bibitem[{\citenamefont{Astrakharchik et~al.}(2004)\citenamefont{Astrakharchik,
  Boronat, Casulleras, and Giorgini}}]{PhysRevLett.93.200404}
\bibinfo{author}{\bibfnamefont{G.~E.} \bibnamefont{Astrakharchik}},
  \bibinfo{author}{\bibfnamefont{J.}~\bibnamefont{Boronat}},
  \bibinfo{author}{\bibfnamefont{J.}~\bibnamefont{Casulleras}},
  \bibnamefont{and} \bibinfo{author}{\bibfnamefont{S.}~\bibnamefont{Giorgini}},
  \bibinfo{journal}{Phys. Rev. Lett.} \textbf{\bibinfo{volume}{93}},
  \bibinfo{pages}{200404} (\bibinfo{year}{2004}),
  \urlprefix\url{http://link.aps.org/doi/10.1103/PhysRevLett.93.200404}.

\bibitem[{\citenamefont{Carlson and Reddy}(2005)}]{PhysRevLett.95.060401}
\bibinfo{author}{\bibfnamefont{J.}~\bibnamefont{Carlson}} \bibnamefont{and}
  \bibinfo{author}{\bibfnamefont{S.}~\bibnamefont{Reddy}},
  \bibinfo{journal}{Phys. Rev. Lett.} \textbf{\bibinfo{volume}{95}},
  \bibinfo{pages}{060401} (\bibinfo{year}{2005}),
  \urlprefix\url{http://link.aps.org/doi/10.1103/PhysRevLett.95.060401}.

\bibitem[{\citenamefont{Lee and Sch\"afer}(2006)}]{PhysRevC.73.015202}
\bibinfo{author}{\bibfnamefont{D.}~\bibnamefont{Lee}} \bibnamefont{and}
  \bibinfo{author}{\bibfnamefont{T.}~\bibnamefont{Sch\"afer}},
  \bibinfo{journal}{Phys. Rev. C} \textbf{\bibinfo{volume}{73}},
  \bibinfo{pages}{015202} (\bibinfo{year}{2006}),
  \urlprefix\url{http://link.aps.org/doi/10.1103/PhysRevC.73.015202}.

\bibitem[{\citenamefont{Bulgac et~al.}(2006)\citenamefont{Bulgac, Drut, and
  Magierski}}]{PhysRevLett.96.090404}
\bibinfo{author}{\bibfnamefont{A.}~\bibnamefont{Bulgac}},
  \bibinfo{author}{\bibfnamefont{J.~E.} \bibnamefont{Drut}}, \bibnamefont{and}
  \bibinfo{author}{\bibfnamefont{P.}~\bibnamefont{Magierski}},
  \bibinfo{journal}{Phys. Rev. Lett.} \textbf{\bibinfo{volume}{96}},
  \bibinfo{pages}{090404} (\bibinfo{year}{2006}),
  \urlprefix\url{http://link.aps.org/doi/10.1103/PhysRevLett.96.090404}.

\bibitem[{\citenamefont{Lee}(2006)}]{PhysRevB.73.115112}
\bibinfo{author}{\bibfnamefont{D.}~\bibnamefont{Lee}}, \bibinfo{journal}{Phys.
  Rev. B} \textbf{\bibinfo{volume}{73}}, \bibinfo{pages}{115112}
  (\bibinfo{year}{2006}),
  \urlprefix\url{http://link.aps.org/doi/10.1103/PhysRevB.73.115112}.

\bibitem[{\citenamefont{Juillet}(2007)}]{1367-2630-9-6-163}
\bibinfo{author}{\bibfnamefont{O.}~\bibnamefont{Juillet}},
  \bibinfo{journal}{New Journal of Physics} \textbf{\bibinfo{volume}{9}},
  \bibinfo{pages}{163} (\bibinfo{year}{2007}),
  \urlprefix\url{http://stacks.iop.org/1367-2630/9/i=6/a=163}.

\bibitem[{\citenamefont{Lee}(2008{\natexlab{a}})}]{springerlink:10.1140/epja/i%
2008-10537-2}
\bibinfo{author}{\bibfnamefont{D.}~\bibnamefont{Lee}}, \bibinfo{journal}{The
  European Physical Journal A - Hadrons and Nuclei}
  \textbf{\bibinfo{volume}{35}}, \bibinfo{pages}{171}
  (\bibinfo{year}{2008}{\natexlab{a}}), ISSN \bibinfo{issn}{1434-6001},
  \bibinfo{note}{10.1140/epja/i2008-10537-2},
  \urlprefix\url{http://dx.doi.org/10.1140/epja/i2008-10537-2}.

\bibitem[{\citenamefont{Lee}(2008{\natexlab{b}})}]{PhysRevC.78.024001}
\bibinfo{author}{\bibfnamefont{D.}~\bibnamefont{Lee}}, \bibinfo{journal}{Phys.
  Rev. C} \textbf{\bibinfo{volume}{78}}, \bibinfo{pages}{024001}
  (\bibinfo{year}{2008}{\natexlab{b}}),
  \urlprefix\url{http://link.aps.org/doi/10.1103/PhysRevC.78.024001}.

\bibitem[{\citenamefont{Bulgac et~al.}(2008)\citenamefont{Bulgac, Drut, and
  Magierski}}]{PhysRevA.78.023625}
\bibinfo{author}{\bibfnamefont{A.}~\bibnamefont{Bulgac}},
  \bibinfo{author}{\bibfnamefont{J.~E.} \bibnamefont{Drut}}, \bibnamefont{and}
  \bibinfo{author}{\bibfnamefont{P.}~\bibnamefont{Magierski}},
  \bibinfo{journal}{Phys. Rev. A} \textbf{\bibinfo{volume}{78}},
  \bibinfo{pages}{023625} (\bibinfo{year}{2008}),
  \urlprefix\url{http://link.aps.org/doi/10.1103/PhysRevA.78.023625}.

\bibitem[{\citenamefont{Abe and Seki}(2009)}]{PhysRevC.79.054003}
\bibinfo{author}{\bibfnamefont{T.}~\bibnamefont{Abe}} \bibnamefont{and}
  \bibinfo{author}{\bibfnamefont{R.}~\bibnamefont{Seki}},
  \bibinfo{journal}{Phys. Rev. C} \textbf{\bibinfo{volume}{79}},
  \bibinfo{pages}{054003} (\bibinfo{year}{2009}),
  \urlprefix\url{http://link.aps.org/doi/10.1103/PhysRevC.79.054003}.

\bibitem[{\citenamefont{Magierski et~al.}(2009)\citenamefont{Magierski,
  Wlaz\l{}owski, Bulgac, and Drut}}]{PhysRevLett.103.210403}
\bibinfo{author}{\bibfnamefont{P.}~\bibnamefont{Magierski}},
  \bibinfo{author}{\bibfnamefont{G.}~\bibnamefont{Wlaz\l{}owski}},
  \bibinfo{author}{\bibfnamefont{A.}~\bibnamefont{Bulgac}}, \bibnamefont{and}
  \bibinfo{author}{\bibfnamefont{J.~E.} \bibnamefont{Drut}},
  \bibinfo{journal}{Phys. Rev. Lett.} \textbf{\bibinfo{volume}{103}},
  \bibinfo{pages}{210403} (\bibinfo{year}{2009}),
  \urlprefix\url{http://link.aps.org/doi/10.1103/PhysRevLett.103.210403}.

\bibitem[{\citenamefont{Gandolfi et~al.}(2011)\citenamefont{Gandolfi, Schmidt,
  and Carlson}}]{PhysRevA.83.041601}
\bibinfo{author}{\bibfnamefont{S.}~\bibnamefont{Gandolfi}},
  \bibinfo{author}{\bibfnamefont{K.~E.} \bibnamefont{Schmidt}},
  \bibnamefont{and} \bibinfo{author}{\bibfnamefont{J.}~\bibnamefont{Carlson}},
  \bibinfo{journal}{Phys. Rev. A} \textbf{\bibinfo{volume}{83}},
  \bibinfo{pages}{041601} (\bibinfo{year}{2011}),
  \urlprefix\url{http://link.aps.org/doi/10.1103/PhysRevA.83.041601}.

\bibitem[{\citenamefont{Forbes et~al.}(2011)\citenamefont{Forbes, Gandolfi, and
  Gezerlis}}]{PhysRevLett.106.235303}
\bibinfo{author}{\bibfnamefont{M.~M.} \bibnamefont{Forbes}},
  \bibinfo{author}{\bibfnamefont{S.}~\bibnamefont{Gandolfi}}, \bibnamefont{and}
  \bibinfo{author}{\bibfnamefont{A.}~\bibnamefont{Gezerlis}},
  \bibinfo{journal}{Phys. Rev. Lett.} \textbf{\bibinfo{volume}{106}},
  \bibinfo{pages}{235303} (\bibinfo{year}{2011}),
  \urlprefix\url{http://link.aps.org/doi/10.1103/PhysRevLett.106.235303}.

\bibitem[{\citenamefont{Li et~al.}(2011)\citenamefont{Li,
  Koloren\ifmmode~\check{c}\else \v{c}\fi{}, and Mitas}}]{PhysRevA.84.023615}
\bibinfo{author}{\bibfnamefont{X.}~\bibnamefont{Li}},
  \bibinfo{author}{\bibfnamefont{J.~c.~v.}
  \bibnamefont{Koloren\ifmmode~\check{c}\else \v{c}\fi{}}}, \bibnamefont{and}
  \bibinfo{author}{\bibfnamefont{L.}~\bibnamefont{Mitas}},
  \bibinfo{journal}{Phys. Rev. A} \textbf{\bibinfo{volume}{84}},
  \bibinfo{pages}{023615} (\bibinfo{year}{2011}),
  \urlprefix\url{http://link.aps.org/doi/10.1103/PhysRevA.84.023615}.

\bibitem[{\citenamefont{Carlson et~al.}(2011)\citenamefont{Carlson, Gandolfi,
  Schmidt, and Zhang}}]{PhysRevA.84.061602}
\bibinfo{author}{\bibfnamefont{J.}~\bibnamefont{Carlson}},
  \bibinfo{author}{\bibfnamefont{S.}~\bibnamefont{Gandolfi}},
  \bibinfo{author}{\bibfnamefont{K.~E.} \bibnamefont{Schmidt}},
  \bibnamefont{and} \bibinfo{author}{\bibfnamefont{S.}~\bibnamefont{Zhang}},
  \bibinfo{journal}{Phys. Rev. A} \textbf{\bibinfo{volume}{84}},
  \bibinfo{pages}{061602} (\bibinfo{year}{2011}),
  \urlprefix\url{http://link.aps.org/doi/10.1103/PhysRevA.84.061602}.

\bibitem[{\citenamefont{Kaplan et~al.}(1998{\natexlab{a}})\citenamefont{Kaplan,
  Savage, and Wise}}]{Kaplan:1998tg}
\bibinfo{author}{\bibfnamefont{D.~B.} \bibnamefont{Kaplan}},
  \bibinfo{author}{\bibfnamefont{M.~J.} \bibnamefont{Savage}},
  \bibnamefont{and} \bibinfo{author}{\bibfnamefont{M.~B.} \bibnamefont{Wise}},
  \bibinfo{journal}{Phys.Lett.} \textbf{\bibinfo{volume}{B424}},
  \bibinfo{pages}{390} (\bibinfo{year}{1998}{\natexlab{a}}),
  \eprint{nucl-th/9801034}.

\bibitem[{\citenamefont{Kaplan et~al.}(1998{\natexlab{b}})\citenamefont{Kaplan,
  Savage, and Wise}}]{Kaplan:1998we}
\bibinfo{author}{\bibfnamefont{D.~B.} \bibnamefont{Kaplan}},
  \bibinfo{author}{\bibfnamefont{M.~J.} \bibnamefont{Savage}},
  \bibnamefont{and} \bibinfo{author}{\bibfnamefont{M.~B.} \bibnamefont{Wise}},
  \bibinfo{journal}{Nucl.Phys.} \textbf{\bibinfo{volume}{B534}},
  \bibinfo{pages}{329} (\bibinfo{year}{1998}{\natexlab{b}}),
  \eprint{nucl-th/9802075}.

\bibitem[{\citenamefont{Chen and Kaplan}(2004)}]{Chen:2003vy}
\bibinfo{author}{\bibfnamefont{J.-W.} \bibnamefont{Chen}} \bibnamefont{and}
  \bibinfo{author}{\bibfnamefont{D.~B.} \bibnamefont{Kaplan}},
  \bibinfo{journal}{Phys.Rev.Lett.} \textbf{\bibinfo{volume}{92}},
  \bibinfo{pages}{257002} (\bibinfo{year}{2004}), \eprint{hep-lat/0308016}.

\bibitem[{\citenamefont{{Stratonovich}}(1957)}]{1957SPhD....2..416S}
\bibinfo{author}{\bibfnamefont{R.~L.} \bibnamefont{{Stratonovich}}},
  \bibinfo{journal}{Soviet Physics Doklady} \textbf{\bibinfo{volume}{2}},
  \bibinfo{pages}{416} (\bibinfo{year}{1957}).

\bibitem[{\citenamefont{Hubbard}(1959)}]{PhysRevLett.3.77}
\bibinfo{author}{\bibfnamefont{J.}~\bibnamefont{Hubbard}},
  \bibinfo{journal}{Phys. Rev. Lett.} \textbf{\bibinfo{volume}{3}},
  \bibinfo{pages}{77} (\bibinfo{year}{1959}).

\bibitem[{\citenamefont{Endres et~al.}(2011{\natexlab{a}})\citenamefont{Endres,
  Kaplan, Lee, and Nicholson}}]{Endres:2011er}
\bibinfo{author}{\bibfnamefont{M.~G.} \bibnamefont{Endres}},
  \bibinfo{author}{\bibfnamefont{D.~B.} \bibnamefont{Kaplan}},
  \bibinfo{author}{\bibfnamefont{J.-W.} \bibnamefont{Lee}}, \bibnamefont{and}
  \bibinfo{author}{\bibfnamefont{A.~N.} \bibnamefont{Nicholson}},
  \bibinfo{journal}{Phys.Rev.} \textbf{\bibinfo{volume}{A84}},
  \bibinfo{pages}{043644} (\bibinfo{year}{2011}{\natexlab{a}}),
  \eprint{1106.5725}.

\bibitem[{\citenamefont{{Chen}}(2012)}]{2011arXiv1109.5327C}
\bibinfo{author}{\bibfnamefont{Q.}~\bibnamefont{{Chen}}},
  \bibinfo{journal}{\pra} \textbf{\bibinfo{volume}{86}}, \bibinfo{eid}{023610}
  (\bibinfo{year}{2012}), \eprint{1109.5327}.

\bibitem[{\citenamefont{{Drut} et~al.}(2011)\citenamefont{{Drut}, {L{\"a}hde},
  {Wlaz{\l}owski}, and {Magierski}}}]{2011arXiv1111.5079D}
\bibinfo{author}{\bibfnamefont{J.~E.} \bibnamefont{{Drut}}},
  \bibinfo{author}{\bibfnamefont{T.~A.} \bibnamefont{{L{\"a}hde}}},
  \bibinfo{author}{\bibfnamefont{G.}~\bibnamefont{{Wlaz{\l}owski}}},
  \bibnamefont{and}
  \bibinfo{author}{\bibfnamefont{P.}~\bibnamefont{{Magierski}}},
  \bibinfo{journal}{ArXiv e-prints}  (\bibinfo{year}{2011}),
  \eprint{1111.5079}.

\bibitem[{\citenamefont{{Privitera} and {Capone}}(2012)}]{2012PhRvA..85a3640P}
\bibinfo{author}{\bibfnamefont{A.}~\bibnamefont{{Privitera}}} \bibnamefont{and}
  \bibinfo{author}{\bibfnamefont{M.}~\bibnamefont{{Capone}}},
  \bibinfo{journal}{\pra} \textbf{\bibinfo{volume}{85}}, \bibinfo{eid}{013640}
  (\bibinfo{year}{2012}), \eprint{1112.0587}.

\bibitem[{\citenamefont{Symanzik}(1983{\natexlab{a}})}]{Symanzik1983187}
\bibinfo{author}{\bibfnamefont{K.}~\bibnamefont{Symanzik}},
  \bibinfo{journal}{Nuclear Physics B} \textbf{\bibinfo{volume}{226}},
  \bibinfo{pages}{187 } (\bibinfo{year}{1983}{\natexlab{a}}), ISSN
  \bibinfo{issn}{0550-3213},
  \urlprefix\url{http://www.sciencedirect.com/science/article/pii/055032138390%
4686}.

\bibitem[{\citenamefont{Symanzik}(1983{\natexlab{b}})}]{Symanzik1983205}
\bibinfo{author}{\bibfnamefont{K.}~\bibnamefont{Symanzik}},
  \bibinfo{journal}{Nuclear Physics B} \textbf{\bibinfo{volume}{226}},
  \bibinfo{pages}{205 } (\bibinfo{year}{1983}{\natexlab{b}}), ISSN
  \bibinfo{issn}{0550-3213},
  \urlprefix\url{http://www.sciencedirect.com/science/article/pii/055032138390%
4698}.

\bibitem[{\citenamefont{{Lee} and {Sch{\"a}fer}}(2005)}]{2005PhRvC..72b4006L}
\bibinfo{author}{\bibfnamefont{D.}~\bibnamefont{{Lee}}} \bibnamefont{and}
  \bibinfo{author}{\bibfnamefont{T.}~\bibnamefont{{Sch{\"a}fer}}},
  \bibinfo{journal}{\prc} \textbf{\bibinfo{volume}{72}},
  \bibinfo{pages}{024006} (\bibinfo{year}{2005}),
  \eprint{arXiv:nucl-th/0412002}.

\bibitem[{\citenamefont{L{\"u}scher}(1986{\natexlab{a}})}]{Luscher:1985dn}
\bibinfo{author}{\bibfnamefont{M.}~\bibnamefont{L{\"u}scher}},
  \bibinfo{journal}{Commun.Math.Phys.} \textbf{\bibinfo{volume}{104}},
  \bibinfo{pages}{177} (\bibinfo{year}{1986}{\natexlab{a}}).

\bibitem[{\citenamefont{L{\"u}scher}(1986{\natexlab{b}})}]{Luscher:1986pf}
\bibinfo{author}{\bibfnamefont{M.}~\bibnamefont{L{\"u}scher}},
  \bibinfo{journal}{Commun.Math.Phys.} \textbf{\bibinfo{volume}{105}},
  \bibinfo{pages}{153} (\bibinfo{year}{1986}{\natexlab{b}}).

\bibitem[{\citenamefont{L{\"u}scher}(1991)}]{Luscher:1990ux}
\bibinfo{author}{\bibfnamefont{M.}~\bibnamefont{L{\"u}scher}},
  \bibinfo{journal}{Nucl.Phys.} \textbf{\bibinfo{volume}{B354}},
  \bibinfo{pages}{531} (\bibinfo{year}{1991}).

\bibitem[{\citenamefont{Beane et~al.}(2004)\citenamefont{Beane, Bedaque,
  Parreno, and Savage}}]{Beane:2003da}
\bibinfo{author}{\bibfnamefont{S.}~\bibnamefont{Beane}},
  \bibinfo{author}{\bibfnamefont{P.}~\bibnamefont{Bedaque}},
  \bibinfo{author}{\bibfnamefont{A.}~\bibnamefont{Parreno}}, \bibnamefont{and}
  \bibinfo{author}{\bibfnamefont{M.}~\bibnamefont{Savage}},
  \bibinfo{journal}{Phys.Lett.} \textbf{\bibinfo{volume}{B585}},
  \bibinfo{pages}{106} (\bibinfo{year}{2004}), \eprint{hep-lat/0312004}.

\bibitem[{\citenamefont{Mandula et~al.}(1983)\citenamefont{Mandula, Zweig, and
  Govaerts}}]{Mandula:1983ut}
\bibinfo{author}{\bibfnamefont{J.~E.} \bibnamefont{Mandula}},
  \bibinfo{author}{\bibfnamefont{G.}~\bibnamefont{Zweig}}, \bibnamefont{and}
  \bibinfo{author}{\bibfnamefont{J.}~\bibnamefont{Govaerts}},
  \bibinfo{journal}{Nucl. Phys.} \textbf{\bibinfo{volume}{B228}},
  \bibinfo{pages}{91} (\bibinfo{year}{1983}).

\bibitem[{\citenamefont{Luu and Savage}(2011)}]{Luu:2011ep}
\bibinfo{author}{\bibfnamefont{T.}~\bibnamefont{Luu}} \bibnamefont{and}
  \bibinfo{author}{\bibfnamefont{M.~J.} \bibnamefont{Savage}},
  \bibinfo{journal}{Phys. Rev.} \textbf{\bibinfo{volume}{D83}},
  \bibinfo{pages}{114508} (\bibinfo{year}{2011}), \eprint{1101.3347}.

\bibitem[{\citenamefont{{Pricoupenko} and
  {Castin}}(2007)}]{2007JPhA...4012863P}
\bibinfo{author}{\bibfnamefont{L.}~\bibnamefont{{Pricoupenko}}}
  \bibnamefont{and} \bibinfo{author}{\bibfnamefont{Y.}~\bibnamefont{{Castin}}},
  \bibinfo{journal}{Journal of Physics A Mathematical General}
  \textbf{\bibinfo{volume}{40}}, \bibinfo{pages}{12863} (\bibinfo{year}{2007}),
  \eprint{0705.1502}.

\bibitem[{\citenamefont{Endres et~al.}(2011{\natexlab{b}})\citenamefont{Endres,
  Kaplan, Lee, and Nicholson}}]{PhysRevLett.107.201601}
\bibinfo{author}{\bibfnamefont{M.~G.} \bibnamefont{Endres}},
  \bibinfo{author}{\bibfnamefont{D.~B.} \bibnamefont{Kaplan}},
  \bibinfo{author}{\bibfnamefont{J.-W.} \bibnamefont{Lee}}, \bibnamefont{and}
  \bibinfo{author}{\bibfnamefont{A.~N.} \bibnamefont{Nicholson}},
  \bibinfo{journal}{Phys. Rev. Lett.} \textbf{\bibinfo{volume}{107}},
  \bibinfo{pages}{201601} (\bibinfo{year}{2011}{\natexlab{b}}),
  \urlprefix\url{http://link.aps.org/doi/10.1103/PhysRevLett.107.201601}.

\bibitem[{\citenamefont{{Tan}}(2004)}]{2004cond.mat.12764T}
\bibinfo{author}{\bibfnamefont{S.}~\bibnamefont{{Tan}}},
  \bibinfo{journal}{ArXiv Condensed Matter e-prints}  (\bibinfo{year}{2004}),
  \eprint{arXiv:cond-mat/0412764}.

\bibitem[{\citenamefont{Griesshammer}(2005)}]{Griesshammer:2005ga}
\bibinfo{author}{\bibfnamefont{H.~W.} \bibnamefont{Griesshammer}},
  \bibinfo{journal}{Nucl.Phys.} \textbf{\bibinfo{volume}{A760}},
  \bibinfo{pages}{110} (\bibinfo{year}{2005}), \eprint{nucl-th/0502039}.

\bibitem[{\citenamefont{Griesshammer}(2006)}]{Griesshammer:2005sj}
\bibinfo{author}{\bibfnamefont{H.~W.} \bibnamefont{Griesshammer}},
  \bibinfo{journal}{Few Body Syst.} \textbf{\bibinfo{volume}{38}},
  \bibinfo{pages}{67} (\bibinfo{year}{2006}), \eprint{nucl-th/0511039}.

\bibitem[{\citenamefont{{Nishida} and {Son}}(2007)}]{2007PhRvD..76h6004N}
\bibinfo{author}{\bibfnamefont{Y.}~\bibnamefont{{Nishida}}} \bibnamefont{and}
  \bibinfo{author}{\bibfnamefont{D.~T.} \bibnamefont{{Son}}},
  \bibinfo{journal}{\prd} \textbf{\bibinfo{volume}{76}},
  \bibinfo{pages}{086004} (\bibinfo{year}{2007}), \eprint{0706.3746}.

\bibitem[{\citenamefont{Bour et~al.}(2011)\citenamefont{Bour, Li, Lee,
  Mei\ss{}ner, and Mitas}}]{PhysRevA.83.063619}
\bibinfo{author}{\bibfnamefont{S.}~\bibnamefont{Bour}},
  \bibinfo{author}{\bibfnamefont{X.}~\bibnamefont{Li}},
  \bibinfo{author}{\bibfnamefont{D.}~\bibnamefont{Lee}},
  \bibinfo{author}{\bibfnamefont{U.-G.} \bibnamefont{Mei\ss{}ner}},
  \bibnamefont{and} \bibinfo{author}{\bibfnamefont{L.}~\bibnamefont{Mitas}},
  \bibinfo{journal}{Phys. Rev. A} \textbf{\bibinfo{volume}{83}},
  \bibinfo{pages}{063619} (\bibinfo{year}{2011}),
  \urlprefix\url{http://link.aps.org/doi/10.1103/PhysRevA.83.063619}.

\bibitem[{\citenamefont{Drut}(2012)}]{2012arXiv1203.2565D}
\bibinfo{author}{\bibfnamefont{J.~E.} \bibnamefont{Drut}},
  \bibinfo{journal}{Phys.Rev.} \textbf{\bibinfo{volume}{A86}},
  \bibinfo{pages}{013604} (\bibinfo{year}{2012}), \eprint{1203.2565}.

\end{thebibliography}

\end{document}